\documentclass[twocolumn]{aastex6}
\usepackage{draftwatermark}
\SetWatermarkLightness{0.5}
\SetWatermarkScale{4}
\SetWatermarkAngle{45}
\usepackage{float}
\usepackage{graphicx}
\usepackage{bmpsize}
\usepackage{ulem}
\usepackage{afterpage}

\newcommand{\kms}{km$~$s$^{-1}$}

\newcommand{\mydir}[1]{{#1}}

\shorttitle{Blueshifts in modelled coronal loops}
\shortauthors{R{\'e}gnier and Walsh}

\begin{document}


\title{Red and Blueshifts in Multi-stranded Coronal Loops: \\
A New Temperature Diagnostic}


\author{S. R\'egnier\altaffilmark{1}, R. W. Walsh\altaffilmark{2} 
	%and J. Pearson\altaffilmark{1}
	}
\altaffiltext{1}{Northumbria University, Newcastle upon Tyne, NE1 8ST, UK}
\altaffiltext{2}{Jeremiah Horrocks Institute, University of Central Lancashire, Preston, Lancashire, PR1 2HE, UK}%% Mark off your abstract in the ``abstract'' environment. In the manuscript

\begin{abstract}

Based on observations from the EUV Imaging Spectrometer (EIS) on board {\em
Hinode}, the existence of a broad distribution of blue and red Dopplershift in
active region loops has been revealed; the distribution of
Dopplershifts depends on the peak temperature of formation of the observed spectral lines.
To reproduce those observations, we use a nanoflare heating model for multi-stranded coronal loops \citep{sar08,
sar09} and a set of spectral lines
covering a broad range of temperature (from 0.25 MK to 5.6 MK). We first show
that red- and blueshifts are ubiquitous in all wavelength ranges;
redshifts/downflows dominating cool spectral lines (from O {\sc v} to Si {\sc
vii}) and blueshifts/upflows dominating the hot lines (from Fe {\sc xv} to Ca
{\sc xvii}). These Dopplershifts are indicative of plasma condensation and evaporation. By computing the average Dopplershift, we derive a new temperature
diagnostic for coronal loops: the temperature at which the average Dopplershift
vanishes estimates the mean temperature of the plasma along the coronal loop and at the footpoints. To compare closely
with observations, we model dense and sparse {\em Hinode}/EIS rasters at the instrument resolution. The temperature diagnostic provides the same temperature estimates as the model whatever the type of raster or the viewing angle. To conclude, we have developed a robust temperature diagnostic to measure the plasma temperature of a coronal loop using a broad range of spectral lines.

%Even if
%the raster reproduce the global features of up and downflows along the loop, we
%show that this type of raster cannot provide information on the heating
%mechanism.
 
\end{abstract}





\section{Introduction} \label{sec:intro}

%% general introduction on loops (short)
The coronal heating problem is a long-standing issue in solar physics. It aims
at explaining the reason why the solar corona has a mean temperature above 1~MK,
while the surface of the Sun has an effective temperature of 5800~K, and also
how this high temperature can be maintained during a solar cycle. Hence the
solar coronal heating problem is to understand how the thermal energy is
continuously and uniformly transported and distributed over a large volume like
the corona. One of the favorite model is the heating by small bursts of magnetic
energy, the so-called nanoflare heating problem as first mentioned by Parker
(\citeyear{par83, par88}): this model explains that the coronal plasma can be
heated to high temperatures with a high rate of occurrences of uniformly
distributed nanoflares explaining the sustainability of the heating. Parker's
idea has been developed for coronal loops in active regions in which the main
source of magnetic energy is located, and with the wealth of observations in
different wavelength ranges, which provide strong observational constraints
\citep{car93, car94, car97, men02, car04}. By extension, the nanoflare model
refers to the heating of loops by a series of energy releases, and does not
refer to any particular mechanisms generating these bursts of energy (e.g.,
magnetic reconnection, wave mode coupling, turbulence). Even if they account for
a small fraction of the total coronal heating budget, the understanding of the
heating of coronal loops is a crucial step for solving the global coronal
heating problem: loops are well observed in a broad range of temperature bands
and thus their thermodynamical properties are well constrained leading to a
detailed study of the physical processes at play. The
state-of-the-art of this field of research has been reviewed in length by
\citet{rea10}. Despite the large number of loop observations, their nature is
still debated: single field line versus multi-stranded flux bundle
\citep[e.g.,][]{cir07}, isothermal versus multithermal \citep[e.g.,][]{sch09}.
In this paper, we define a loop as a multistranded flux bundle implying a
multithermal plasma. The loop temperature is sustained by a series of small,
short releases of energy mimicking the nanoflare model. The multi-strandedness of coronal structures has been recently revealed by high-resolution instrumentation \citep[e.g.,][]{kon10,bro13,scu14}.

%% may add here part on previous work like CArgill, Bradshaw, Spadaro, Antiochos
%{\bf Discussion about monolithic and multistranded models and the importnat physics
%involved in the different computations including EBTEL ones.}

There is a long history of modelling coronal loop using 0D/1D hydrodynamic
models, 3D mhd models, assuming a monolithic or multistranded loop, including or
not thermal conduction, considering different radiative loss functions and/or
sources of heating, and considering the ionisation equilibrium. These models
have led to a better understanding of the observed loops and their diversity.
All modelled loop have a peak of temperature at or near the apex of loop
depending on the degree of asymmetry of the loop.   

Downflows have been commonly observed in quiet-Sun and active regions since
\cite{dos76} using {\em Skylab} spectroscopic observations. The authors
concluded that, for temperatures between 70000 K and 200000 K, the plasma
responsible for the downflows was producing more emission in the UV than the
plasma responsible for upflows. However, these observations did not shed light
on the nature of those Dopplershifts (transient or persistent). Our work is
motivated by recent observations of high blueshift patches in active region
outskirts reported by \cite{del08} and \cite{bak09} lasting for several hours.
\cite{mur10} have simulated the emergence of an active region similar to the one
studied by \cite{bak11}. The authors found that the blueshifts appear in the
surrounding of the active region during the expansion phase of the emergence
process and are also owed to the presence of a coronal hole interacting with the
emerging flux. \cite{bak11} showed that the active region exhibits an
enhancement of the blueshifts in the Fe {\sc xii} line at 194~\AA\ prior to the
eruption. Hence, the increase in blueshifts has been considered as a precursor
of the Coronal Mass Ejection (CME) associated with this active region. The
complex topology of the magnetic field is also a crucial ingredient of the
scenario developed by \cite{del11} to explain blueshifts observed at the edges
of an active region. The authors have combined EUV and radio observations to
determine that the blueshifts are occurring at the interface of open and close
magnetic field lines, and are sustained by the continuous magnetic reconnection
thanks to the growth of the active region. Based on a nanoflare model, our goal is
to determine if such a model allows us to show the ubiquitous existence of
blueshifts and what are their intrinsic properties at different wavelengths
compared to the observations above mentioned. We also rely on the observations
by \cite{war11} showing that, within an active region, the amount of blue- and
redshifts is strongly dependent on the peak temperature of the spectral line
observed: cooler is the spectral line, redder is the Dopplershift distribution,
and reciprocally, hotter is the spectral line, bluer is the Dopplershift
distribution. \cite{war11} have used the capabilities of {\em Hinode}/EIS to
obtain simultaneous rasters in several EUV spectral lines covering a wide range
of coronal temperatures (from Si {\sc vii} at 0.63~MK to Fe {\sc xv} at 2.2~MK).
The authors found that the Si {\sc vii} observations of an active region are
dominated by redshifts while hotter lines are dominated by blueshifts. It is
also worth mentioning that it seems that the Fe {\sc xii} and Fe {\sc xiii}
lines correspond to a peak in the distribution of blueshifts, and hence less
blueshifts are observed at higher temperatures while blueshifts are still
dominating the distribution of Dopplershifts in the active region. Similar
observations have also been reported by \cite{del08} and \cite{tri09}. Flows in
moss have also attracted a lot of interest. The moss observed in active regions
corresponds supposedly to the emission of hot, core coronal loops at the
transition region \citep{ber99, dep99, mar00, tri10}. In terms of Dopplershifts,
the moss is dominated by redshifts for ions from C {\sc iv} to Fe {\sc xiii},
that is to say, from the transition region to the corona at 1.78 MK 
\citep{tri12, win13}.  

%% our study 
%%%%%%%%%%%%%%%%%%%%%%%%%%%%%%%%%%%%%%%%
%%%	Fig: setup loop 	%%%%%%%%
%%%%%%%%%%%%%%%%%%%%%%%%%%%%%%%%%%%%%%%%
\begin{figure}[!h]
	\centering
	\includegraphics[width=1.\linewidth]{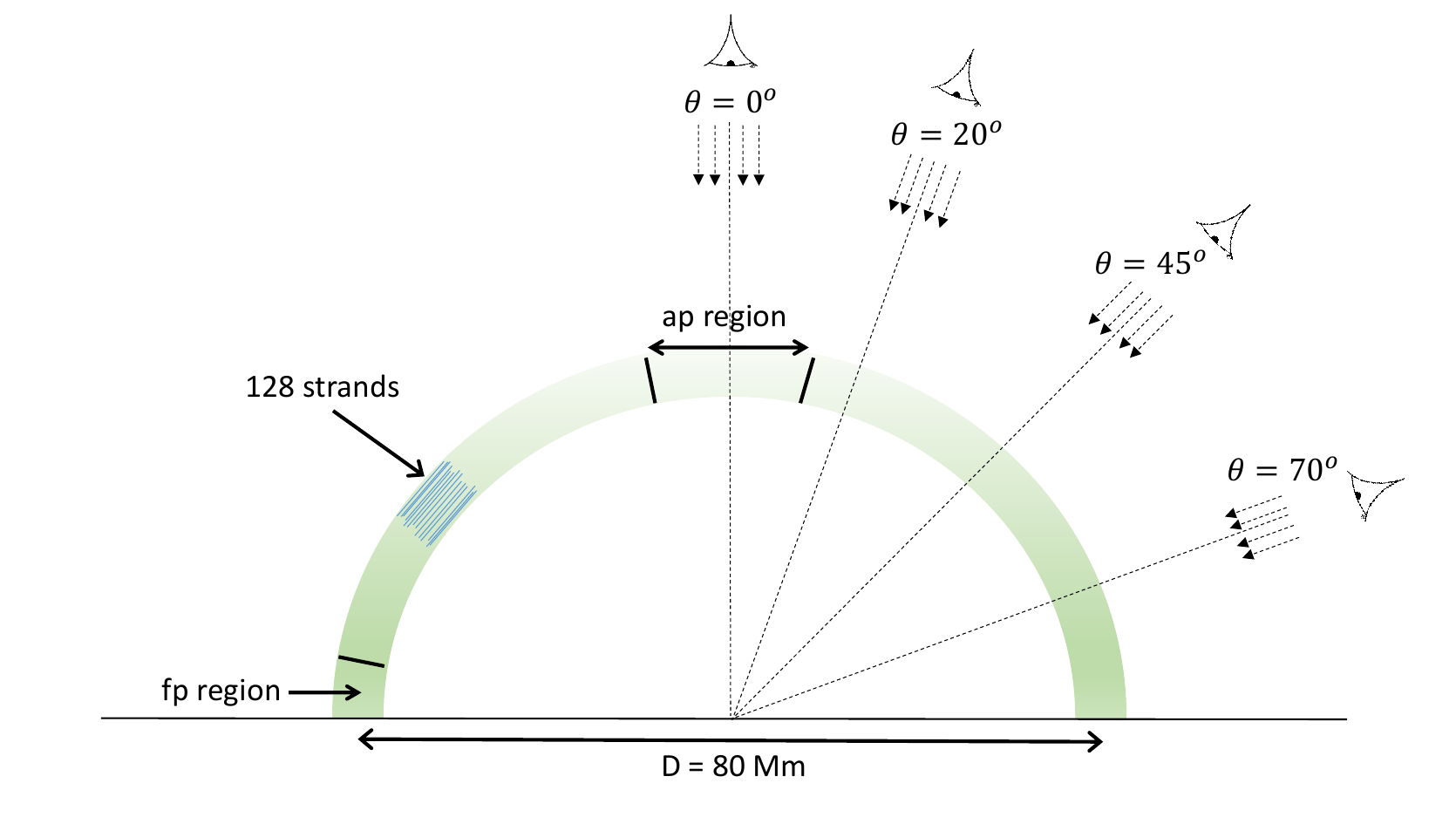}
	\caption{Setup of the multi-stranded loop as a semi-circular loop with 128 strands. The integration regions for the footpoint ($fp$) and apex ($ap$) heating is specified. Line-of-sight integration at different angles is also specified (see \ref{sec:appa}).The radiative loss function $Q(T)$ between 10000 K and 10 MK.}
	\label{fig:setuploop}
\end{figure}
%%%%%%%%%%%%%%%%%%%%%%%%%%%%%%%%%%%%%%%%%

In Section~\ref{sec:model}, we describe the nanoflare heating model used to
determine Dopplershifts in coronal loops. We describe the different
spectral lines used to simulate the observed Dopplershifts
(Section~\ref{sec:spectra}). In Section~\ref{sec:doppler}, we present the
spatial distribution of Dopplershifts for the different spectral lines. In
Section~\ref{sec:temp}, we present a new diagnostic method to estimate the
temperature of coronal loops based on multi-spectral observations. In
Section~\ref{sec:concl}, we discuss the properties of blue- and red-shifts for
{\bf two simulated} loops as seen by an imaging spectrometer rastering a region of the
Sun such as {\em Hinode}/EIS. The conclusions are drawn in
Section~\ref{sec:concl2}. In the Appendice~\ref{sec:appa},
we describe the effects of changing the observer point of view on the
measurements of Dopplershifts and the temperature diagnostic.

%%%%%%%%%%%%%%%%%%%%%%%%%%%%%%%%%%%%%%%%%%%%%%%%%%%%%%%%%%%%%%%%%%
\section{Multi-stranded model for coronal loops} \label{sec:model}
%%%%%%%%%%%%%%%%%%%%%%%%%%%%%%%%%%%%%%%%%%%%%%%%%%%%%%%%%%%%%%%%%%

	%%%%%%%%%%%%%%%%%%%%%%%%%%%%%%%
	\subsection{Setup of the Model}
	%%%%%%%%%%%%%%%%%%%%%%%%%%%%%%%

%%%%%%%%%%%%%%%%%%%%%%%%%%%%%%%%%%%%%%%%
%%%	Fig: radiative loss 	%%%%%%%%
%%%%%%%%%%%%%%%%%%%%%%%%%%%%%%%%%%%%%%%%
\begin{figure}[!h]
\centering
\includegraphics[width=\linewidth]{\mydir{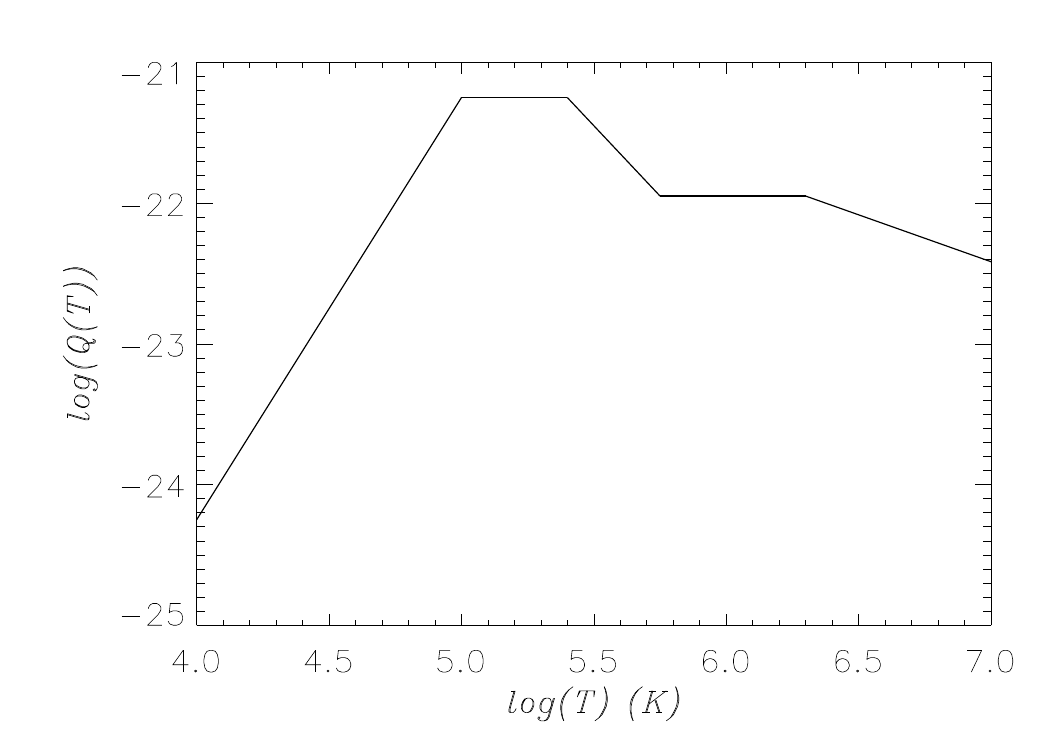}}
\caption{The radiative loss function $Q(T)$ between 10000 K and 10 MK.}
\label{fig:rad_loss}
\end{figure}
%%%%%%%%%%%%%%%%%%%%%%%%%%%%%%%%%%%%%%%%%

The corona is a highly conducting medium with a low plasma $\beta$ (ratio of
the plasma pressure to the magnetic pressure). Therefore one can assume that the plasma dynamics
occur along magnetic field lines, and that there is no or limited feedback between
the magnetic field and associated thermodynamic parameters. These particular physical
conditions justify the 1D modeling of coronal loops using the hydrodynamic equations
projected along the loop (curvilinear abscissa) and, neglecting thermal conduction across magnetic field lines. Despite the increasing number of observations of coronal loops, there is no definitive evidence that the cross section of a coronal loop increases significantly with height \citep{wat00,lop06}. We thus neglect the expansion of the loop radius with height, which should occur along a coronal loop as a consequence of the solenoidal condition and the decrease of the magnetic field strength with altitude.

A coronal loop is defined as a collection of 128 strands. Each individual strand
($i$) evolves following the time-dependent one-dimensional (1D) hydrodynamic
model satisfying mass, momentum, and energy conservation as well as the perfect
gas law:

% mass conservation
\begin{equation}
\frac{D\rho_i}{Dt} + \rho_i \frac{\partial v_i}{\partial s} = 0, 
\end{equation}
% momentum equation
\begin{equation}
\rho_i \frac{Dv_i}{Dt} = -\frac{\partial p_i}{\partial s} + \rho_i g + \rho_i \nu
\frac{\partial^2 v_i}{\partial s^2},
\end{equation}  
% energy equation
\begin{equation}
\frac{\rho_i^{\gamma}}{\gamma - 1} \frac{D}{Dt}\left( \frac{p_i}{\rho_i^{\gamma}} \right)
= \frac{\partial}{\partial s} \left( \kappa \frac{\partial T_i}{\partial s} 
\right) - n_i^2 Q(T_i) + H_i(s, t),
\end{equation}
% perfect gas law  
\begin{equation}
p_i = \frac{R}{\tilde{\mu}} \rho_i T_i,
\end{equation}  
% particle derivative
\begin{displaymath}
\textrm{with} \quad \frac{D}{Dt} \equiv \frac{\partial}{\partial t} + v_i 
	\frac{\partial}{\partial s},
\end{displaymath}
where $\rho_i$, $p_i$, and $T_i$ are the density, pressure and temperature of
the strand $i$ ($i = 1,\dots, 128$), $v_i$ is the plasma flow along the strand, $s$ is the
curvilinear abscissa along the semi-circular loop ($\gamma = 5/3$ and
$\tilde \mu = 0.6$). The number
density $n_i$ (cm$^{-3}$) is $10^{21}\,\rho_i$ (kg m$^{-3}$). The numerical code is based
on the Lagrangian remap method developed by \cite{arb01}, and further developed
for multi-stranded coronal loops by \cite{sar08,sar09}. 

%%%%%%%%%%%%%%%%%%%%%%%%%%%%%%%%%%%%%
%%%	Fig: burst prop		%%%%%
%%%%%%%%%%%%%%%%%%%%%%%%%%%%%%%%%%%%%
\begin{figure*}[!ht]
\centering
\includegraphics[width=.8\linewidth]{\mydir{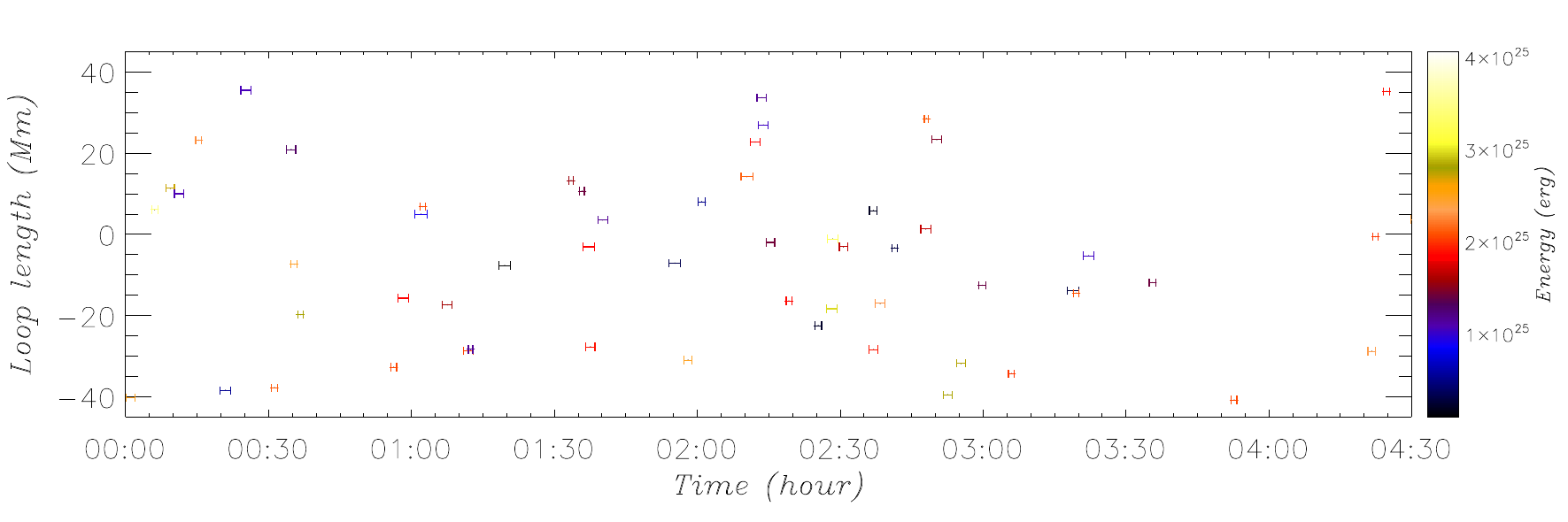}}
\caption{Distribution of bursts along a single strand (\#54) with a minimum
energy of 10$^{24}$ erg ({\it Loop~\sc{ii}}) and a uniform heating. Each line segment indicates the
duration of the burst, and the color-coding the amount of energy injected.}
\label{fig:burst_prop}
\end{figure*}
%%%%%%%%%%%%%%%%%%%%%%%%%%%%%%%%%%%%%%

The loop has a total length of 100 Mm, and a radius of 2
Mm, and results from the amalgamation of 128 individual strands. Assuming that the loop is semi-circular, the height of the loop above the solar surface is thus 32
Mm, which is less than the typical pressure scale-height ($\sim$50 Mm) of a
coronal plasma at 1 MK. This fact justifies the use of a constant gravity, $g$. The 100 Mm
loop is divided into two main areas: a transition region (width of 5 Mm) in
which the temperature is increasing rapidly and the density decreasing rapidly,
and a corona. Therefore the coronal part of the loop has a length of 90 Mm. In
addition, there is a reservoir of chromospheric plasma at each footpoint, which has a depth of 5 Mm. In combining the strands to form a loop, we assume that the radius of the loop is small compared to the length, and thus all strands have the same length of 100 Mm; in the geometry defined above, however, the inner (resp. outer) loop should be of length 94 Mm (resp. 107 Mm). The latter does not modify the physical prosses analysed in this study. We compute the thermodynamic evolution of the loop during a total time of 4 hours 30 min and taking snapshots every 1 s. The time cadence of the snapshots does not define the time-scales used within the MSHD code; the time-steps used in the Lagrangian remap code are changing with the thermal conduction and radiative losses.      

The energy equation contains three terms on the right hand side: the thermal
conduction with $\kappa(T) = \kappa_0 T^{5/2} = 9.2\times 10^{-7} T^{5/2}$ being
a function of the temperature following Spitzer formula, the radiative cooling
term where $Q(T)$ is a piecewise function (see Figure~\ref{fig:rad_loss})
adapted from \citet{coo89}, and the heating source term $H_i(s, t)$. 

The latter is defined as a distribution of successive heating bursts. Individual bursts have an energy level comparable to the energy of nanoflares (between $10^{23}$-$10^{25}$ erg). The heating source function has the following properties:
\begin{itemize}
\item[(i)]{Location: each strand is subject to 64 bursts which are randomly 
distributed in
time, location, duration, and amount of energy injected. An example of such 
random distribution is given in Figure~\ref{fig:burst_prop}. }
\item[(ii)]{Frequency: we consider a high-frequency heating mechanism. The 
heating rate is estimated to an average of 6.4~10$^{-4}$ erg cm$^{-3}$ s$^{-1}$ 
per strand and the frequency is 3.8 mHz per strand. The high frequency regime 
justifies the quick establishment of a steady state (see Section 
\ref{sec:thermo}).}
\item[(iii)]{Energy distribution: the geometry, location, and frequency of 
energy bursts are imposing an energy distribution which has power law from 
small to large energy with a negative slope of $-2.57$.}
\end{itemize}

To analyse the behaviour of Dopplershifts in multi-stranded coronal loops subject to high-frequency heating, we perform two experiments:
\begin{itemize}
\item[]{{\it Loop \sc{i}}: the minimum of energy per bursts is $10^{23}$ erg giving a total energy injected of 5.13$\times$10$^{27}$ erg;}
\item[]{{\it Loop \sc{ii}}: the minimum of energy per bursts is $10^{24}$
erg giving a total energy injected of 5.13$\times$10$^{28}$ erg.} 
\end{itemize}

\begin{figure}[!ht]
\centering
\includegraphics[width=\linewidth]{\mydir{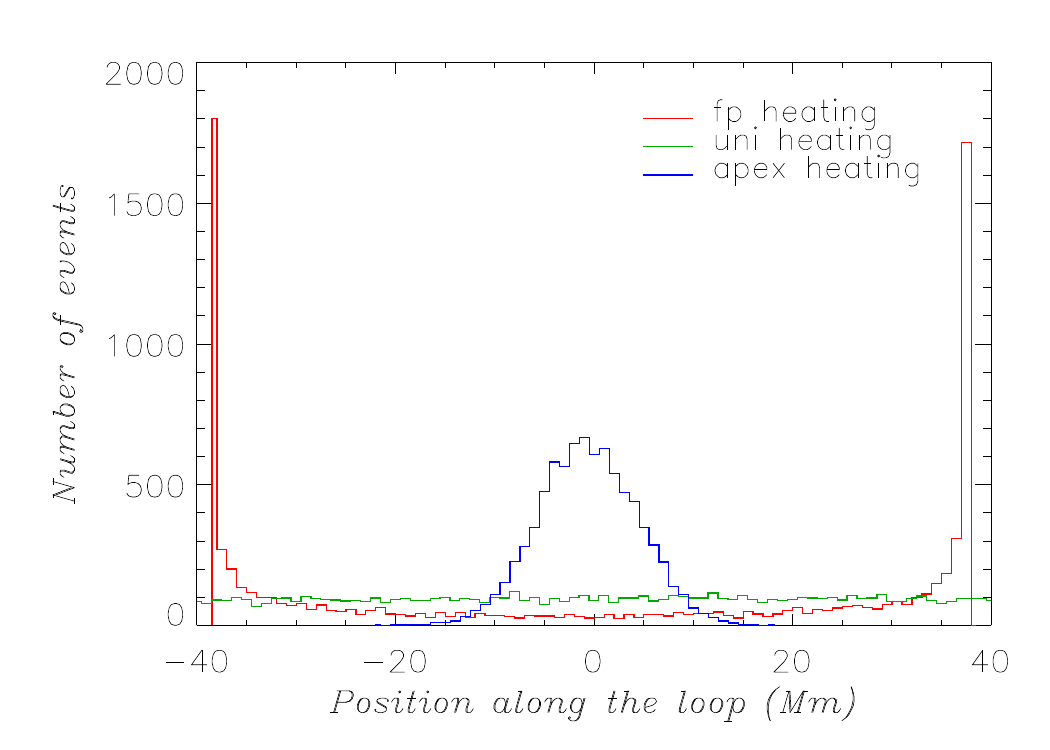}}
\caption{Distribution of the bursts along the loop for $fp$ (red), $uni$ (green)
and $ap$ (blue) heating locations.}
\label{fig:rad_burst}
\end{figure}

We also solve the 1D hydrodynamics equations for three particular energy injection location: at the footpoints ($fp$), uniformly distributed along the
strand ($uni$), and at the apex of the strand ($ap$). These three distributions are depicted in Figure~\ref{fig:rad_burst}. We use the three different distributions in Section \ref{sec:thermo} to show that our multi-stranded code is able to reproduce the well-known properties of modelled loops \cite[e.g.,][]{rea02}.

%{\bf Considering the case of the hot loop with footpoint heating, the heating rate is estimated to an average of 6.4~10$^{-4}$ erg cm$^{-3}$ s$^{-1}$ per strand and the frequency is 3.8 mHz per strand. The high frequency justifies the establishment of a steady state.}

%{\bf      , and compare with power
%laws determined by Aschwanden, include the monolithic case (single strand),
%intro for a discussion of the effects of averaging on the thermodynamic
%parameters, discussion on the viscous term in the energy equation} 

%%%%%%%%%%%%%%%%%%%%%%%%%%%%%%%%%%%%%%%%
%%%	Fig: bursts distribution     %%%
%%%%%%%%%%%%%%%%%%%%%%%%%%%%%%%%%%%%%%%%
%%%%%%%%%%%%%%%%%%%%%%%%%%%%%%%%%%%%%%%%%

	%%%%%%%%%%%%%%%%%%%%%%%%%%%%%%%%%%%%%%%%%%%%%%%%%
	\subsection{Thermodynamic Properties of the Loop} \label{sec:thermo}
	%%%%%%%%%%%%%%%%%%%%%%%%%%%%%%%%%%%%%%%%%%%%%%%%%

%%%%%%%%%%%%%%%%%%%%%%%%%%%%%%%%%%%%%
%%%	Fig: density init	%%%%%
%%%%%%%%%%%%%%%%%%%%%%%%%%%%%%%%%%%%%
\begin{figure}[!h]
\centering
\includegraphics[width=1\linewidth]{\mydir{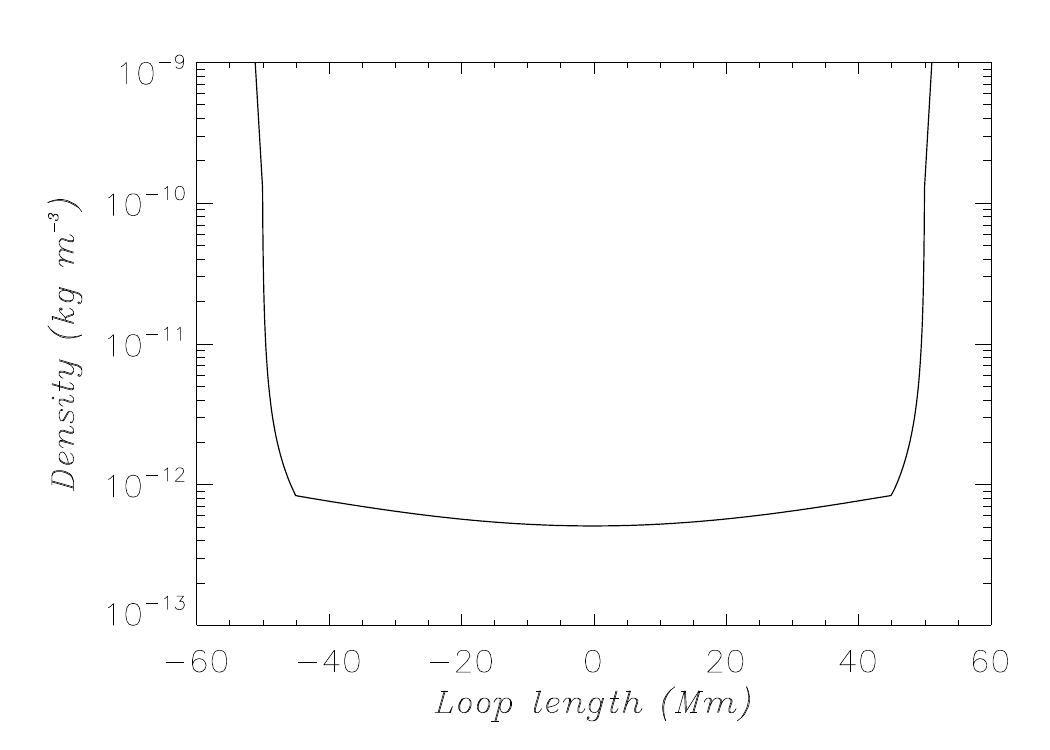}}
\caption{Initial profile of density (in kg m$^{-3}$) along the loop: the
coronal density profile is between -45 and 45 Mm.}
\label{fig:dens}
\end{figure}
%%%%%%%%%%%%%%%%%%%%%%%%%%%%%%%%%%%%%%

\begin{figure}[!h]
\centering
\begin{tabular}{c}
\includegraphics[width=0.8\linewidth]{\mydir{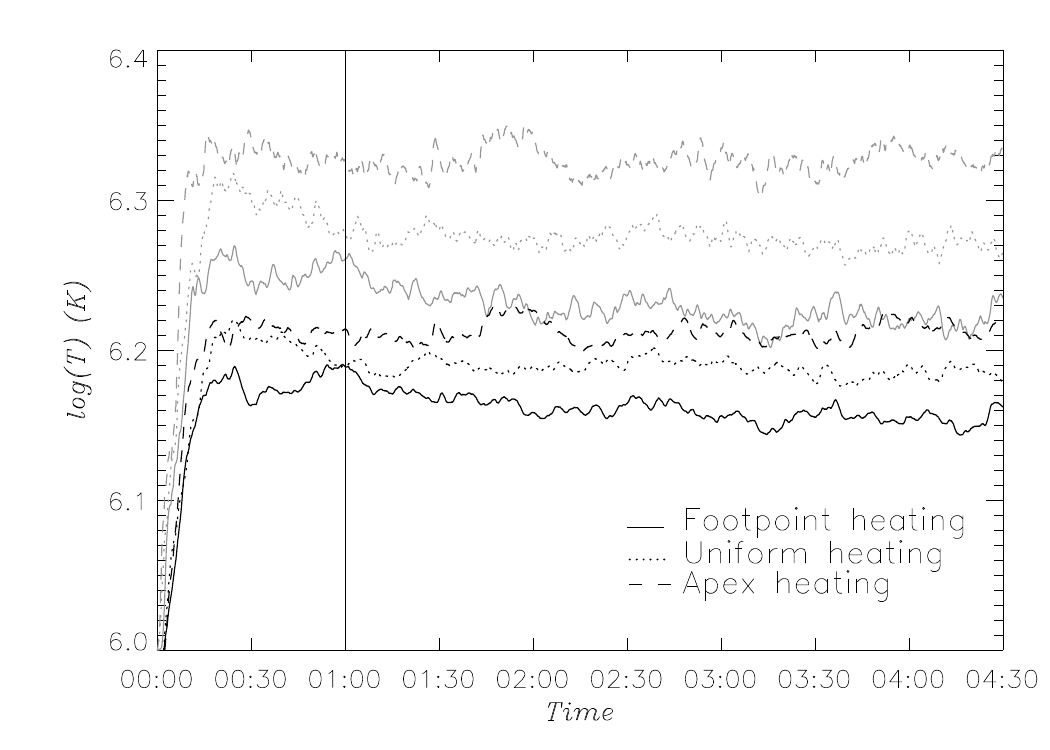}} \\
\includegraphics[width=0.8\linewidth]{\mydir{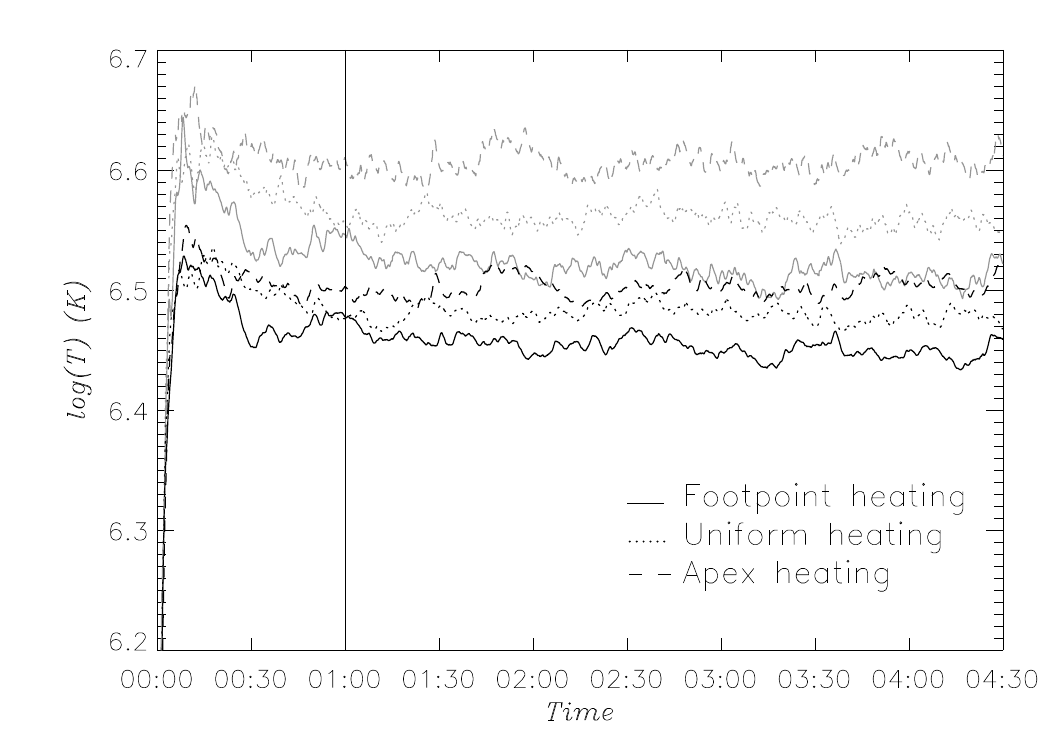}}
\end{tabular}
\caption{Characteristic temperature evolution of {\it Loop \sc{i}} ({\em top}) and {\it Loop \sc{ii}}
({\em bottom}). We distinguish between the mean temperature along the loop
(dark curves) and at the apex (light-gray curves) for the different heating
location ($fp$: solid curves; $uni$: dot curves; $ap$: dashed curves).}
\label{fig:temp}
\end{figure}

The thermodynamic parameters are computed independently for individual strands, 
and then combined to produce averaged values. The initial temperature profile for a single strand is linearly increasing from the chromospheric reservoir at about 10000 K to reach the bottom of the corona at 1 MK, and then remains a constant throughout the corona. In Figure~\ref{fig:dens}, we plot the density profile along the loop which
describe the initial density as an hydrostatic equilibrium (exponential decay
with height with a constant pressure-scale height). The thermodynamic evolution
of the loop is defined by the evolution of the density $\rho$ (sum of strand
density $\rho_i$) and the temperature $T_{EM}$. The multi-stranded loop is
characterised by an emission measure weighted  temperature \citet{sar08}:
\begin{displaymath}
T_{EM}(s, t) = \frac{\displaystyle \sum_{i=1}^{128} \rho_i^2(s,t)\,T_i(s,t)\,dl(s)}
	{\displaystyle \sum_{i=1}^{128} \rho_i^2(s,t)\,dl(s)}.
\end{displaymath}
{\em Loop \sc{i}} and {\em Loop \sc{ii}} are classified by their thermal properties: 
{\em Loop \sc{i}} (warm loop) has a mean
temperature of 1.5 MK and an apex temperature of at most 2.2 MK, and 
{\em Loop \sc{ii}} (hot loop) has a mean temperature around 3~MK and an apex 
temperature of about 4~MK. Those two experiments have been chosen to study the
similarities and differences of warm and hot loops in a steady state, and thus
to provide observational constraints. In Figure~\ref{fig:temp}, we plot the time
evolution of $T_{EM}$ for {\em Loop \sc{i}} (top) and {\em Loop \sc{ii}}
(bottom) averaged along the coronal section of the loop (solid black lines) and
at the apex (light-gray lines). All three different heating locations are also
considered for the sake of completeness. In each case,
the temperature reaches a steady state rapidly after the start of the
computation; we will consider that, after one hour, the loops have reached
a steady state (solid
vertical line in Figure~\ref{fig:temp}). We also note that the loops are filled 
by the heated plasma at a different rate: as expected, it takes a
much longer time (15$-$20 minutes) for the {\em Loop \sc{i}} to reach 
reasonable
thermodynamic values, compared to less than 10 minutes for the {\em Loop
\sc{ii}}. This is justified by the improved efficiency of thermal conduction when
the total amount of energy injected is increased ({\em Loop \sc{ii}}) 
considering the same initial equilibrium at 1 MK for both
cases. Changing the location 
of the heating sources does not
significantly modify the time taken to reach a steady state.

\begin{figure}[!h]
\centering
\begin{tabular}{c}
\includegraphics[width=0.8\linewidth]{\mydir{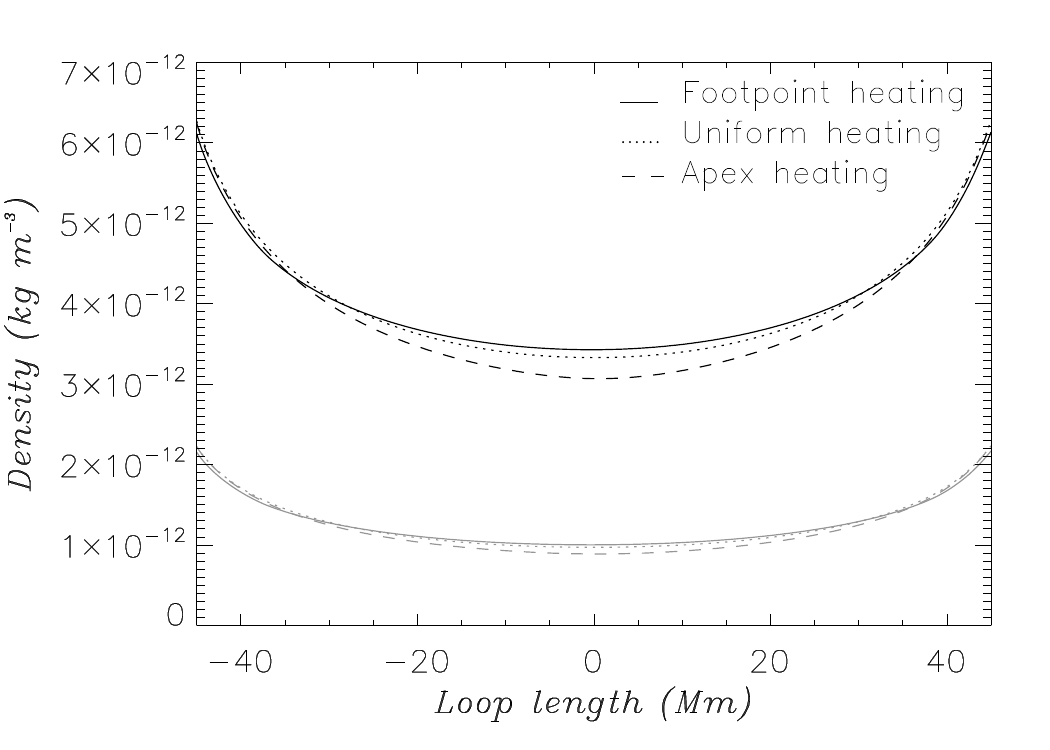}} \\
\includegraphics[width=0.8\linewidth]{\mydir{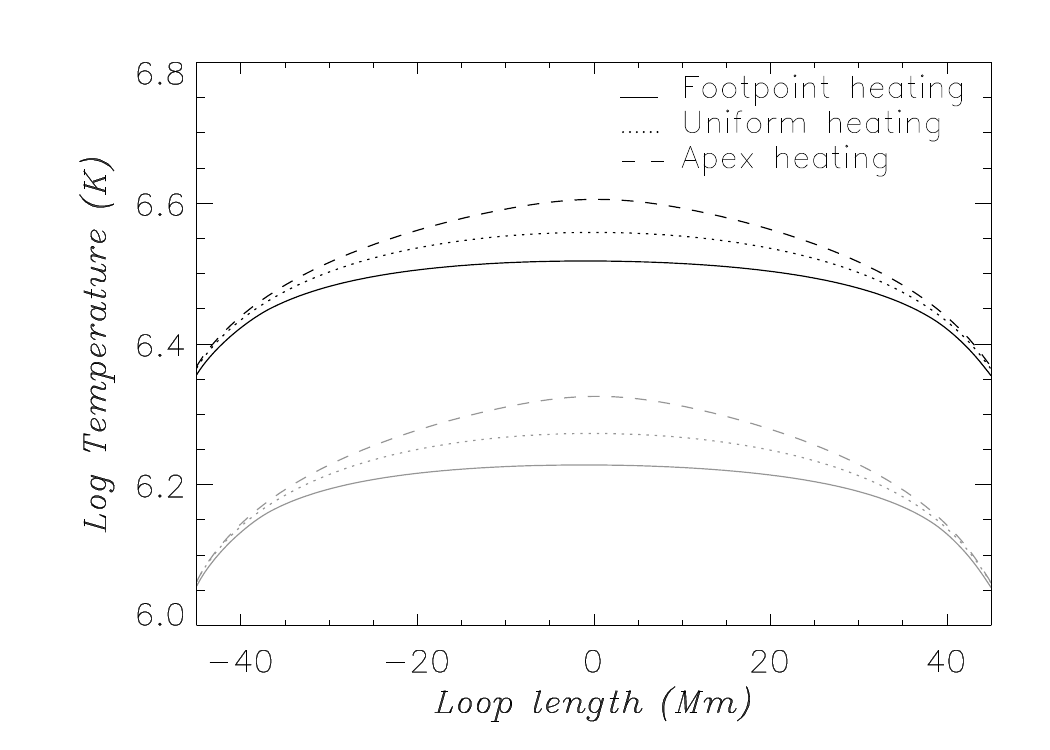}}
\end{tabular}
\caption{Average density (top) and temperature (bottom) variation along the
coronal loop: ({\em gray}) {\it Loop \sc{i}}, ({\em black}) {\it Loop \sc{ii}}. The quantities are
integrated during two hours from 3600 s of the simulation.}
\label{fig:dt}
\end{figure}

We plot the profiles of density (Figure~\ref{fig:dt} top) and temperature
(Figure~\ref{fig:dt} bottom) average over two hours for the two experiments
({\em Loop \sc{i}} in black, {\em Loop \sc{ii}} in gray) and for the different heating locations ($fp$:
solid line, $uni$: dotted line, and $ap$: dashed line). While both
profiles are fairly symmetric with respect to the apex of the loop, individual
strands exhibit asymmetric profiles owed to the randomness in the deposition of
energy. The loop symmetry is thus a consequence of the collective behaviour of the
strands and the time averaging. Keeping in mind that the thermodynamics
quantities are averaged in time, the
average density at the apex of {\em Loop \sc{ii}} decreases from 3.4$\times$10$^{-12}$ kg m$^{-3}$ for the
$fp$ heating to 3.1$\times$10$^{-12}$ kg m$^{-3}$ for the $ap$ heating, while
the average temperature increases from 3.3 MK to 4 MK. As noted by \citet{rea00} in
simulating a multi-stranded loop, the $fp$ heating can mimick a coronal loop with
a constant temperature due to the flatness of the temperature profile. The
temperature profiles are similar to those modelled by \citet{gal99} in a
numerical experiment of flux braiding. Although the change in density and
temperature is 10$-$20\%\ between the different heating locations, the
observation of coronal loops does not permit a clear identification of the
heating source \citep[see e.g.,][]{pri98,mac00,rea02}. It is also important to
notice the difference in temperature at the apex of the loop depending on the
location of the heating deposition: the apex temperature for {\em Loop \sc{ii}} 
is 3.3
MK for the $fp$ heating and 4 MK for the $ap$ heating. This difference
results from the small temperature gradients at the apex compared to the
coronal footpoints near the transition to chromospheric temperatures, and also
from the bidirectional flows generated by the energy deposition which over-heats
the loop top.

As there is no remarkable difference in the physical processes at play between the three heating locations, we only consider the $fp$ heating in the following sections. The $fp$ heating is also the most favorable location for heating the corona: the complexity (braiding/twisting/tangling) of the magnetic field which can lead to continuous nanoflare occurrence is most likely to be located near the chromosphere \citep{reg08} making magnetic reconnection and wave mode
coupling two mechanisms highly efficient in this region.

%%%%%%%%%%%%%%%%%%%%%%%%%%%%%%%%%%%%%
%%%	Fig:temperature		%%%%%
%%%%%%%%%%%%%%%%%%%%%%%%%%%%%%%%%%%%%
%%%%%%%%%%%%%%%%%%%%%%%%%%%%%%%%%%%%%%

%%%%%%%%%%%%%%%%%%%%%%%%%%%%%%%%%%%%%
%%%	Fig: density temp	%%%%%
%%%%%%%%%%%%%%%%%%%%%%%%%%%%%%%%%%%%%
%%%%%%%%%%%%%%%%%%%%%%%%%%%%%%%%%%%%%%

%%%%%%%%%%%%%%%%%%%%%%%%%%%%%%%%%%%%
%%%	Table: spectral line	%%%%
%%%%%%%%%%%%%%%%%%%%%%%%%%%%%%%%%%%%
\begin{table}[!h]
\begin{center}
\caption{Characteristic parameters of the selected spectral lines}
\label{tab:lines}
\begin{tabular}{ccc}
\tableline\tableline \\[-0.2cm]
Spectral & Wavelength & $T_e$  \\
Line & (\AA) & log(K) \\[0.1cm]
\tableline \\[-0.2cm]
O {\sc v} & 248.46 & 5.4  \\[0.1cm]
Mg {\sc v} & 276.58 & 5.5 \\[0.1cm]
Si {\sc vii} & 275.36 & 5.8 \\[0.1cm]
Fe {\sc x} & 184.54 & 6.05 \\[0.1cm]
Fe {\sc xii} & 195.12 & 6.2 \\[0.1cm]
Fe {\sc xiii} & 202.24 & 6.25 \\[0.1cm]
Fe {\sc xv} & 284.16 & 6.35 \\[0.1cm]
Fe {\sc xvi} & 262.98 & 6.4 \\[0.1cm]
Ca {\sc xv} & 200.97 & 6.65 \\[0.1cm]
Ca {\sc xvii} & 192.85 & 6.75 \\[0.1cm]
\tableline
\end{tabular}
\end{center}
\end{table}
%%%%%%%%%%%%%%%%%%%%%%%%%%%%%%%%%%%%%

%%%%%%%%%%%%%%%%%%%%%%%%%%%%%%%%%%%%%%%%%%%
%%%	 Fig: contribution functions	%%%
%%%%%%%%%%%%%%%%%%%%%%%%%%%%%%%%%%%%%%%%%%%
\begin{figure}[!h]
\centering
\includegraphics[width=\linewidth]{\mydir{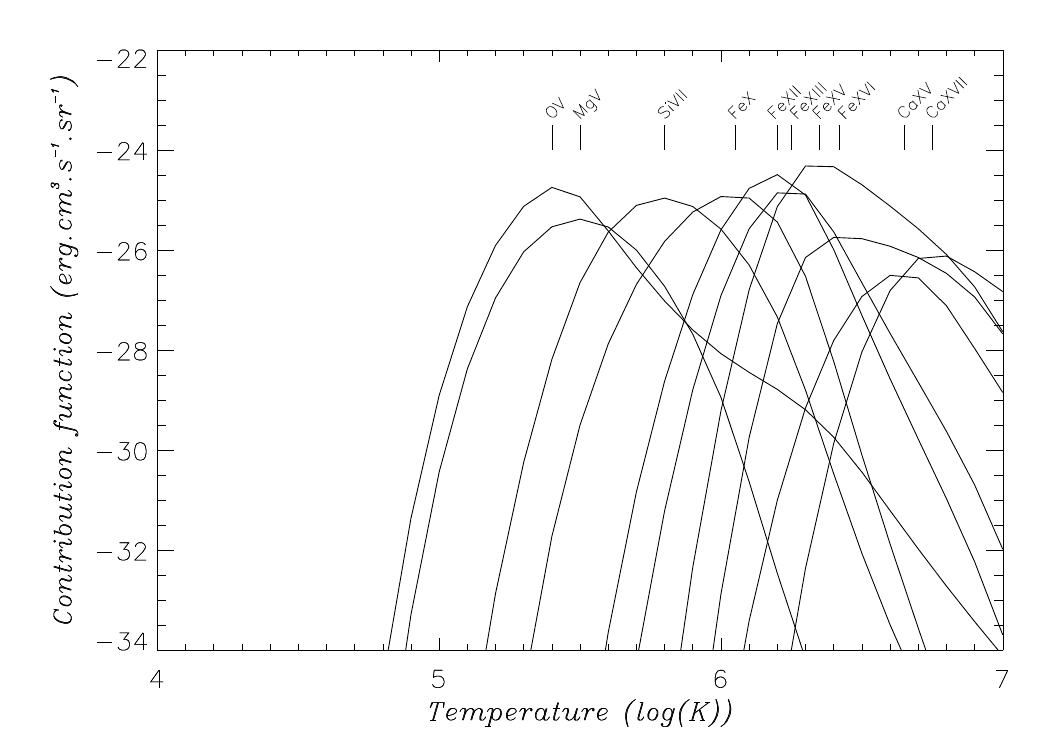}}
\caption{Contribution functions $C_{sp}$ (logarithmic scale) for the EUV spectral lines considered
in this paper. The temperature range is the same as for the radiative loss
function (see Figure~\ref{fig:rad_loss})}
\label{fig:contrib}
\end{figure}
%%%%%%%%%%%%%%%%%%%%%%%%%%%%%%%%%%%%%%%%%%%%
%%%%%%%%%%%%%%%%%%%%%%%%%%%%%%%%%%%%%%%%%%%
\section{Modelling Synthetic Observations}
%%%%%%%%%%%%%%%%%%%%%%%%%%%%%%%%%%%%%%%%%%%

	%%%%%%%%%%%%%%%%%%%%%%%%%%%%%%%%%%%%%%%%%%%%%%%%
	\subsection{Spectral Lines} \label{sec:spectra}
	%%%%%%%%%%%%%%%%%%%%%%%%%%%%%%%%%%%%%%%%%%%%%%%%
	
In order to compare the behavior of the simulated Dopplershifts with those
observed, ten spectral lines with a peak emission temperature
$T_e$ between 250\,000 K and 5.6 MK are selected (see Table~\ref{tab:lines}) along with
the following criteria:
\begin{itemize}
\item[-]{the spectral lines are referenced in \cite{war11};}
\item[-]{the spectral lines are also on the list of {\em Hinode}/EIS spectral
lines suggested by \cite{you07} to ensure that the selected lines
are not or only weakly blended: for instance, we do not consider the Fe {\sc
xi} line at 188.21~\AA\ studied by \cite{war11} due to its complex blend. Some
properties of these spectral lines observed in active regions or quiet-Sun
regions can be found in \cite{bro08};}
\item[-]{we extend the thermal coverage of the spectral lines, in particular, by
adding two relatively cool lines (O {\sc v} and Mg {\sc v}) and two hot lines
(Ca {\sc xv} and Ca {\sc xvii}). }
\end{itemize}

Relying on the CHIANTI 6.0 database \citep{der09}, the contribution functions
$C_{sp}$ for each transition are depicted in Figure~\ref{fig:contrib}.

	%%%%%%%%%%%%%%%%%%%%%%%%%%%%%%%%%%%%%%%%%%%%%%%%%%%%%%%%%
	\subsection{Dopplershift Measurement} \label{sec:doppler}
	%%%%%%%%%%%%%%%%%%%%%%%%%%%%%%%%%%%%%%%%%%%%%%%%%%%%%%%%%

%%%%%%%%%%%%%%%%%%%%%%%%%%%%%%%%%%%%%%%%%%%
%%%	Fig: warm loop Dopplershift	%%%
%%%%%%%%%%%%%%%%%%%%%%%%%%%%%%%%%%%%%%%%%%%

%\afterpage{
%\clearpage
\begin{figure*}[t]
\centering
% dopplershifts O5-Fe12
\includegraphics[width=0.9\linewidth]
	{\mydir{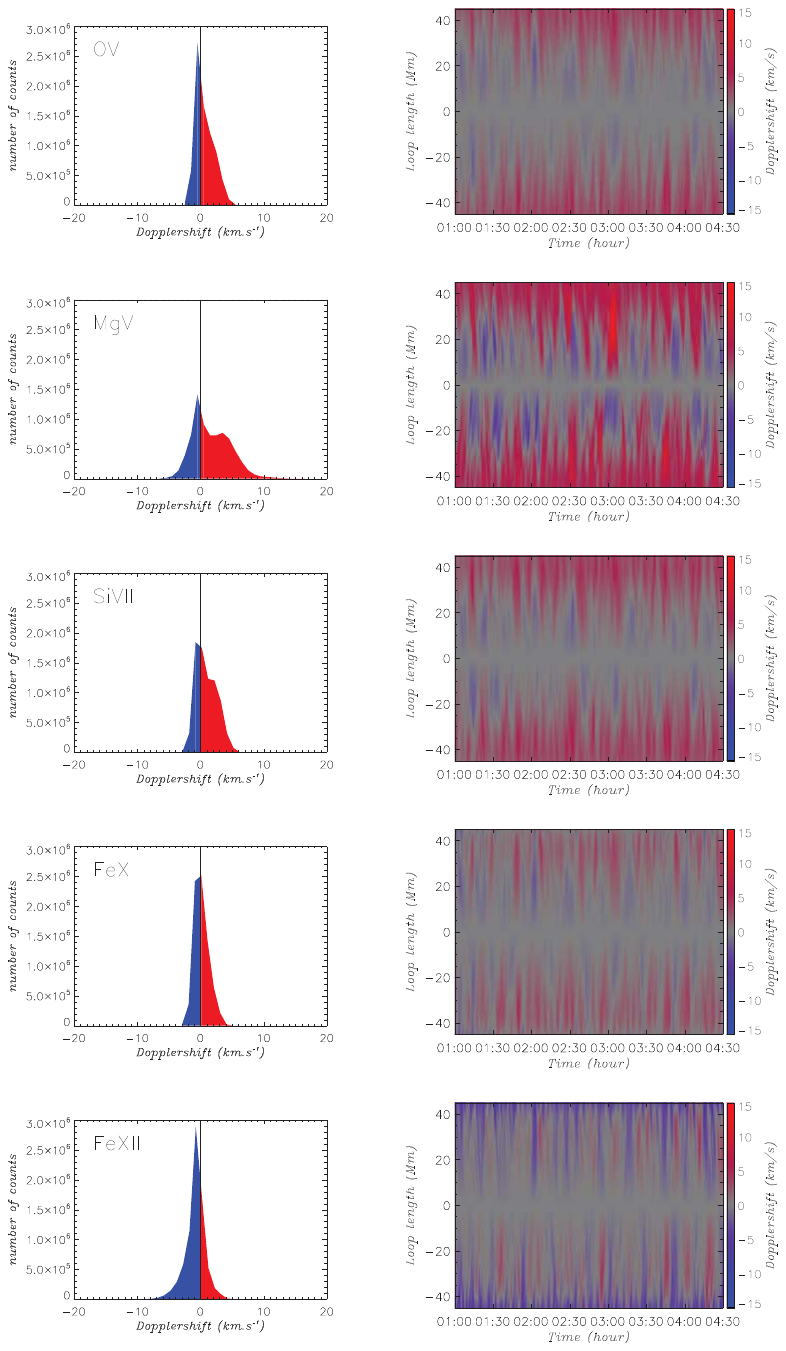}}
\caption{Statistical and spatial distributions of Dopplershift velocities for
{\it Loop {\sc i}} ($fp$ heating) once the equilibrium is established for
wavelengths from O {\sc v} to Fe {\sc xii} (see Table~\ref{tab:lines}).}
\end{figure*}

\addtocounter{figure}{-1}
\begin{figure*}[h]
\centering
% dopplershift Fe13-Ca17
\includegraphics[width=0.9\linewidth]
	{\mydir{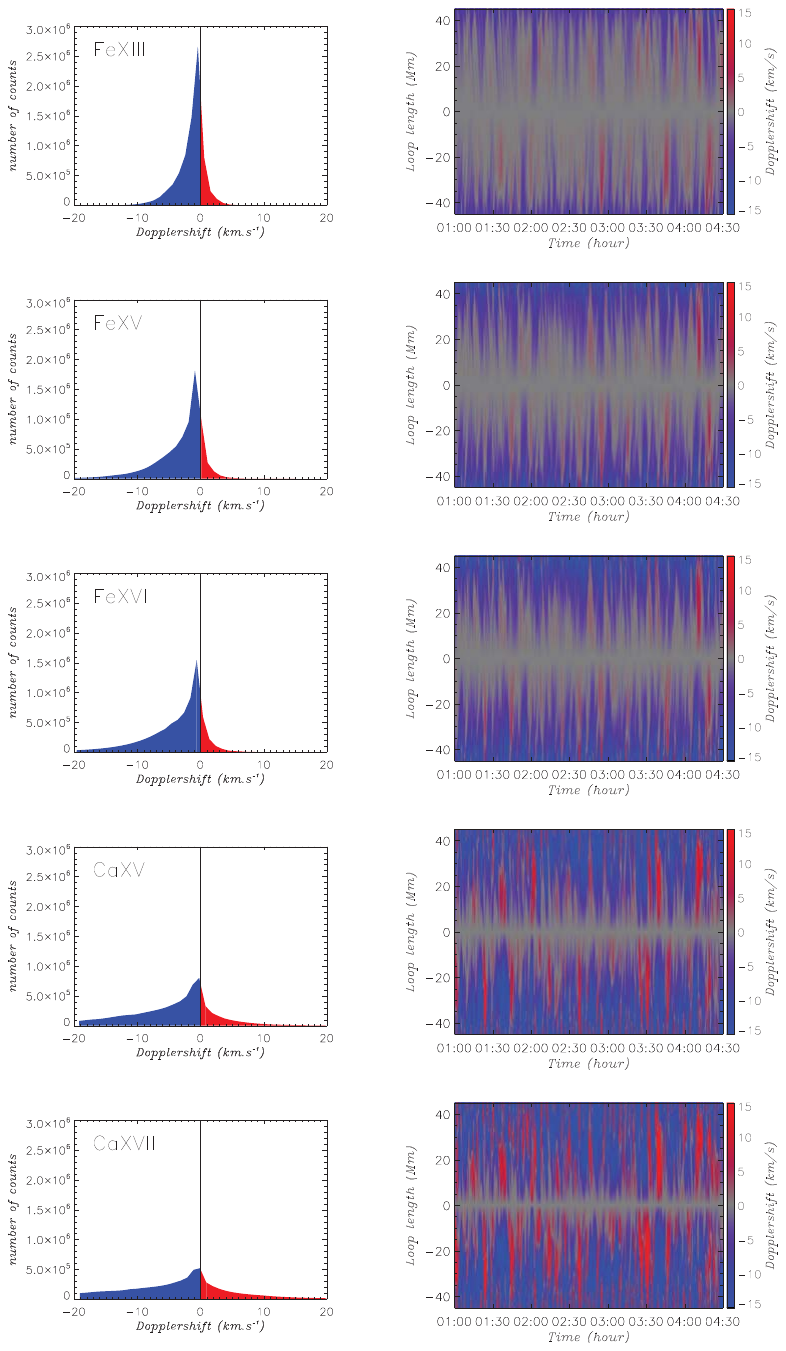}}
\caption{{\em (Continued)} from Fe {\sc xiii} to Ca {\sc xvii}}
\label{fig:warm_vel}
\end{figure*}
%%%%%%%%%%%%%%%%%%%%%%%%%%%%%%%%%%%%%%%%%%%

%%%%%%%%%%%%%%%%%%%%%%%%%%%%%%%%%%%%%%%%%%%
%%%	Fig: hot loop Dopplershift	%%%
%%%%%%%%%%%%%%%%%%%%%%%%%%%%%%%%%%%%%%%%%%%
\begin{figure*}[h]
\centering
%\begin{tabular}{cc}
% Dopplershift O5-Fe12
\includegraphics[width=0.9\linewidth]
	{\mydir{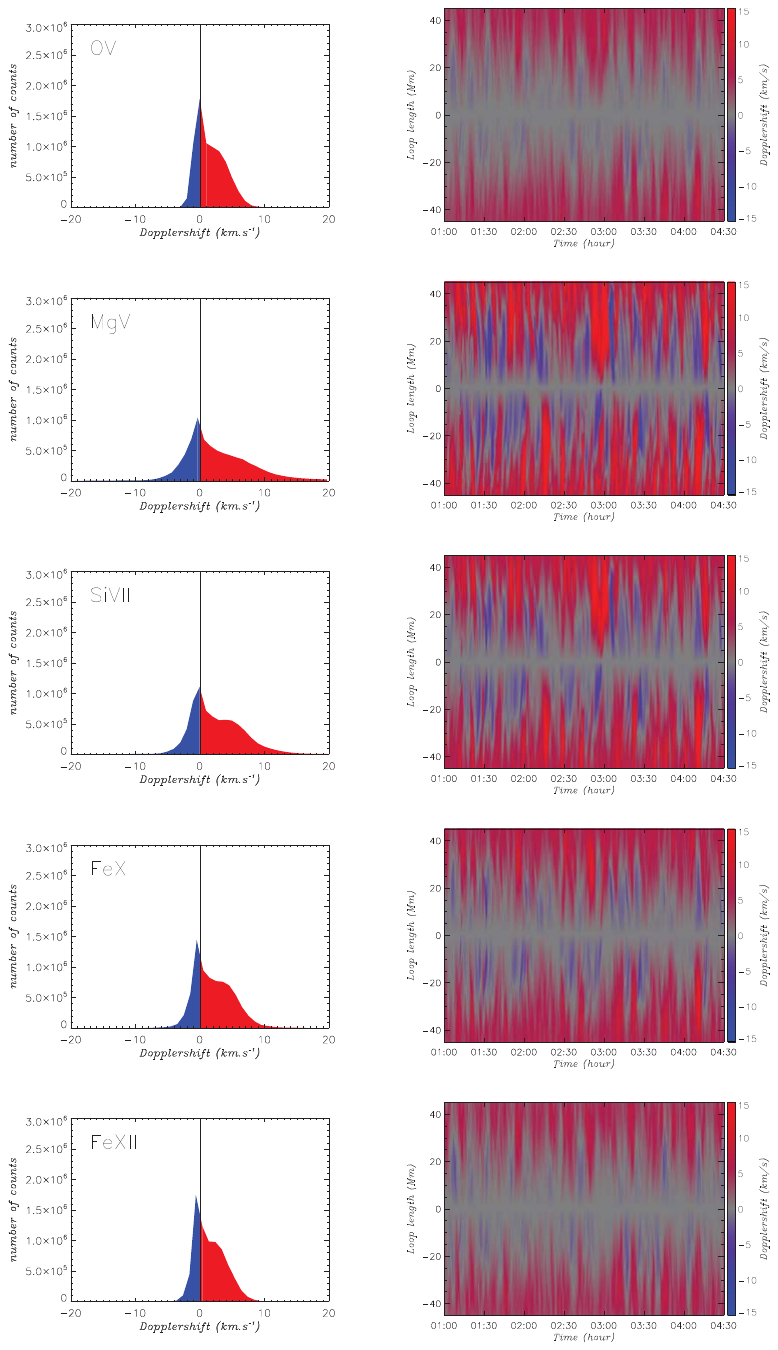}}
\caption{Same as Figure~\ref{fig:warm_vel} for {\it Loop \sc{ii}} from O {\sc v} to Fe {\sc xii}.}
\end{figure*}

\addtocounter{figure}{-1}
\begin{figure*}[h]

\centering
% Dopplershift Fe13-Ca17
\includegraphics[width=0.9\linewidth]
	{\mydir{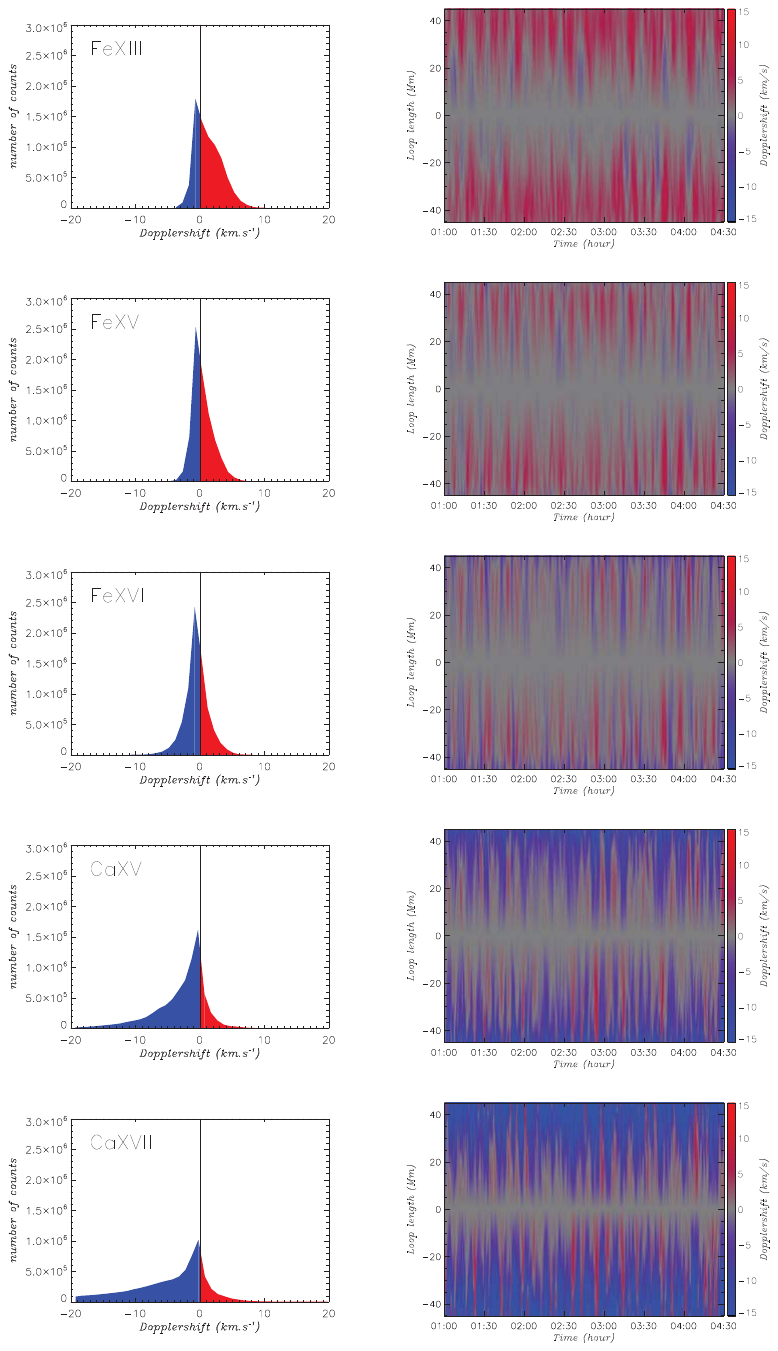}}
\caption{{\em (Continued)} from Fe {\sc xiii} to Ca {\sc xvii}.}
\label{fig:hot_vel}
\end{figure*}
%%%%%%%%%%%%%%%%%%%%%%%%%%%%%%%%%%%%%%%%%%%%%
%\clearpage}

The Doppler velocity for a given spectral line in computed in two steps. The first step is to compute the velocity for a given spectral line with a contribution $C_{sp}$ using the following equation:
%\begin{displaymath}
\begin{equation}\label{eq:vdop}
v_{sp}(s, t) = \frac{\displaystyle \sum_{i=1}^{128}
	\rho_i^2(s,t)\,v_i(s,t)\,C_{sp}(T_i(s, t))\,dl(s)} {\displaystyle \sum_{i=1}^{128}
	\rho_i^2(s,t)\,C_{sp}(T_i(s, t))\,dl(s)}.
%\end{displaymath}
\end{equation}

As indicated in the previous sub-section, the $C_{sp}$ contributions are depicted in Figure~\ref{fig:contrib} for the ten spectral lines employed in this paper. 

In the second step, the geometry of
the loop and the observer point-of-view are defined to obtain the projected velocity along the line-of-sight from 
which the red (downflows) and blue (upflows) Dopplershifts are derived (see Figure~\ref{fig:setuploop}).

%From the nanoflare, multistranded model described in Section~\ref{sec:model}, we
%extract the flow field along the semi-circular loop. Combined with the
%contribution functions of 10 different spectral lines (see
%Figure~\ref{fig:contrib}), we derive the Dopplershifts as they will be seen by an
%observer looking above the loop (typical observation from 1 AU). 

Assuming that the loop is semi-circular, the line-of-sight is defined perpendicular to the apex of loop and corresponds to a viewing angle of 0$^{o}$ (see Figure~\ref{fig:setuploop}). For the sake of completeness, the effects of changing the viewing angle on the distributions of Dopplershift and temperature diagnostic are descirbed in Section~\ref{sec:appa}.

%In Figure~\ref{fig:warm_vel}, we plot the Dopplershift distributions for {\it Loop \sc{i}}and
%their associated spatial distributions in the ten spectral lines described
%above. The Dopplershift distribution describes the number of occurrences of a Doppler speed %once the steady-state has been reached to the end of the simulation. The spatial distribution is showing the variation in time of the Dopplershift speed along the loop from -45 Mm to 45 Mm (the apex being at the origin) once the steady-state has been established. The red and blue colors indicate the red and blueshifts respectively. The
%Dopplershift distributions are plotted between -40 and 40 \kms, while the
%spatial distributions are scaled between -20 and 20 \kms. It is clear from both
%types of distributions that there is an obvious behavior for the Dopplershift to
%be bluer when the temperature $T_e$ of the spectral line increases. Blue- and redshifts are ubiquitous in all spectral lines. 

Figure~\ref{fig:warm_vel} plots the statistical  and spatial map of the Dopplershift distributions for all ten spectral lines under consideration for {\it Loop \sc{i}} where all 128 sub-strands are amalgamated together and where the simulation has reached a quasi-steady equilibrium. The statistical Dopplershift distribution (left column for each wavelength) describes the number of occurrences of a specific Doppler speed during the simulation whereas the spatial map (right column for each wavelength) displays the variation in time of the Dopplershift speed along the loop from -45 Mm to 45 Mm (the apex is at the origin). The red and blue colours indicate the red- and blue-shifts respectively. Note that the Dopplershift statistical distributions are plotted between -20 and 20 \kms, while the spatial maps are scaled between -15 and 15 \kms. It is clear from both types of distributions that blue- and redshifts are ubiquitous in all spectral lines. However, there is an obvious change in behaviour of the Dopplershift as the temperature is altered; specifically, from red to 
blue dominance as the temperature $T_e$ of the spectral line increases.  

\begin{figure}[!t]
	\centering
	\includegraphics[width=1.\linewidth]{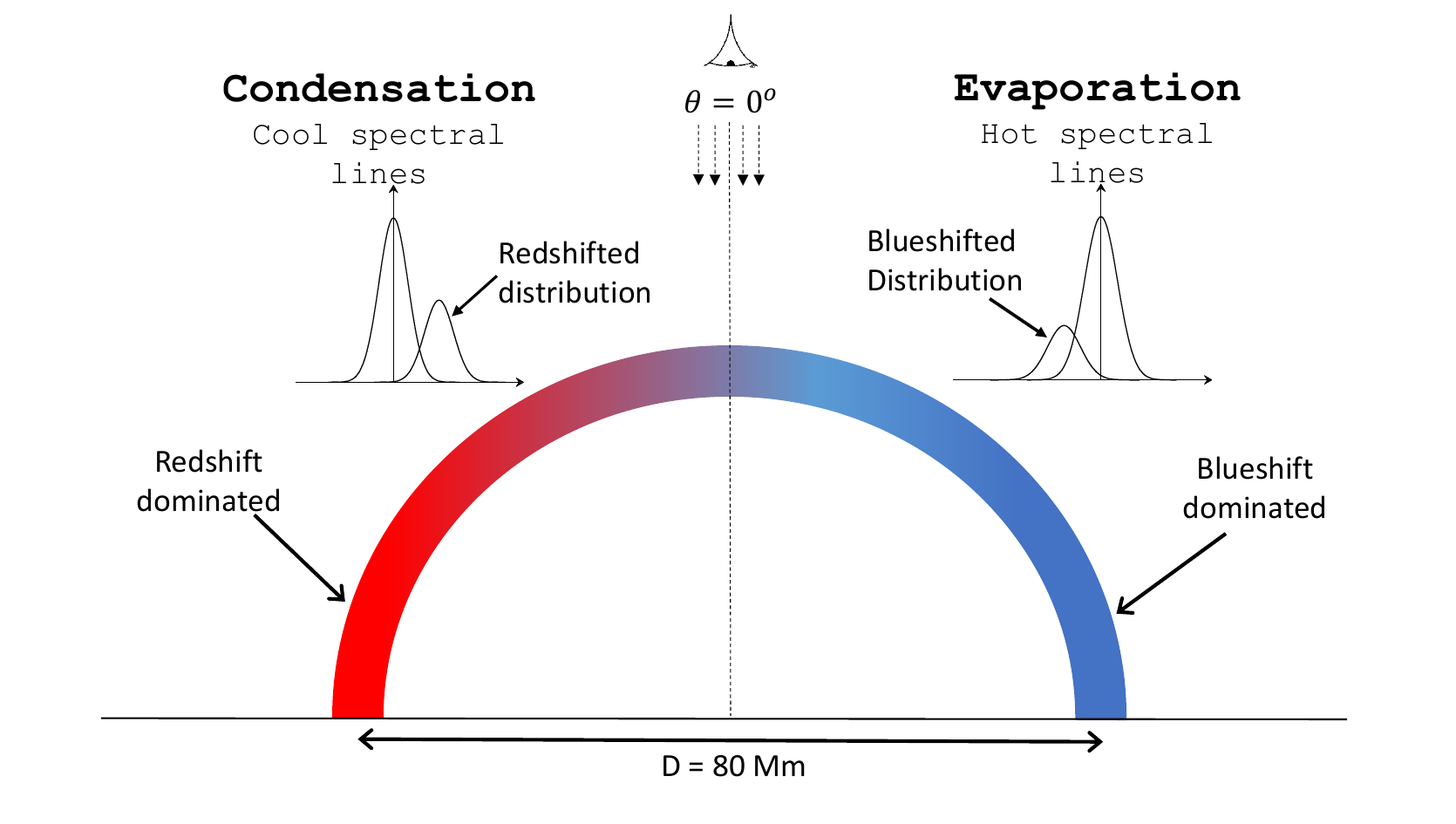}
	\caption{Schematics of the Dopplershift distributions describing the plasma condensation and evaporation mechanisms as they are observed using the 1D MSHD model.}
	\label{fig:vel_distrib}
\end{figure}

There are three physical process operating in the system that can be identified within the results of Figure ~\ref{fig:warm_vel} and summarised in Figure~\ref{fig:vel_distrib}. There is (i) the localised, short duration red and blue shift distribution centred on zero velocity that arises from the randomised heating bursts along the strands; (ii) a separate red-shift profile resulting from plasma cooling down, condensing and being observed in the cooler lines;
and (iii) a separate component due to the evaporation of material into the loop from the chromospheric reservoir at its base.
In Figure~\ref{fig:vel_distrib}, the velocity distributions are depicted as Gaussian distributions: the plasma condensation is described on the left-hand side with a redshifted distribution corresponding to positive velocity going away from the observer, and the plasma evaporation is described on the right-hand side with a blueshifted distribution. The relative importance of the peak distribution is just indicative. 

Considering {\it Loop {\sc i}} further (see Figure~\ref{fig:warm_vel}), the Dopplershift statistical distributions and the spatial Dopplermaps are clearly dominated by redshifts for cooler lines from O {\sc v} to Fe {\sc x}. Also, the overall Dopplershift statistical distribution is clearly double-peaked for the Mg {\sc v} and Si {\sc vii} lines and corresponds to the localised, red and blue shift distribution  from the randomised heating events 
as well as the condensation of hotter material to cooler temperatures. The downflow near the footpoints at about 3-5 \kms. On the other hand, blueshifts start to become more dominant above the Fe {\sc xii} line with the evaporation process observed as a long distribution tail. It should be noted that for all spectral lines, the maximum of redshift velocity 
is around 10 \kms, while for the blueshift distribution, the tail towards high Doppler velocity develops from a maximum of 10 \kms\ in Fe {\sc xii} to 30 \kms\ in Ca {\sc xvii}.

%The distribution of Dopplershift is double-peaked for the Mg {\sc v} and
%Si {\sc vii} lines, which corresponds to a downflow near the footpoints at about 3-5 \kms\. %The Dopplershift distribution and the Dopplermaps are
%clearly dominated by redshifts for cool lines from O {\sc v} to Fe {\sc x},
%while blueshifts start to appear in the Fe {\sc xii} lines and
%developed in hotter lines. For all spectral lines, the maximum of redshift velocity 
%is about 10
%\kms\, while for the blueshift distribution, the tail towards high Doppler velocity is
%getting more and more prominent when the spectral line temperature $T_e$ increases,
%from a maximum of 10 \kms\ in Fe {\sc xii} to 30 \kms\ in Ca {\sc xvii}.  

%%%%%%%%%%%%%%%%%%%%%%%%%%%%%%%%%%%%%%%%
%%%	Fig: blue/red distrib 	%%%%%%%%
%%%%%%%%%%%%%%%%%%%%%%%%%%%%%%%%%%%%%%%%
%%%%%%%%%%%%%%%%%%%%%%%%%%%%%%%%%%%%%%%%%

%%%%%%%%%%%%%%%%%%%%%%%%%%%%%%%%%%%
%%%	Fig: percentage		%%%
%%%%%%%%%%%%%%%%%%%%%%%%%%%%%%%%%%%
\begin{figure}[!ht]
\centering
\includegraphics[width=.49\linewidth]
	{\mydir{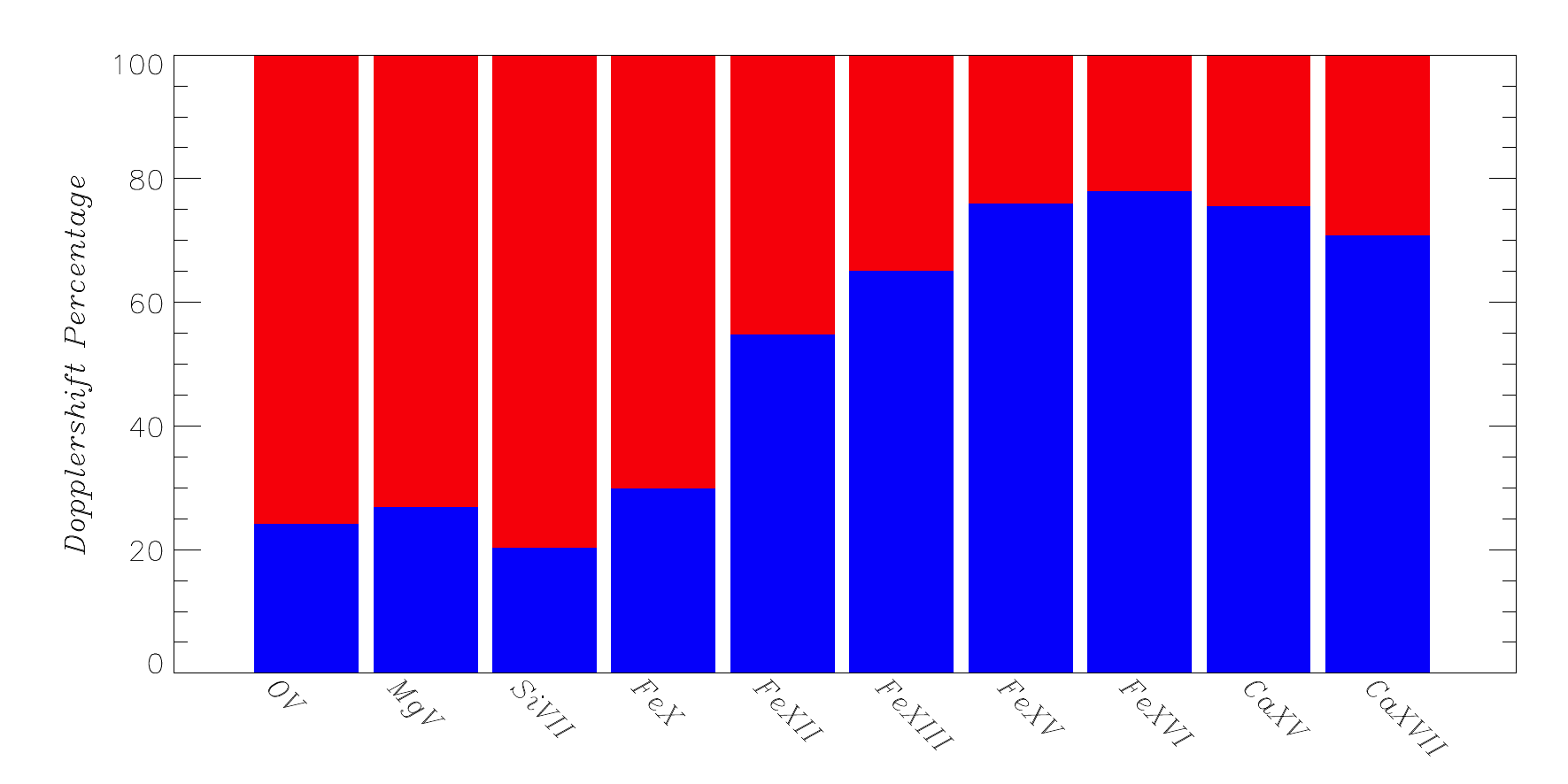}}
\includegraphics[width=.49\linewidth]
	{\mydir{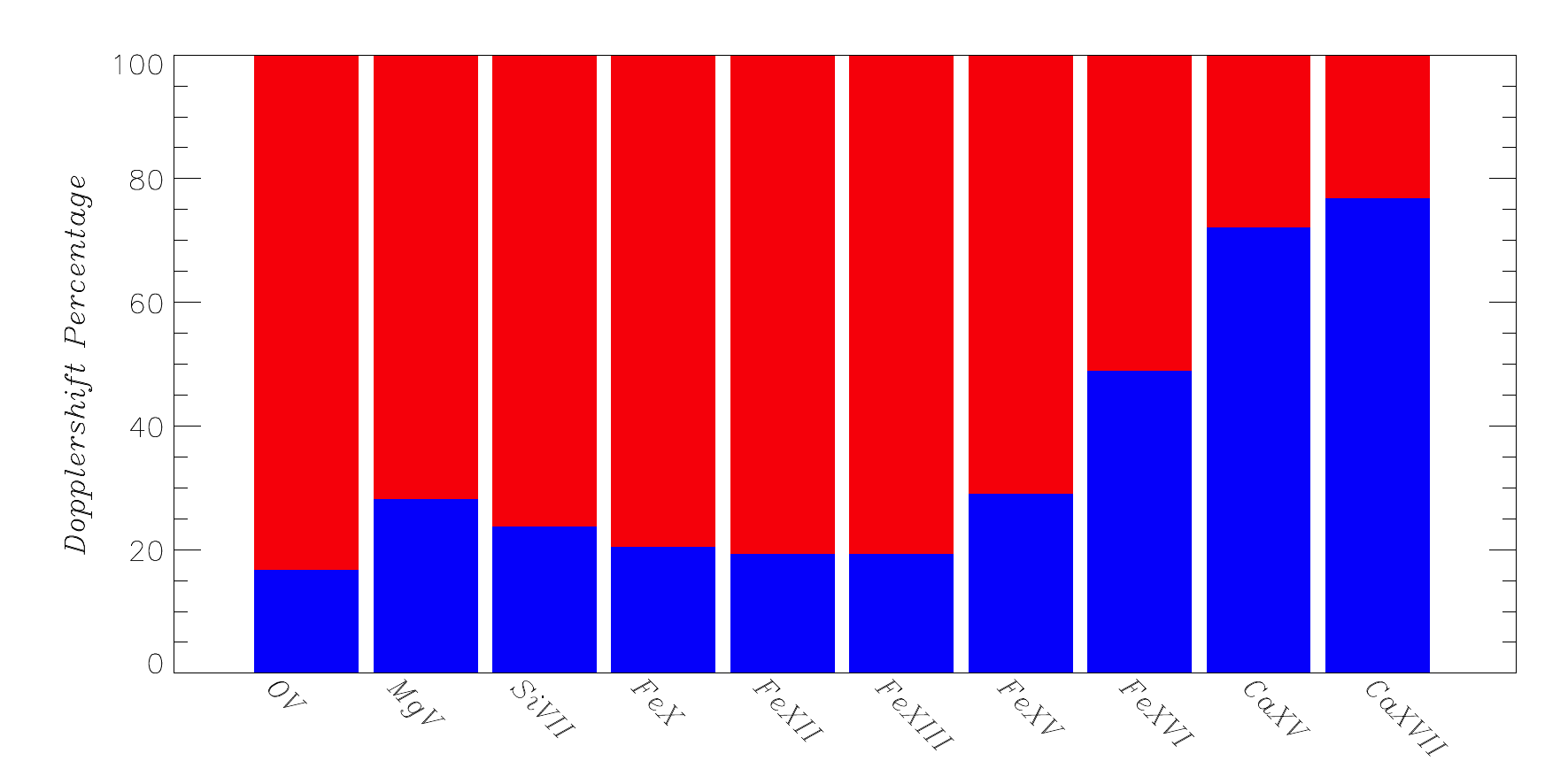}}
\includegraphics[width=.49\linewidth]
	{\mydir{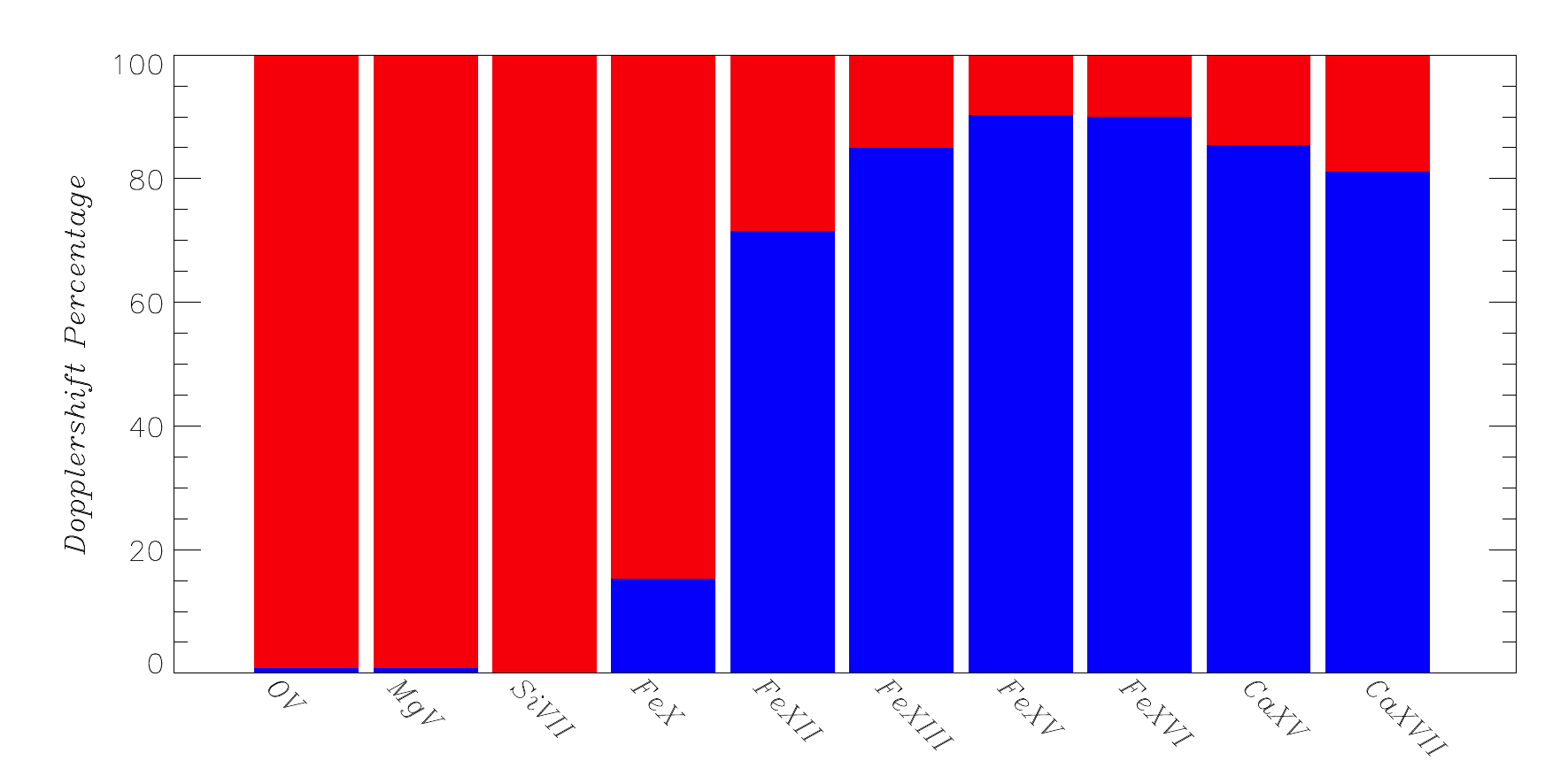}}
\includegraphics[width=.49\linewidth]
	{\mydir{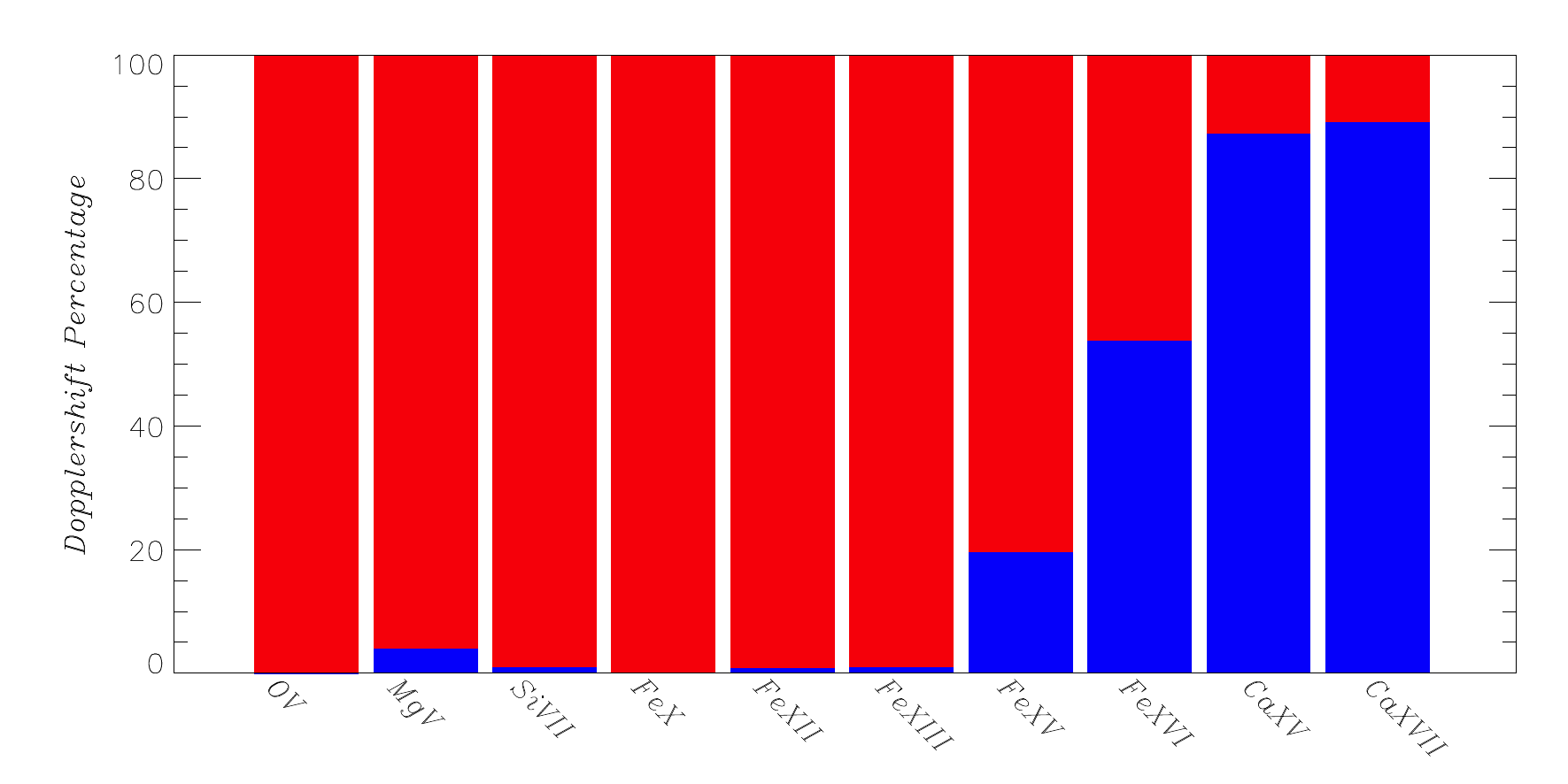}}
\includegraphics[width=.49\linewidth]
	{\mydir{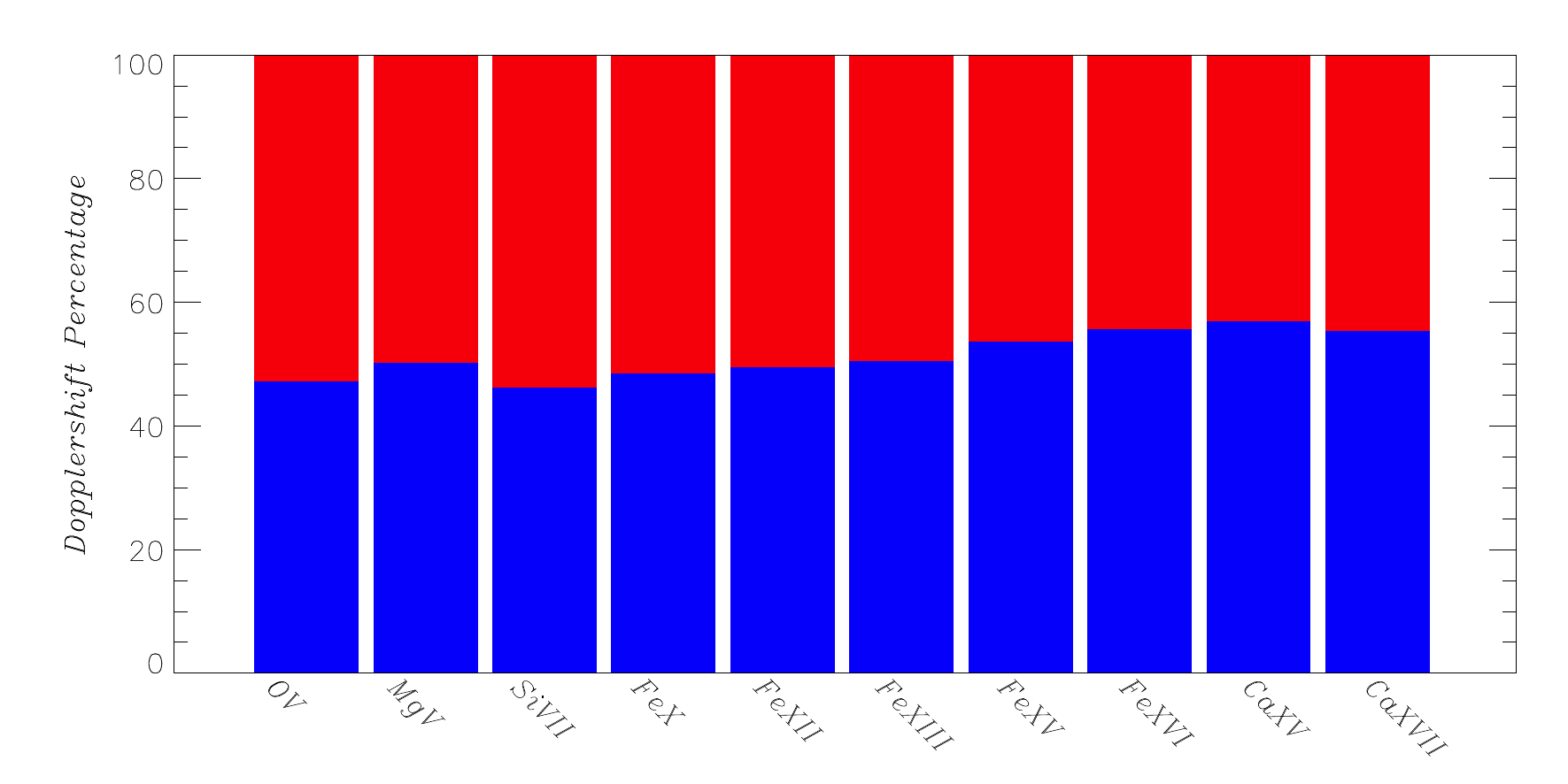}}
\includegraphics[width=.49\linewidth]
	{\mydir{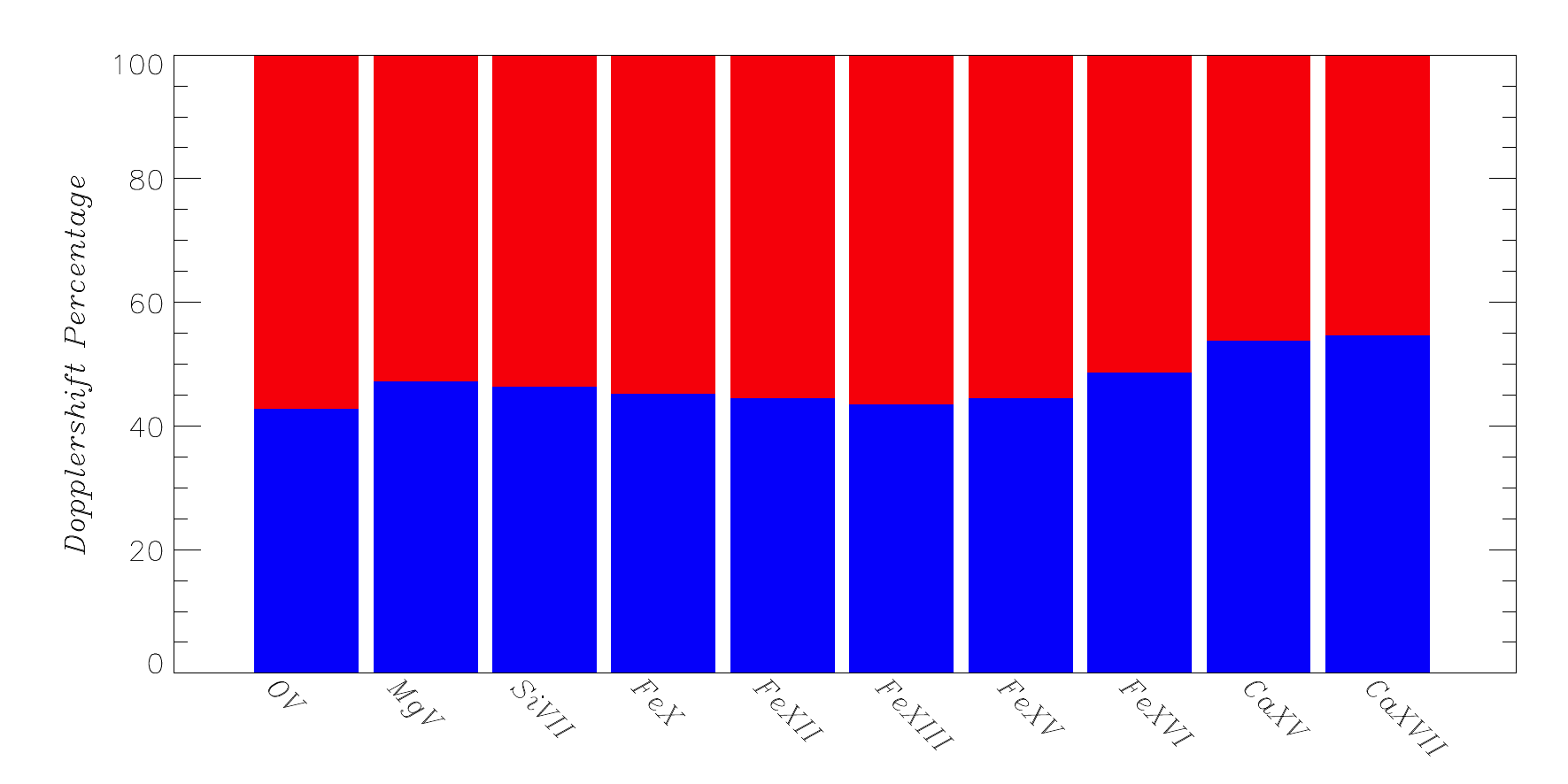}}
\caption{Percentage of blue- and redshift along the loop (top row), at the footpoints
(middle row) and at the apex (bottom row) for all spectral lines: ({\em left}) for {\it Loop \sc{i}}, ({\em right}) for {\it Loop \sc{ii}}. }
\label{fig:stat}
\end{figure}
%%%%%%%%%%%%%%%%%%%%%%%%%%%%%%%%%%%%%

Figure~\ref{fig:hot_vel} displays the same quantities as
in Figure~\ref{fig:warm_vel} but for {\it Loop {\sc ii}}. To facilitate the comparison between the two loop models, we use the same ranges of velocities as in Figure~\ref{fig:warm_vel}.

As for {\it Loop {\sc i}}, red- and blueshifts are ubiquitous at all wavelengths. 
The Dopplershift statistical distributions show a similar double-peak to that in Figure~\ref{fig:warm_vel} but now 
from the O {\sc v} line to the Fe {\sc xii} line, which is again an evidence of a downflow/condensation at about 4-6 \kms\ (see Figure~\ref{fig:vel_distrib}). 
For the cooler lines (below Fe {\sc xiii}), the redshifts are concentrated near the footpoints, while for hotter lines (from Fe {\sc xv} to
Ca {\sc xvii}) the blueshifts dominate in this region. The physical processes at play are identical to the ones described for {\it Loop {\sc i}} but shitfed towards higher temperatures as the transition between red- and blueshift dominated plasma occurs for Fe {\sc x} for {\it Loop {\sc i}} and for Fe {\sc xii} for {\it Loop {\sc ii}}. This is reasonable
given the overall increase in total energy into the system of {\it Loop {\sc ii}} compared to {\it Loop {\sc i}}.

%The Dopplershift distributions show a double-peak
%distribution from the O {\sc v} line to the Fe {\sc xii} line, which is again an evidence of a downflow at about 4-6 \kms. As for {\it Loop \sc{i}}, blue- and redshifts are ubiquitous at all wavelengths. For the
%cooler lines (below Fe {\sc xiii}), the redshifts are concentrated near the foopoints, while for hotter lines (from Fe {\sc xv} to
%Ca {\sc xvii}) the blueshifts dominates near the footpoints. The maximum
%velocity in absolute value oscillates between 6 \kms\ and 30 \kms. However, the
%maximum of the blueshift velocity as in the Ca {\sc  xvii} line can be greater
%than 40 \kms, while the maximum of the redshift velocity (Mg {\sc v} line) is
%about 30 \kms. For all spectral lines, the maximum redshift velocity is around 10-15 \kms. 

%%%%%%%%%%%%%%%%%%%%%%%%%%%%%%%%%%%%%%
\section{Temperature Diagnostic} \label{sec:temp}
%%%%%%%%%%%%%%%%%%%%%%%%%%%%%%%%%%%%%%

	%%%%%%%%%%%%%%%%%%%%%%%%%%%%%
	\subsection{Hot vs Warm Loop}
	%%%%%%%%%%%%%%%%%%%%%%%%%%%%%

Given that the synthesised Doppler-shift observations can be related back directly to known physical parameters in the loops (and their sub-strands), it
is instructive to further compare the properties of the warm ({\it Loop {\sc i}}) and hot ({\it Loop {\sc ii}}) loops in terms of their Dopplershift
distributions. Therefore, Figure~\ref{fig:stat} plots the percentage of red- and blue-shifts occurring within the loop over the simulation period for the ten spectral lines sorted by increasing peak formation temperature $T_e$. This percentage is calculated in different parts of the loop (along the entire coronal loop, near the loop base and finally around the apex as defined in Figure~\ref{fig:setuploop}) once the quasi-steady state has been established. As examined in Figures~\ref{fig:hot_vel} and~\ref{fig:warm_vel}, although red- and blue-shifts are ubiquitous in all wavelengths, it is very clear that red (blue) shifts dominate in the cool (hotter) lines. 

Focusing on {\it Loop {\sc i}} only and considering the entire loop plot (Figure~\ref{fig:stat} top left), the blue shift component begins to increase in
percentage from the Fe {\sc x} line.  A maximum in the percentage of blue-shifts (about 80\%) is reached for the Fe {\sc xvi} line.The distribution of blue-shift for {\it Loop \sc{i}} then decreases after the Fe {\sc xvi} line. 

In Figure~\ref{fig:stat} is also plotted the red- and blueshift contribution at the loop-base (middle row, left) and at the loop apex
(bottom row, left) for {\it Loop {\sc i}} . 
The Dopplershifts at the loop base are highly redshifted dominant for cooler lines and abruptly increase in blue-shift to 80-90\% after 
Fe {\sc x} and subsequently peaking at the Fe {\sc xii} line. In contrast, the percentage of Dopplershift around the apex is relatively flat for this 
loop with a very slight blue dominance in the hotter lines.

The same thermodynamic processes are operating {\it Loop {\sc ii}} but are scaled upwards in temperature due to the increase in injected total energy into this loop.
As shown in In Figure~\ref{fig:stat} (right column), the entire loop blue shift component begins to increase from Fe {\sc xv} line with a maximum in blue-shift
calculated for Ca{\sc xvii} line (though we need to note that this is the highest temperature line under consideration). Similarly, the loop base blue raises from
the Fe {\sc xv} line and peaks in the Ca {\sc xvii} line. The distribution red and blue shifts in the {\it Loop {\sc ii}} apex region is once again approximately evenly split.

\begin{figure*}[!ht]
\centering
\begin{tabular}{cc}
\multicolumn{1}{l}{(a)} & \multicolumn{1}{l}{(b)} \\
\includegraphics[width=0.45\linewidth]
	{\mydir{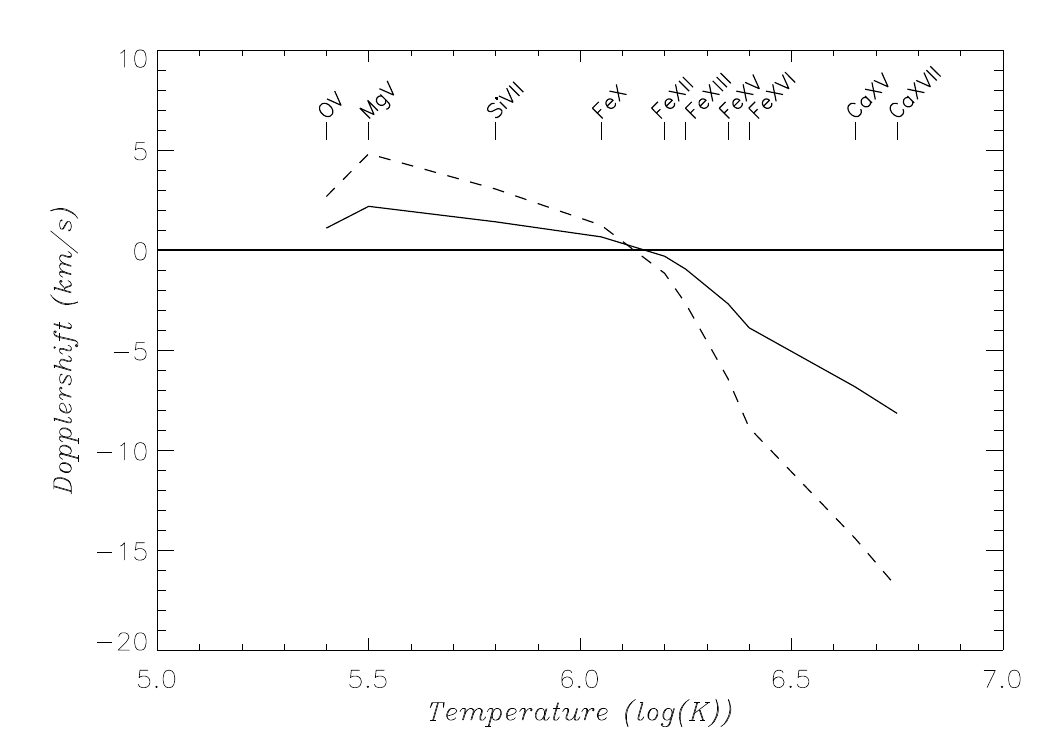}}
&\includegraphics[width=0.45\linewidth]
	{\mydir{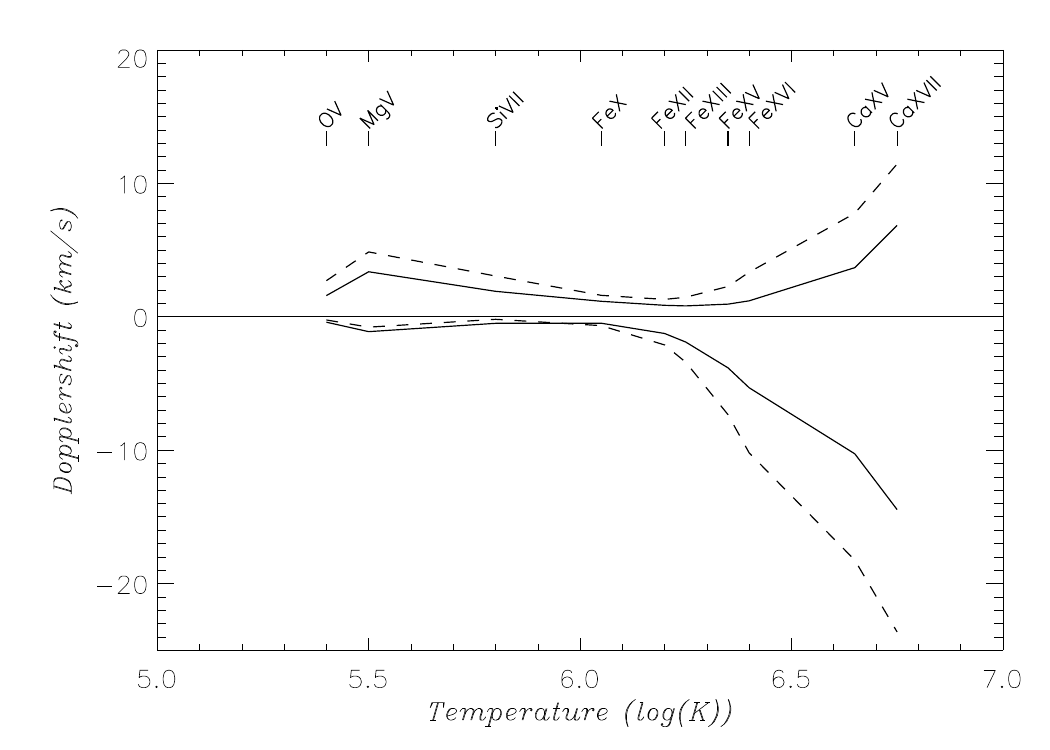}} \\
\multicolumn{1}{l}{(c)} & \multicolumn{1}{l}{(d)} \\
\includegraphics[width=0.45\linewidth]
	{\mydir{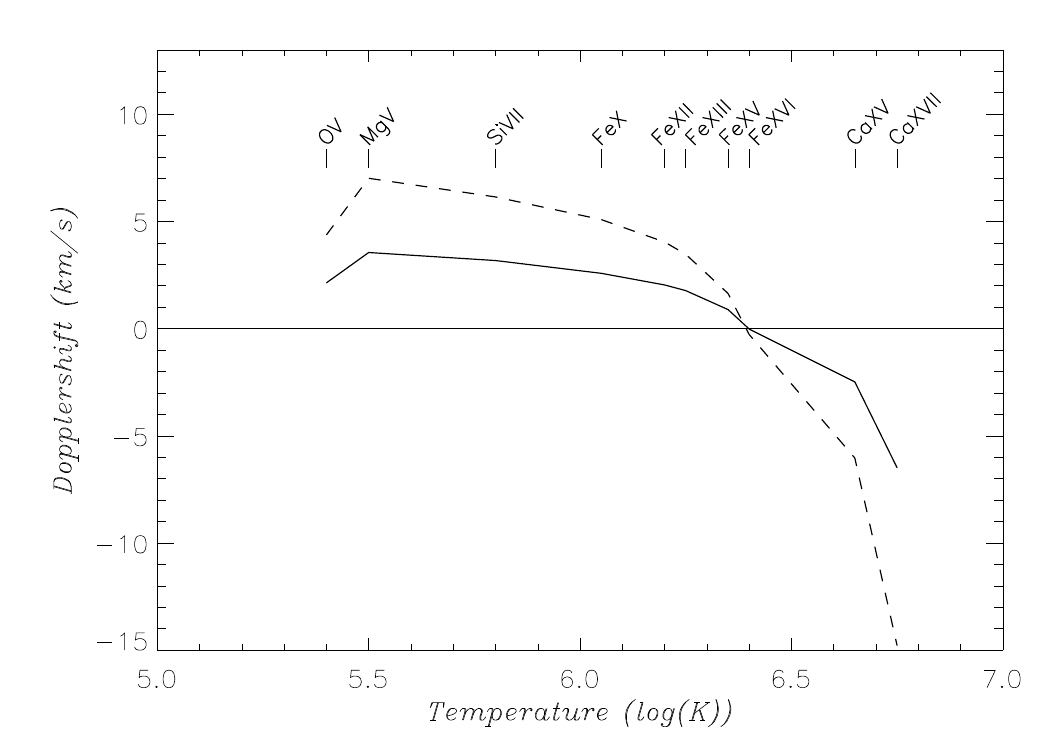}}
&
\includegraphics[width=0.45\linewidth]
	{\mydir{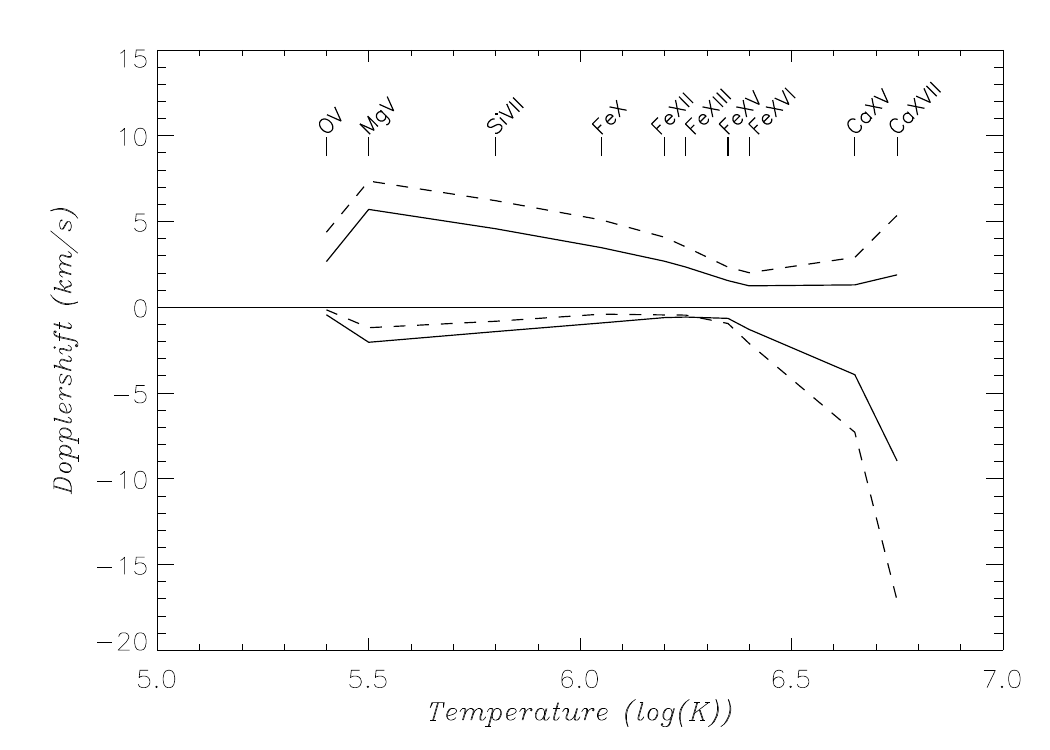}}
\end{tabular}
\caption{{\it Loop {\sc i}}: (a) average Dopplershift along the loop (solid line) and at
the footpoint (dash line), and (b) average redshifts (positive) and blueshifts
(negative) along the loop (solid line) and at the footpoints (dash line). (c) and (d): same as (a) and (b) respectively for {\it Loop {\sc ii}}}
\label{fig:av_dop}
\end{figure*}

Another way of demonstrating this but focusing upon actual values for the average Dopplershifts over the simulation period can be found in Figure~\ref{fig:av_dop} (a)-(b) for {\it Loop {\sc i}} and Figure~\ref{fig:av_dop} (c)-(d) for {\it Loop {\sc ii}}. In these Figures the average Dopplershift velocity (left column) plus that average split into its positive and negative Dopplershift velocity components (right column) for
the entire loop (solid line) and at the footpoint area (dash line, see \ref{fig:setuploop}) are plotted against the spectral line peak formation temperature. 

For {\it Loop {\sc i}}, the average Dopplershift approximately crosses the zero velocity point between the Fe {\sc x} and Fe {\sc xii} lines for the entire loop and at the loop base. Similarly for {\it Loop {\sc ii}}, the average Dopplershift vanishes between the Fe {\sc xvi} and Ca {\sc xv} lines for both the entire loop and for the loop base.

For the average velocity above the Fe {\sc xvi} in Figure~\ref{fig:av_dop} (b), the contribution of the
footpoints (dash line) is large with a maximum of -24 \kms\ for blueshift and 11 \kms\ for
redshift in the Ca {\sc xvii} line, compared to -15 \kms\ and 7 \kms\ respectively for the entire loop (solid line). For {\it Loop {\sc ii}}
(Figure~\ref{fig:av_dop} (d)), the distribution is shifted towards
higher temperatures, and the distribution at the footpoints exhibits a maximum
average redshift in the Ca {\sc xvii} line at 5 \kms\ while the average blueshift
reaches -17 \kms.

%%%%%%%%%%%%%%%%%%%%%%%%%%%%%%%%%%%%%%%%%%%%

%%%%%%%%%%%%%%%%%%%%%%%%%%%%%%%%%%%%%%%%%%%
%%%	Fig: temperature diagnostic	%%%
%%%%%%%%%%%%%%%%%%%%%%%%%%%%%%%%%%%%%%%%%%%
\begin{figure*}[!ht]
\centering
\includegraphics[width=0.49\linewidth]{\mydir{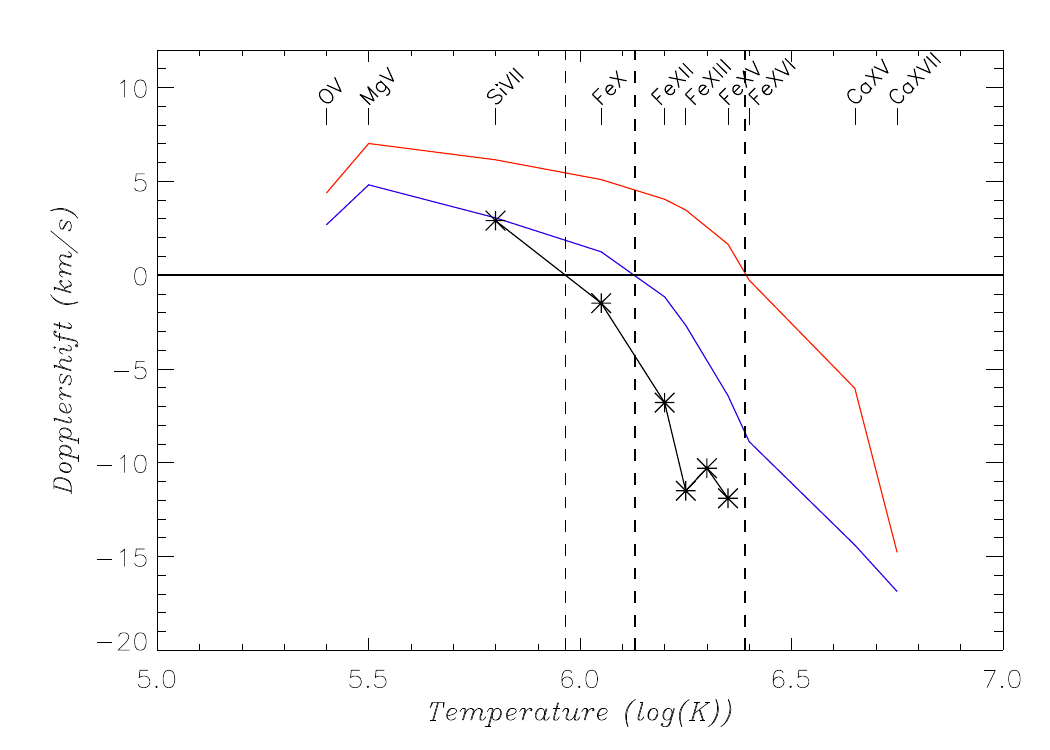}}
\includegraphics[width=0.49\linewidth]{\mydir{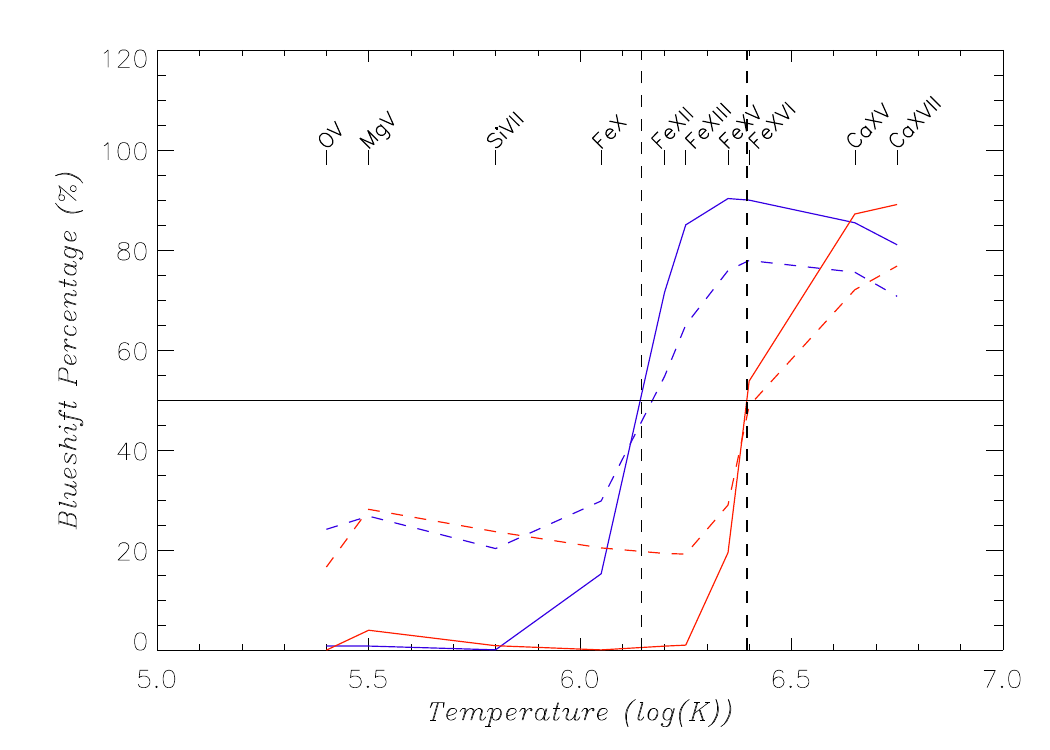}}
\caption{Temperature diagnostic tool. (Left) Average Dopplershifts at the footpoint of {\it Loop \sc{i}} (blue) and
{\it Loop \sc{ii}} (red) for $fp$ heating, and the average Dopplershift values (black curve) obtained by
\citet{tri09}; (Right) Percentage of blueshift at the footpoint (solid) and
along the loop (dash) for {\it Loop \sc{i}} (blue) and {\it Loop \sc{ii}} (red) cases. The dashed black
lines indicate the crossing of the curves with a vanishing average 
Dopplershift or a 50\% level.}
\label{fig:temp_diag}
\end{figure*}
%%%%%%%%%%%%%%%%%%%%%%%%%%%%%%%%%%%%%%%%%%%%%

Therefore the temperature at which the average
Dopplershift vanishes depends on the total energy injected in the loop; it corresponds to the temperature at which there is a balance between red- and blueshifts, 
similar to an equilibrium position. The evolution of the average positive and negative Dopplershifts is consistent with
the evolution of the width of the Dopplershift distributions as depicted in Figures~\ref{fig:hot_vel} and~\ref{fig:warm_vel}. 

Thus, there are a number of aspects to note from Figures~\ref{fig:stat} to~\ref{fig:av_dop}. Firstly, the effect of the increase in the total energy into
the loop model is clear - the qualitative behaviour from {\it Loop {\sc i}} to {\it Loop 
{\sc ii}} scales up in temperature, across the spectral lines under consideration.
Secondly, the modelling qualitatively reproduces the behaviour observed in loops where cool lines are dominated by red-shifts while hotter lines are
predominantly bluer. Thirdly, while the entire loop statistics are usefully employed in the simulation (where all spatial positions are known), this would not be 
as applicable to actual observations. Thus, the localised loop base or loop apex analyses are better observational proxies. However, it should be noted that 
in this model simulation where the random heating bursts are localised at the loop based, the loop apex Dopplershift measurements in Figures~\ref{fig:stat} 
(bottom panels) are merely a symptom of the motions generated after these bursts have occurred (and hence has a weak dependence on temperature). 

Consequently, as will be outlined in the next section, this new approach could be employed to determine the mean temperature of an observed loop bundle.

	\subsection{A Temperature Diagnostic Tool}\label{sec:tdiag}
	%%%%%%%%%%%%%%%%%%%%%%%%%%%%%

Using the analysis performed in the previous Section, we develop a new tool to
approximate the temperature of a coronal loop which can be used in conjunction
with observations.

By plotting the evolution of the average Dopplershift at the footpoints with
increasing temperature (see Figure~\ref{fig:temp_diag} left), the mean temperature of the loop is obtained by finding the temperature at which the average Dopplershift vanishes.
In Figure~\ref{fig:temp_diag} left, we plot the average Dopplershift curves for
{\it Loop {\sc i}} (blue) and {\it Loop {\sc ii}} (red). The estimated temperatures (indicated by the
vertical dash lines) are $T_{\textrm{\it Loop {\sc i}}} = 1.35$ MK, and $T_{\textrm{\it Loop {\sc ii}}} = 2.45$ MK, which
have to be compared to the temperatures mentioned in Section~\ref{sec:thermo},
i.e, $T_{\textrm{\it Loop {\sc i}}}^{mean} = 1.5$ MK, and $T_{\textrm{\it Loop {\sc ii}}}^{mean} = 3$ MK. It is worth noting
that our estimate of the temperatures relies on piecewise linear interpolation
between consecutive points, which, in this case, tends to give a lower value of
the mean temperature. 

We also notice that the mean temperature of the loop can be estimated from the
percentage of blueshifts at the footpoint or along a coronal loop. In
Figure~\ref{fig:temp_diag} right, we plot the percentage of blueshift as a
function of temperature for {\it Loop {\sc i}} (blue) and {\it Loop {\sc ii}} (red) cases computed at the
footpoint of the loop (solid curve) and along the loop (dash curve). The 50\%
level of blueshift crosses the different curves at an approximation of the mean
temperature similar to the average Dopplershifts at the footpoints (see
Figure~\ref{fig:temp_diag} left). That is to say, $T_{\textrm{\it Loop {\sc i}}} = 1.4$ MK, and
$T_{\textrm{\it Loop {\sc ii}}} = 2.5$ MK. However, the percentage of blueshifts is not a quantity
accessible with the current observations due to the spatial resolution of the
instruments. One possible proxy will be the variation in time of the
Dopplershift at a given location assuming that the loop is stable, and
considering that the exposure time is small (typically of the order of
1--2 s).     

As an example, we use the analysis if \cite{tri09} of an active region observed by {\em Hinode}/EIS. \cite{tri09} have studied two areas observed by {\em
Hinode}/EIS and thus have computed the average Dopplershift in these regions.
They show that the average Doppler velocity decreases with increasing temperature
(with the exception of a Fe {\sc xiv} line). In their examples, the average
Dopplershift vanishes for a temperature between the Si {\sc vii} and Fe {\sc x}
line. Thus following our model these regions will correspond to warm loops (i.e., corresponding the {\it Loop {\sc i}}). To confirm our temperature diagnostic, we  also plot the
results obtained by \citet{tri09} as a black solid curve in Figure~\ref{fig:temp_diag}
left for the footpoints of coronal loop using the {\em Hinode}/EIS
spectrometer. The estimated temperature of the coronal loop is $T = 923000$ K.
Using an EM loci method, \citet{tri09} have estimated that the temperature of
coronal loops in the observed region is between 800000 K and 1.5 MK depending on
the height of the loop. Therefore, our temperature diagnostic is a good
approximation of the mean temperature of coronal loops in this example. As the loop length and viewing angle are different from the modelled loops, we conclude that the temperature diagnostic seems to be robust (see also Appendix~\ref{sec:appa} for a discussion on the viewing angle).  

This new temperature
diagnostic can only be performed from a series of spectral lines covering a
broad range of temperature, typically between 10$^5$ and 10$^7$ K.

%%%%%%%%%%%%%%%%%%%%%%%%%%%%%%%%%%%%%%
\section{Discussion} \label{sec:concl}
%%%%%%%%%%%%%%%%%%%%%%%%%%%%%%%%%%%%%%

	%%%%%%%%%%%%%%%%%%%%%%%%%%%%%%%%%%%%%%%%%%%%%%
	\subsection{Simulated {\em Hinode}/EIS Raster}
	%%%%%%%%%%%%%%%%%%%%%%%%%%%%%%%%%%%%%%%%%%%%%%
	
%%%%%%%%%%%%%%%%%%%%%%%%%%%%%%%%%%%
%%%	Fig: raster warm	%%%
%%%%%%%%%%%%%%%%%%%%%%%%%%%%%%%%%%%
\begin{figure*}[!ht]
\centering
\begin{tabular}{cccc}
\multicolumn{1}{l}{(a)} & \multicolumn{1}{l}{(b)} 
	& \multicolumn{1}{l}{(c)} & \multicolumn{1}{l}{(d)} \\
\includegraphics[width=0.24\linewidth]
	{\mydir{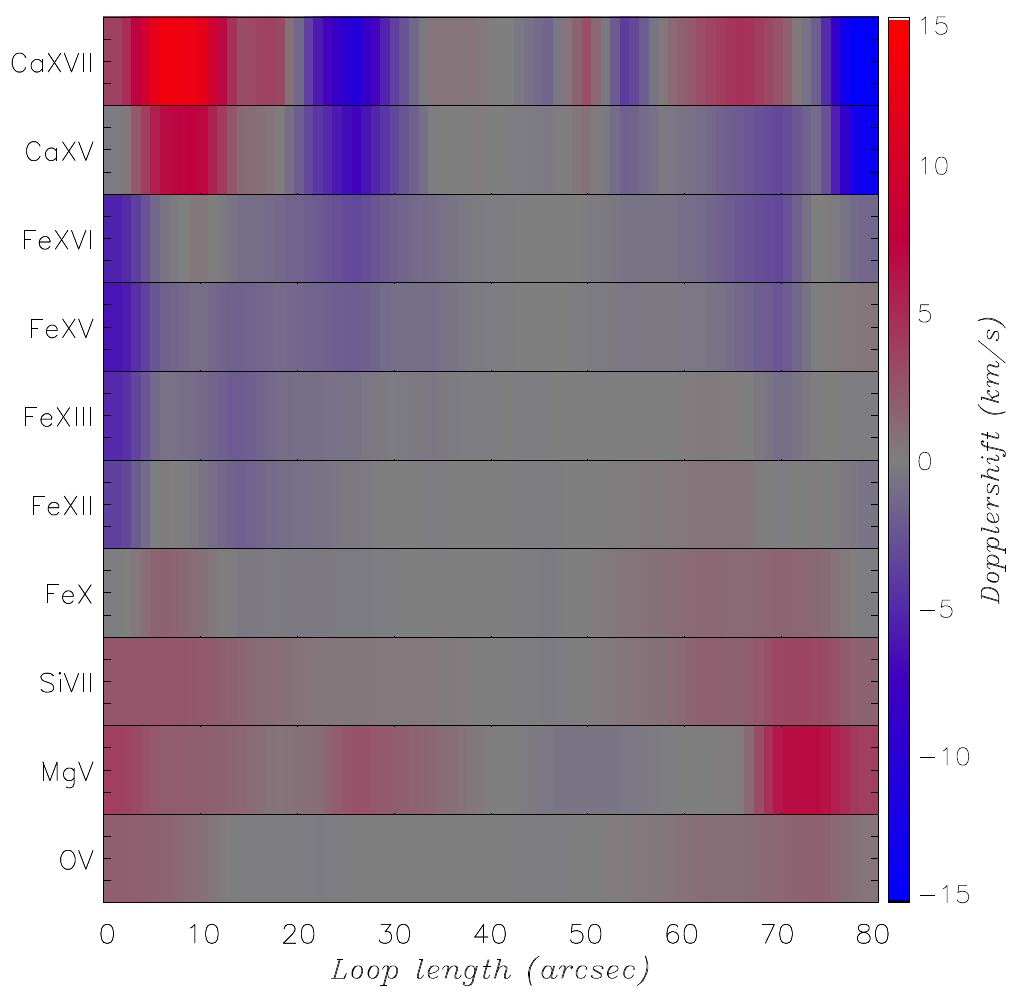}}
&
\includegraphics[width=0.24\linewidth]
	{\mydir{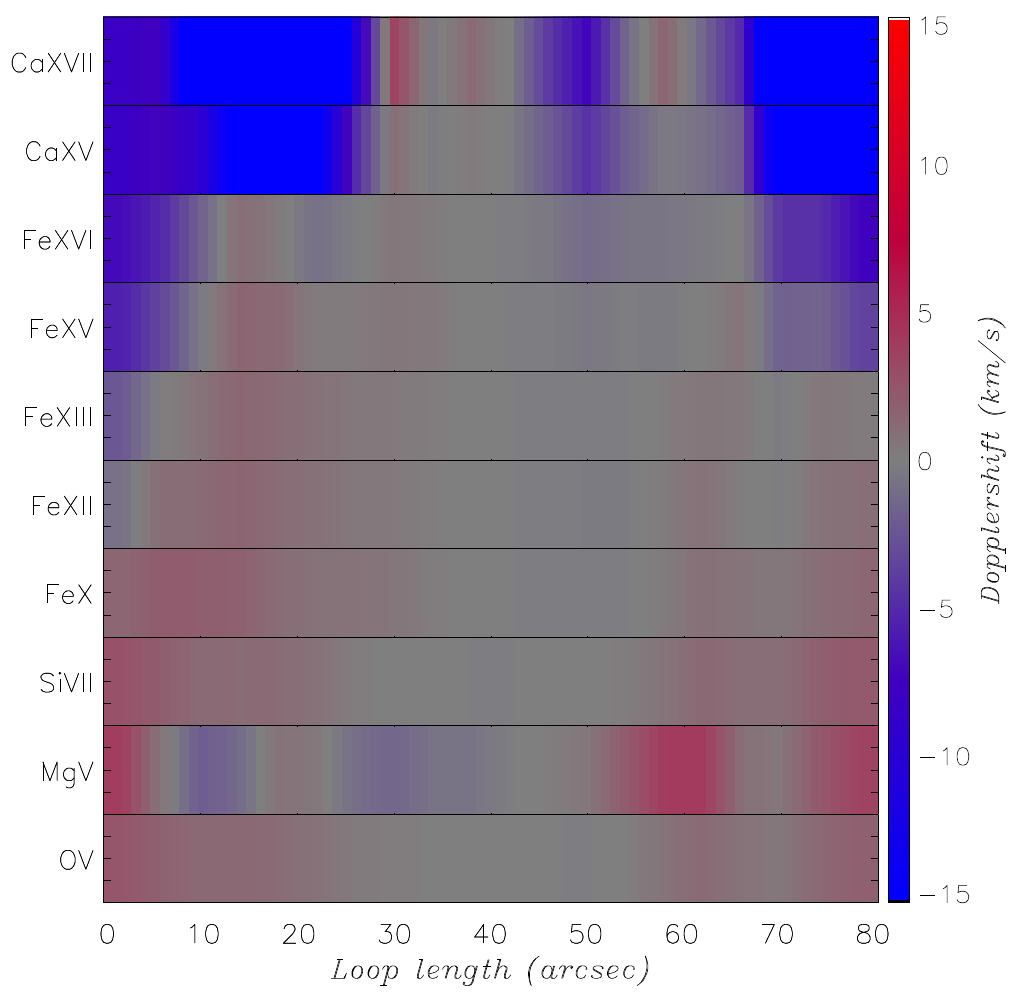}}
&
\includegraphics[width=0.24\linewidth]
	{\mydir{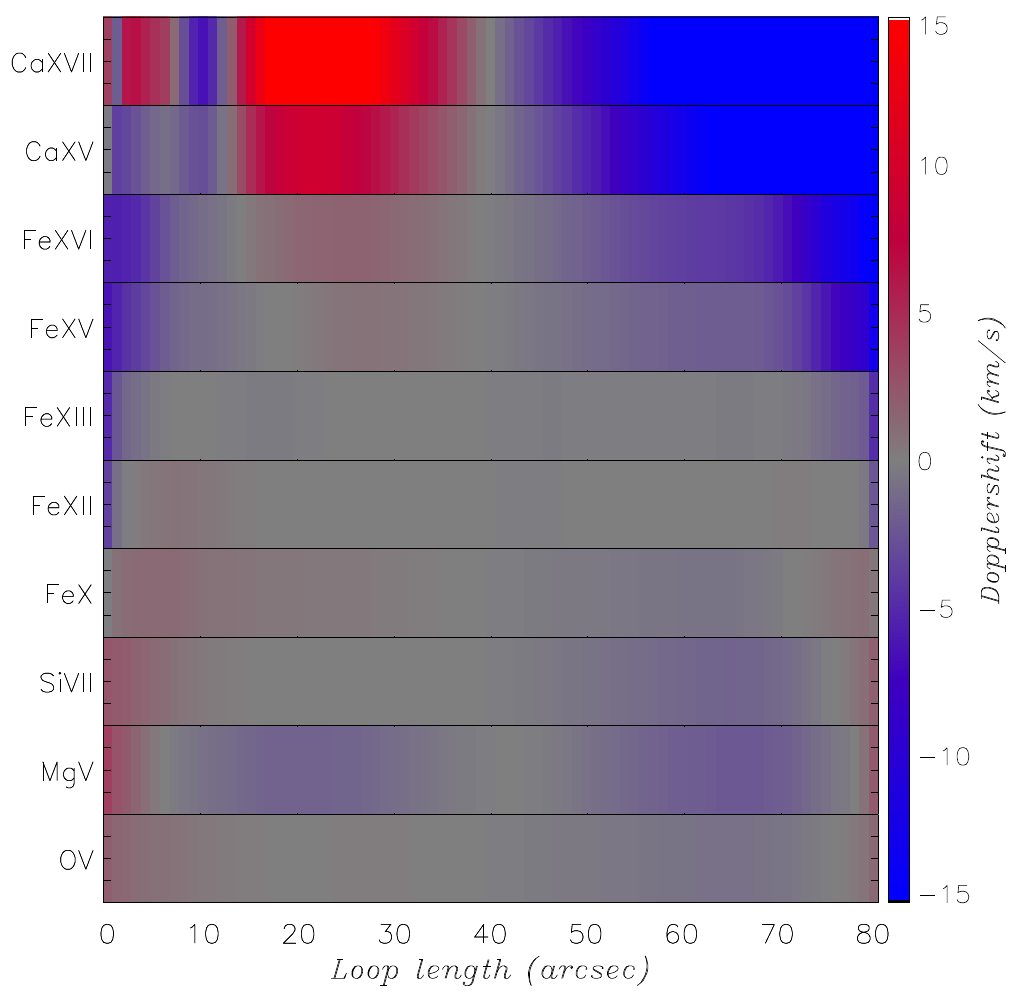}}
&
\includegraphics[width=0.24\linewidth]
	{\mydir{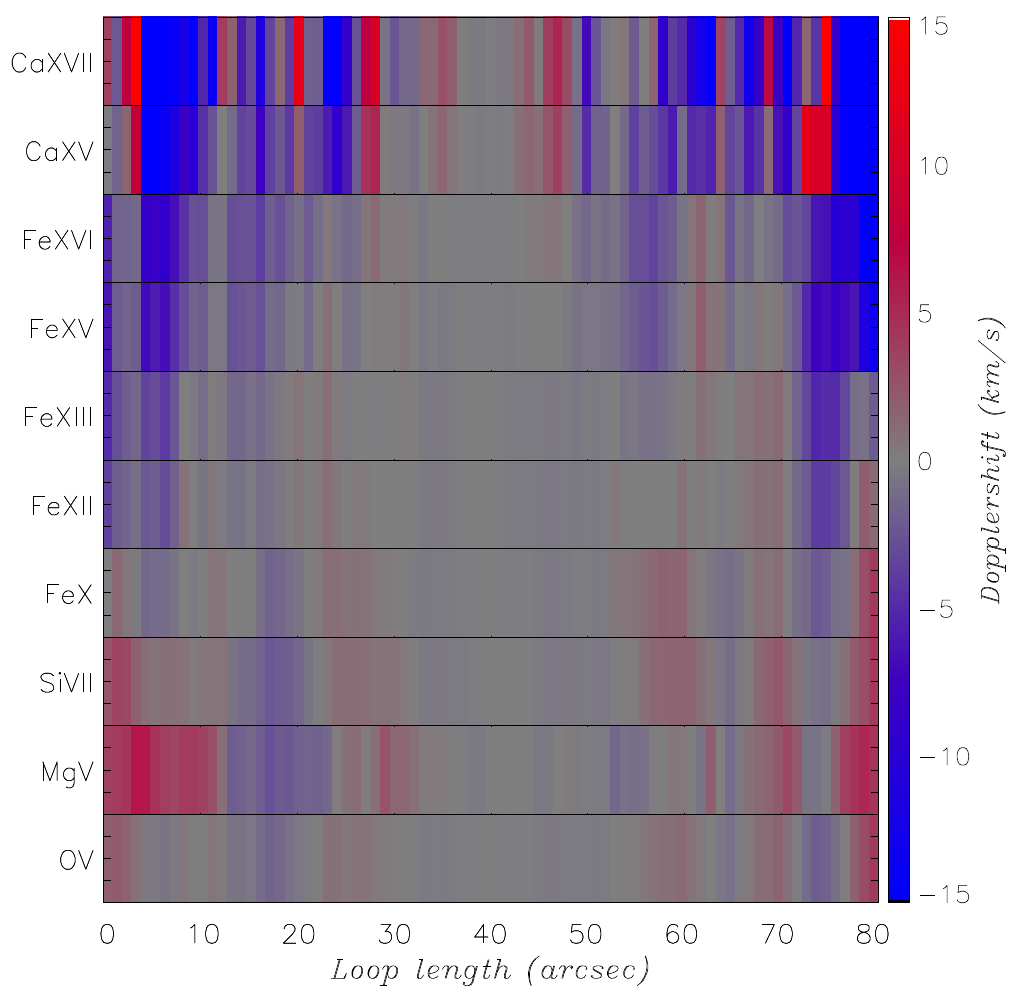}}
\end{tabular}
\caption{{\it Loop {\sc i}}: (a) and (b) Dopplershift for two consecutive simulated {\em Hinode}/EIS sparse rasters with a 1\arcsec~resolution, 50 s exposure time and a 3\arcsec~step
between exposures. The data are interpolated between exposures. (c) Dopplershift for simulated {\em Hinode}/EIS instantaneous observations (no raster); (d) Dopplershift for simulated {\em Hinode}/EIS dense raster (duration of $\sim$2 hours).  (c) and (d) have a
1\arcsec~resolution and a 50 s exposure time.}
\label{fig:raster_fp_warm_step}
\end{figure*}

%%%%%%%%%%%%%%%%%%%%%%%%%%%%%%%%%%%%%
%%%%%%%%%%%%%%%%%%%%%%%%%%%%%%%%%%%
%%%	Fig: raster hot		%%%
%%%%%%%%%%%%%%%%%%%%%%%%%%%%%%%%%%%
\begin{figure*}[!ht]
\centering
\begin{tabular}{cccc}
\multicolumn{1}{l}{(a)} & \multicolumn{1}{l}{(b)} 
	& \multicolumn{1}{l}{(c)} & \multicolumn{1}{l}{(d)} \\
\includegraphics[width=0.24\linewidth]
	{\mydir{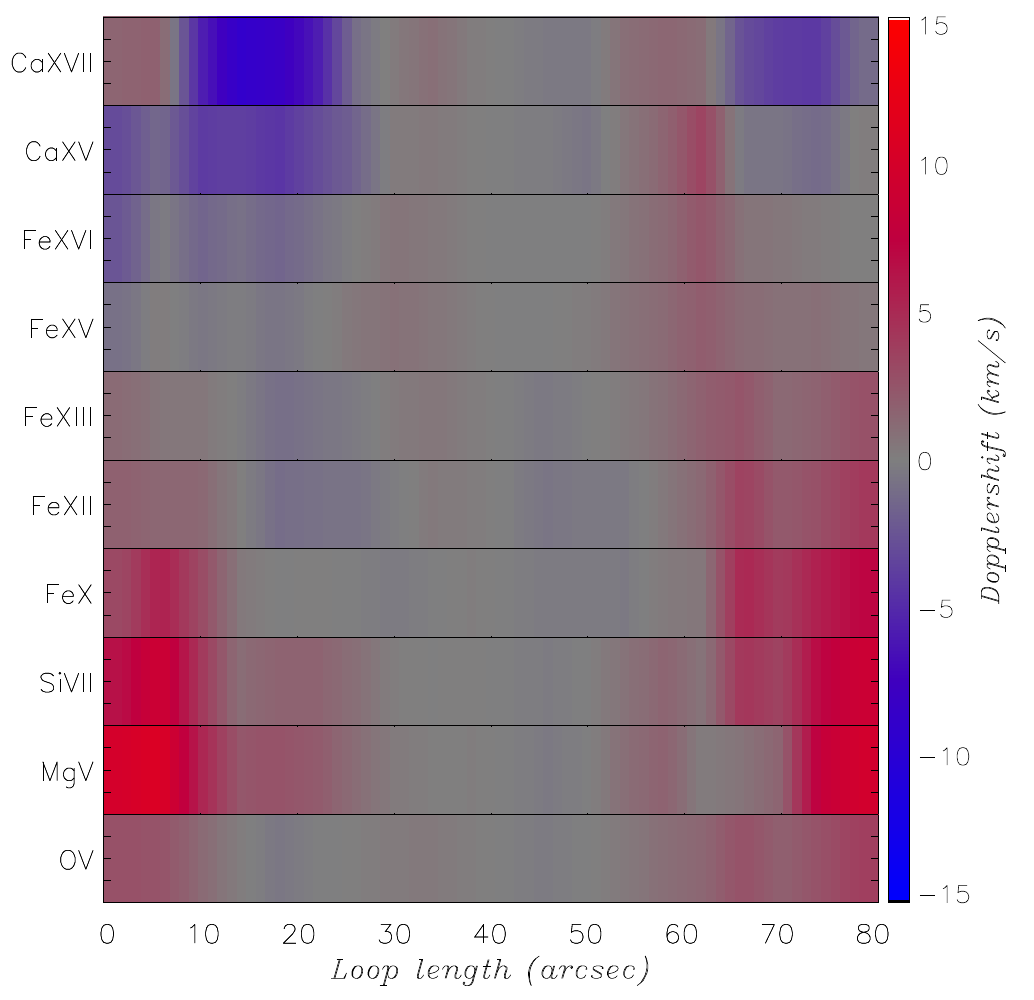}}
&
\includegraphics[width=0.24\linewidth]
	{\mydir{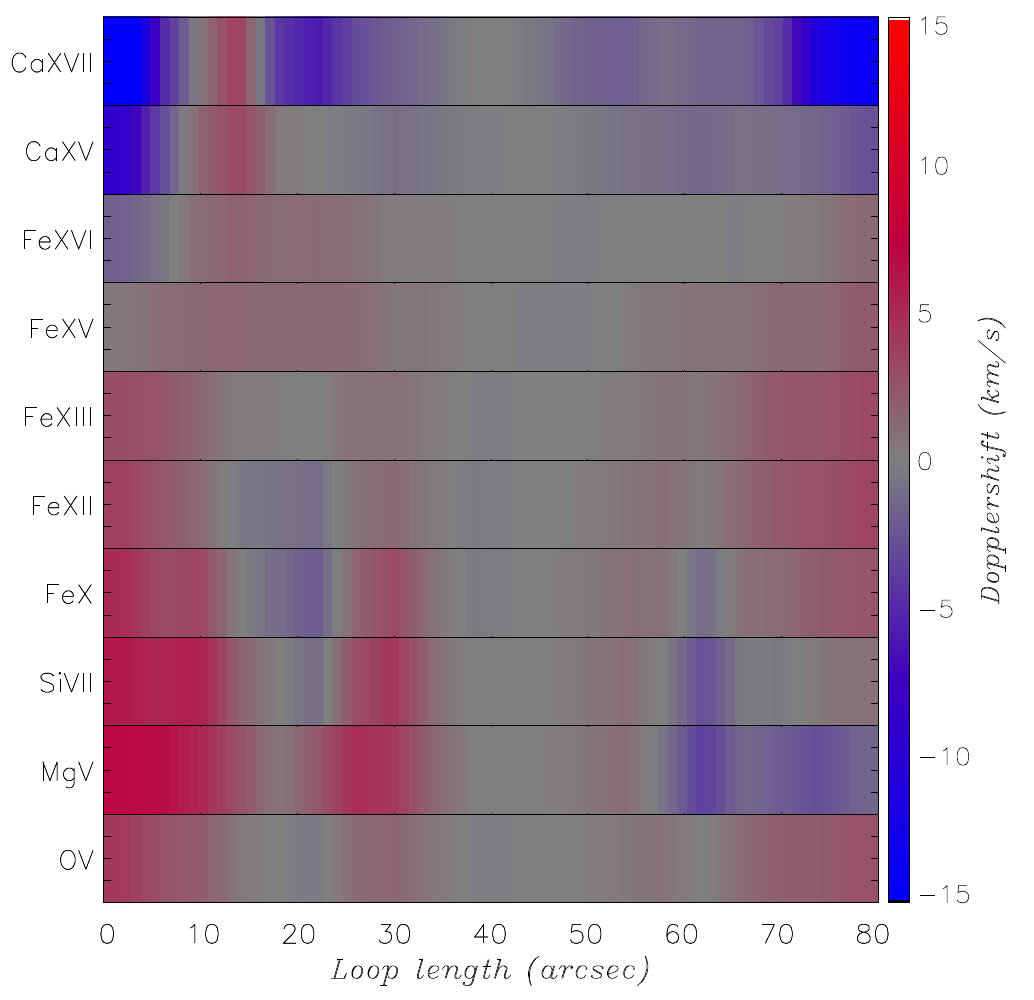}}
&
\includegraphics[width=0.24\linewidth]
	{\mydir{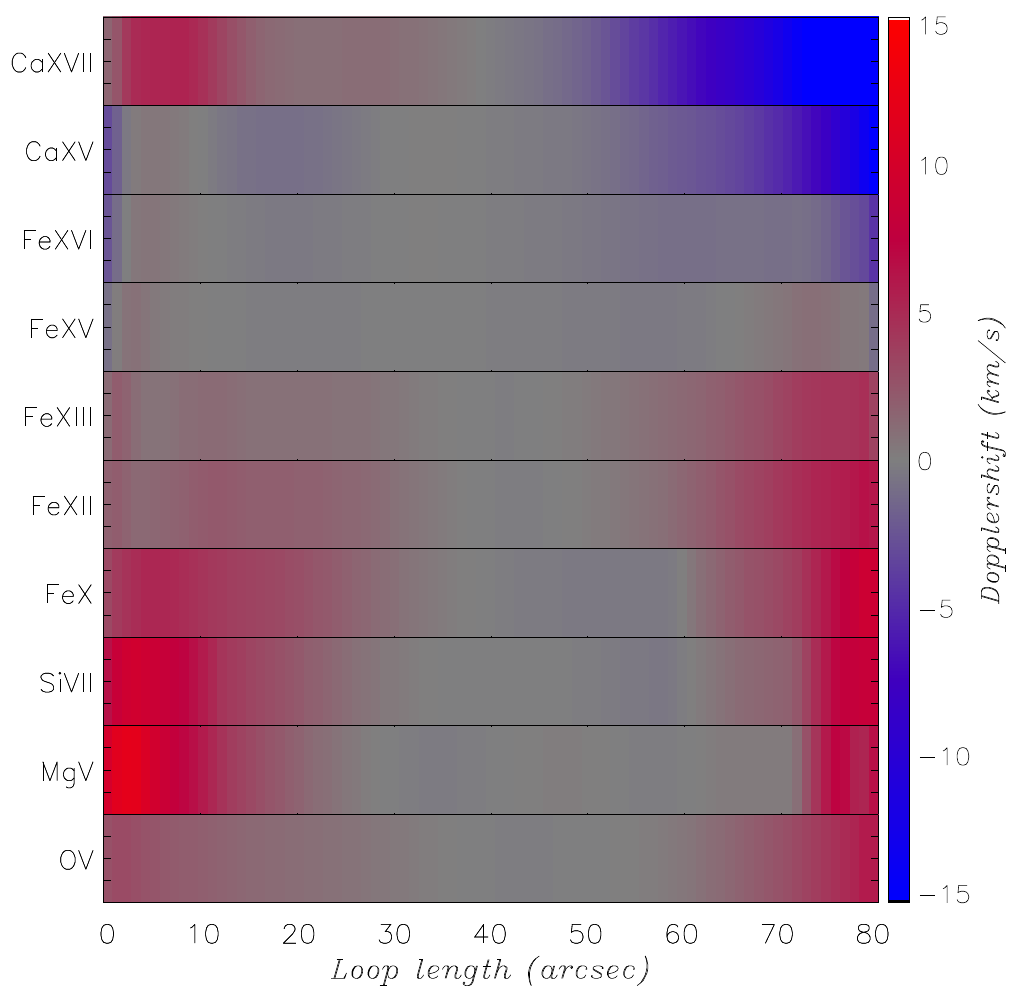}}
&
\includegraphics[width=0.24\linewidth]
	{\mydir{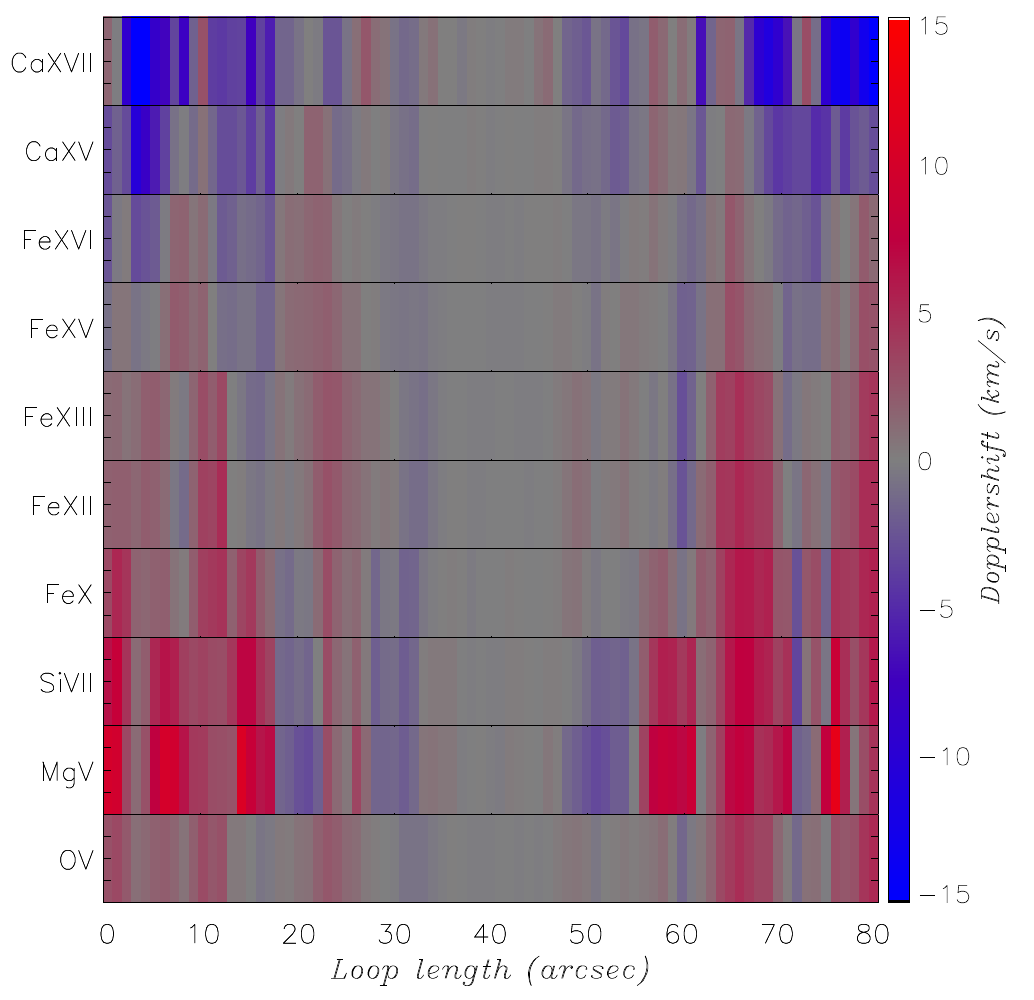}}
\end{tabular}
\caption{Same as in Figure~\ref{fig:raster_fp_warm_step} for {\it Loop \sc{ii}}.}
\label{fig:raster_fp_hot_step}
\end{figure*}
%%%%%%%%%%%%%%%%%%%%%%%%%%%%%%%%%%%%

To compare our simulation results with the {\em Hinode}/EIS observations, we
first construct a raster in a similar way to the EIS spectrometer. We raster the loop
assuming a viewing angle of 0$^\circ$ (see Figure~\ref{fig:setuploop}). In these simulated
observations, the 100 Mm long semi-circular loop correspond to an observed coronal structure of projected length 80 Mm. 
We
resample the simulated loop in order to obtain a pixel size of 1\arcsec\ and an
exposure time of 50 s. We neglect the CCD camera reading time. We simulate the
observations by \cite{war11} by constructing a raster with a 3\arcsec\ step
between successive observations. The stepping is introduced to speed up the
scanning of the region. As mentioned in \cite{war11}, the EIS raster takes about
52 minutes for scanning a length of 180\arcsec\ (about 130 Mm). For our
simulated loop of 80 Mm long in the corona, it will take about 30 min.  The data
are then interpolated between two consecutive exposure times as in \cite{war11}.
The resulting rasters for each spectral lines are shown in
Figure~\ref{fig:raster_fp_warm_step} for {\it Loop \sc{i}}, and in
Figure~\ref{fig:raster_fp_hot_step} for {\it Loop \sc{ii}}. In both
Figures~\ref{fig:raster_fp_warm_step} and \ref{fig:raster_fp_hot_step} (a) and (b), we
construct two consecutive rasters. The first raster starts at t = 3600 s and
lasts for 30 min, so the second one starts at t = 5400 s for the same duration.
The Dopplershift distribution depends on the time the raster has started. For
instance, while the cool lines exhibit only redshifts for the first raster, the 
second raster evidence blueshifts in the Mg {\sc
v} and Si {\sc vii} lines. The distributions are smooth due to the interpolation
between consecutive exposure times. Qualitatively, the rasters indicate the same
behavior as the modelled loop: redshift dominated for cool lines below Fe {\sc
xii} and Fe {\sc xv} for both modelled loops respectively, and blueshift
dominated above Fe {\sc xii} and Fe {\sc xv} for both modelled loops
respectively. The strong blue and redshifts are located at the footpoints of the
loop owe to a combined effect of high density and integration along the
line-of-sight. There is no symmetry between the two footpoints of the loop
due to the randomness of the energy deposition in individual strands. 

\cite{war11} reported {\em Hinode}/EIS observations of two different active regions
with a broad temperature coverage from Si {\sc vii} to Fe {\sc xv} (nine different
lines). Both observations show evidence of temperature dependence of the downflow and
upflow regions: downflows (redshifts) are dominating cool lines such as the Si {\sc
vii}, while upflows (blueshifts) dominate the hot lines (from Fe {\sc xi} to Fe {\sc
xv}). Their first observation (see Figure~3 of \citeauthor{war11} \citeyear{war11})
shows a clear sudden increase of the blueshift in the Fe {\sc xi} line, while their
second observation (see Figure~4 of \citeauthor{war11} \citeyear{war11}) shows an
increase in the Fe {\sc x} line. According to the above results for the multistranded
model (see Figures~\ref{fig:raster_fp_warm_step} and~\ref{fig:raster_fp_hot_step}), we
can complement \cite{war11} article by concluding that the active region loops are
warm loops with a characteristic temperature at the apex of about 2 MK, and that
the overall temperature of the first active region is larger than the temperature of
the second active region.

%%%%%%%%%%%%%%%%%%%%%%%%%%%%%%%%%%%%%%%%%%%
%%%	Fig: types of warm rasters	%%%
%%%%%%%%%%%%%%%%%%%%%%%%%%%%%%%%%%%%%%%%%%%
%\begin{figure}[!ht]
%\centering
%\includegraphics[width=0.32\linewidth]
%	{Images/raster_fp_warm_inst_0-eps-converted-to.pdf}
%\includegraphics[width=0.32\linewidth]
%	{Images/raster_fp_warm_0-eps-converted-to.pdf}
%\includegraphics[width=0.32\linewidth]
%	{Images/raster_fp_warm_step_0-eps-converted-to.pdf}
%\caption{Dopplershift for simulated {\em Hinode}/EIS along {\it Loop \sc{i}} with a
%1\arcsec~resolution and a 50 s exposure time: ({\em left}) instantaneous observations
%(no raster), ({\em middle}) raster of duration $\sim$2 hours, ({\em right}) raster
%with a 3\arcsec~step and of duration about 30 minutes.}
%\label{fig:raster_fp_warm}
%\end{figure}
%%%%%%%%%%%%%%%%%%%%%%%%%%%%%%%%%%%%%%%%%%%%%

%%%%%%%%%%%%%%%%%%%%%%%%%%%%%%%%%%%%%%%%%%%
%%%	Fig: types of hot rasters	%%%
%%%%%%%%%%%%%%%%%%%%%%%%%%%%%%%%%%%%%%%%%%%
%\begin{figure}[!ht]
%\centering
%\includegraphics[width=0.32\linewidth]
%	{Images/raster_fp_hot_inst_0-eps-converted-to.pdf}
%\includegraphics[width=0.32\linewidth]
%	{Images/raster_fp_hot_0-eps-converted-to.pdf}
%\includegraphics[width=0.32\linewidth]
%	{Images/raster_fp_hot_step_0-eps-converted-to.pdf}
%\caption{Same as in Figure~\ref{fig:raster_fp_warm} for {\it Loop \sc{ii}}.}
%\label{fig:raster_fp_hot}
%\end{figure}
%%%%%%%%%%%%%%%%%%%%%%%%%%%%%%%%%%%%%%%%%%%
We now compare the typical {\em Hinode}/EIS raster with a 3\arcsec\ step with two
other possible observations: (i) the instantaneous observation with an exposure time
of 50 s in Figures~\ref{fig:raster_fp_warm_step} and~\ref{fig:raster_fp_hot_step} (c), (ii) the dense raster without the stepping of 3\arcsec\ (consecutive 50 s
exposures) in Figures~\ref{fig:raster_fp_warm_step} and~\ref{fig:raster_fp_hot_step} (d). The latter raster takes about 2 hours to be completed. All three types of observations are
reproducing the main features of the Dopplershifts: (i) cool spectral lines dominated
by redshifts and hot spectral lines by blueshifts, (ii) both modelled loops have a blueshift
distribution dependent on their characteristic temperature that is to say on the total amount of energy injected in the loop. However, the Dopplershift
distribution for the instantaneous observation is smooth owe to the continuity
of the different physical quantities at a given time, the distribution of 3\arcsec\
step raster is also smooth due to the interpolation, while the distribution of the
1\arcsec\ continuous raster is patchy due to the different times of observations and
the bursty nature of the heating mechanism (and the possible occurrence of shocks).
Thus, only a series of instantaneous observations or the dense raster with sufficient
time and spatial resolutions evidence the nature of the heating mechanism along the
loop.

		%%%%%%%%%%%%%%%%%%%%%%%%%%%%%%%%%%%%%%%%%%%%%%%%%%%%%%%%%%
		\subsection{Temperature diagnostic from simulated rasters}
		%%%%%%%%%%%%%%%%%%%%%%%%%%%%%%%%%%%%%%%%%%%%%%%%%%%%%%%%%%

\begin{figure}[!t]
	\centering
	\includegraphics[width=1.\linewidth]{\mydir{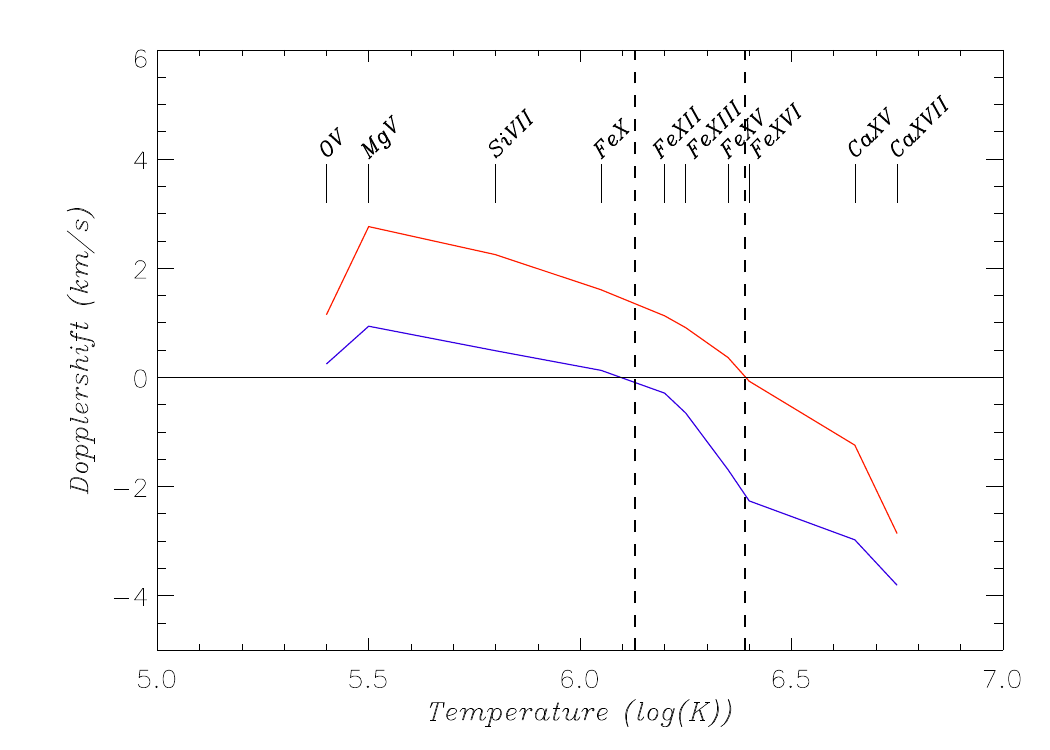}}
	\caption{Temperature diagnostic from a sparse raster (see Figure~\ref{fig:raster_fp_warm_step}(a) and Figure~\ref{fig:raster_fp_hot_step}(a)) for {\it Loop {\sc i}} (blue) and {\it Loop {\sc ii}} (red). }
	\label{fig:tdiag_raster}
\end{figure}

We apply the temperature diagnostic described in Section~\ref{sec:tdiag} by computing the average Dopplershift along the loop. Note that, at the resolution of {\em Hinode}/EIS, there are not enough points to obtain a statistically significant average Dopplershift at the footpoints of the loop. In Figure~\ref{fig:tdiag_raster}, we plot the avearge Dopplershift for {\it Loop {\sc i}} (blue) and {\it Loop {\sc ii}} (red) ass a function of $T_e$ the peak temperature of a spectral line. We only plot the temperature diagnostic for the sparse raster (with the 3\arcsec\ step); we have obtained identical results for the dense raster, the instantaneous observation. 

It is clear that the proposed temperature diagnostic method also works for a raster observation as we obtain the same estimates for the plasma temperature of the coronal loop. 

%%%%%%%%%%%%%%%%%%%%%%%%%%%%%%%%%%%%%%%%
%%%	Fig: temp diag. raster 	%%%%%%%%
%%%%%%%%%%%%%%%%%%%%%%%%%%%%%%%%%%%%%%%%
%%%%%%%%%%%%%%%%%%%%%%%%%%%%%%%%%%%%%%%%%

%%%%%%%%%%%%%%%%%%%%%%%%%%%%%%%%%%%%%%%
\section{Conclusions}\label{sec:concl2}
%%%%%%%%%%%%%%%%%%%%%%%%%%%%%%%%%%%%%%%

We have defined a new temperature diagnostic tool relying on the measure of the avearge Dopplershift for a broad range of observed spectral lines. 

First, we have complemented the work of \cite{sar08, sar09} regarding the thermodynamic
properties of coronal loops defined a collection of strands. We have computed
the physical properties of a 100 Mm loop with two different energy inputs with a
particular focus on the Dopplershifts distribution for different spectral lines.
The recent {\em Hinode}/EIS observations have put forward the existence of both
blueshift and redshift velocities in active region loops \citep{del08,war11},
the amount of Dopplershifts depending on the peak temperature of the observed
spectral line. The multistranded coronal loop model has reproduce the main
observed features: (1) red and blueshifts exist in active region loops of
realistic length (100 Mm); (2) the amount of red and blueshifts depends on the
peak temperature of the observed spectral line: cooler lines being redder,
hotter lines being bluer; (3) the amount of red and blueshifts depends on the
mean/average temperature of the loop (as an amalgamation of strands). These
properties are now well-known for coronal loops and have been reproduced with
other numerical experiments (see, for instance, \citeauthor{tar14}
\citeyear{tar14}). 

Second, the 1D multistranded hydrodynamic model offers more information on
the spatial and temporal evolution of heated coronal loops, and thus on the
physical processes at play. The comparison between the multistranded model and
modelled observations leads to the following physical interpretations:

\begin{itemize}
\item[-]{redshifts in the cooler spectral lines evidence the condensation of the
plasma mostly located near the footpoints of the loop: dense plasma cooling;}
\item[-]{blueshifts in the hotter spectral lines is a signature of the
evaporation of the plasma: hot plasma evacuate towards less dense regions;}
\item[-]{the coexistence of condensation and evaporation for the entire coronal
loop is owed to the multistrandedness nature of the loop.}
\end{itemize}

More importantly for future analysis of multi-spectral observations, this study
suggests that it is possible to estimate the mean/average temperature of a
coronal loop: the transition from redshift-dominated loop to blueshift-dominated
loop gives an estimate of the mean temperature of the coronal loop. 

As an
example, we refer to the study by \citet{li14} of two flare loops observed by
{\em SDO}/AIA and {\em Hinode}/EIS. The authors measured the Dopplershifts at
both footpoints of two coronal loops related to the flaring process. The {\em
Hinode}/EIS lines used in their study are Fe {\sc x} at 184.54\AA\
(log($T$)=6.0), Fe {\sc xii} at 192.39\AA\ (log($T$)=6.09), Fe {\sc xiii} at
202.04\AA\ (log($T$)=6.17), and Fe {\sc xv} at  284.16\AA\ (log($T$)=6.3). From their
Figure 2 showing the average Dopplershifts for the spectral lines mentioned
above, we can derive from the temperature diagnostic that (assuming a view from
the top of the loop):

\begin{itemize}
\item[-]{the first set of loops has a temperature close to 1 MK for the first
half an hour ofthe observation as the average Dopplershift in all spectral lines
is negative;}
\item[-]{the first set of loops is then heated to a temperature above 2 MK for
about 1h45min for the northern footpoints and for about 15min for the southern
footpoints;}
\item[-]{the second set of loops shows a very different behavior: the heating
of the loop is seen as less and less blueshift dominated curve for the northern
footpoints; whilst the southern footpoints exhibit two consecutive heating
events.}
\item[-]{the ordering of the spectral lines is consistent with a top view of the
loops for the northern footpoints of both sets of loops; whilst the ordering for
the southern footpoints is characteristic of an highly inclined field (see
Appendix~\ref{sec:appa}).}
\end{itemize}

Our temperature diagnostic allows us to deduce the same properties of the loop
plasma than \citeauthor{li14} (\citeyear{li14}). The temperature diagnostic from
average Dopplershift measured along the loop or at the footpoints is thus a
robust diagnostic of the temperature of the loop. As a study case, the active
region moss is observed to be dominated by redshift velocities for spectral
lines of peak temperature less than 1.78 MK (Fe {\sc xiii}) according to
\citet{tri12} and \citet{win13}. From the temperature diagnostic detailed above,
we then conclude that the temperature of the coronal loop associated with moss
is above 1.78 MK, which is consistent with the moss being the transition region counter-part
of hot core coronal active region loops \citep{mar00, tri10}. 

In a forthcoming paper, we discuss the temperature diagnostic discussed here applied to {\em IRIS}
and {\em Solar Orbiter}/SPICE characteristic spectral lines. 
%As notice in
%Appendix~\ref{sec:appb}, the temperature diagnostic will be useful for coronal
%loops having a temperature less than 2 MK due to the limitation in the
%spectral lines selected.   

Knowing that the coronal loop temperature can be derived using multi-spectral
observations in the frame of this multi-stranded model, the future work will be
to produce temperature maps within active regions as well as further estimate of
the energy release. The temperature diagnostic will fail at locations where the
heating mechanism is different than the impulsive release of energy considered
in this paper. Therefore, the multi-standed hydrodynamic model can give us more
information on the plausible heating mechanisms at play in active regions. 

\acknowledgments
CHIANTI is a collaborative project involving George Mason University, the
University of Michigan (USA) and the University of Cambridge (UK).

%\bibliographystyle{natbib}
%\bibliography{bib_heating}

%\newpage

\appendix
\section{Viewing Angle Effects} \label{sec:appa}

\begin{figure}[!ht]
\centering
\begin{tabular}{ccccc}
\includegraphics[width=0.18\linewidth, bb= 10 0 560 360]
	{\mydir{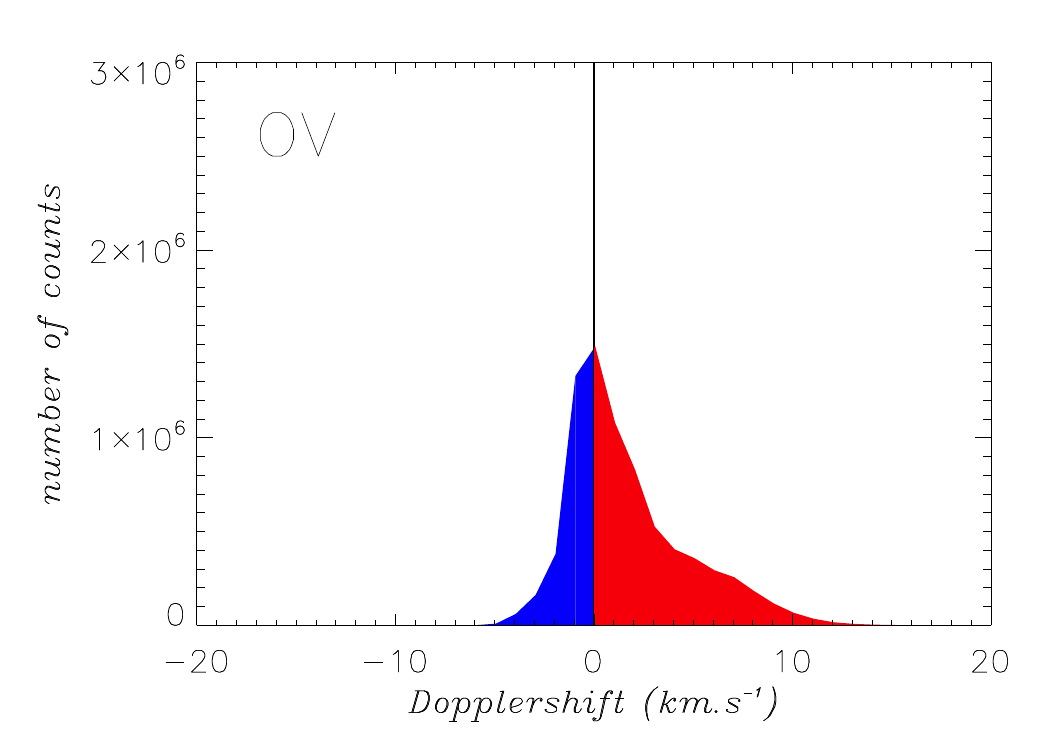}}
&
\includegraphics[width=0.18\linewidth, bb= 10 0 560 360]
	{\mydir{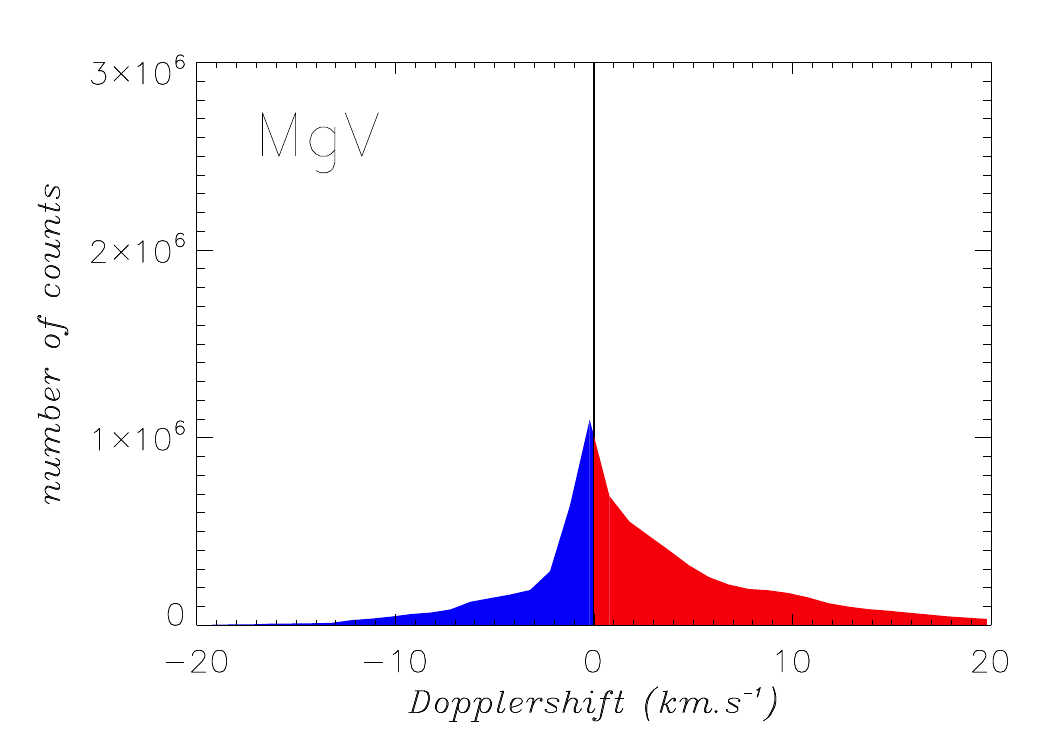}}
&
\includegraphics[width=0.18\linewidth, bb= 10 0 560 360]
	{\mydir{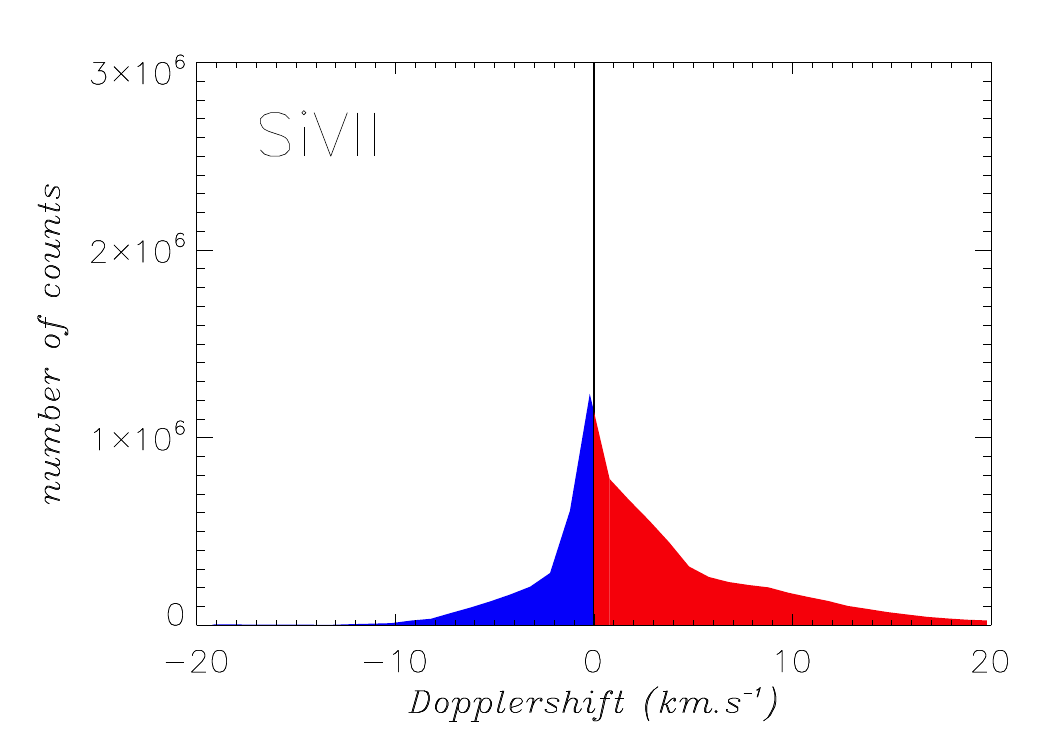}}
&
\includegraphics[width=0.18\linewidth, bb= 10 0 560 360]
	{\mydir{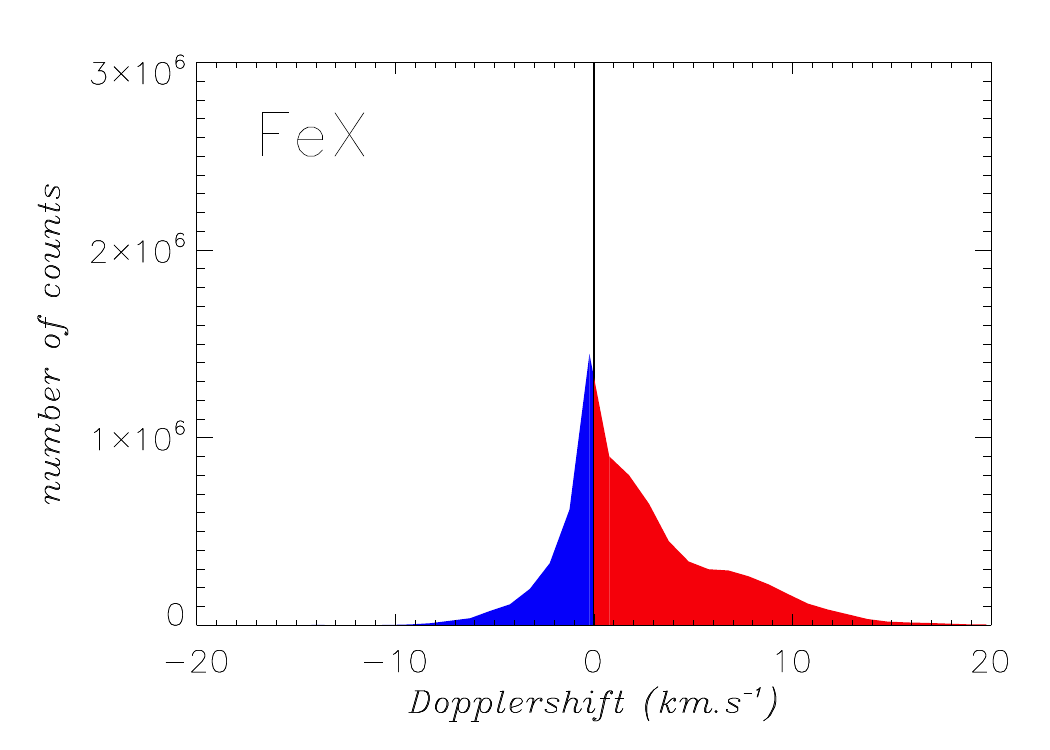}}
&
\includegraphics[width=0.18\linewidth, bb= 10 0 560 360]
	{\mydir{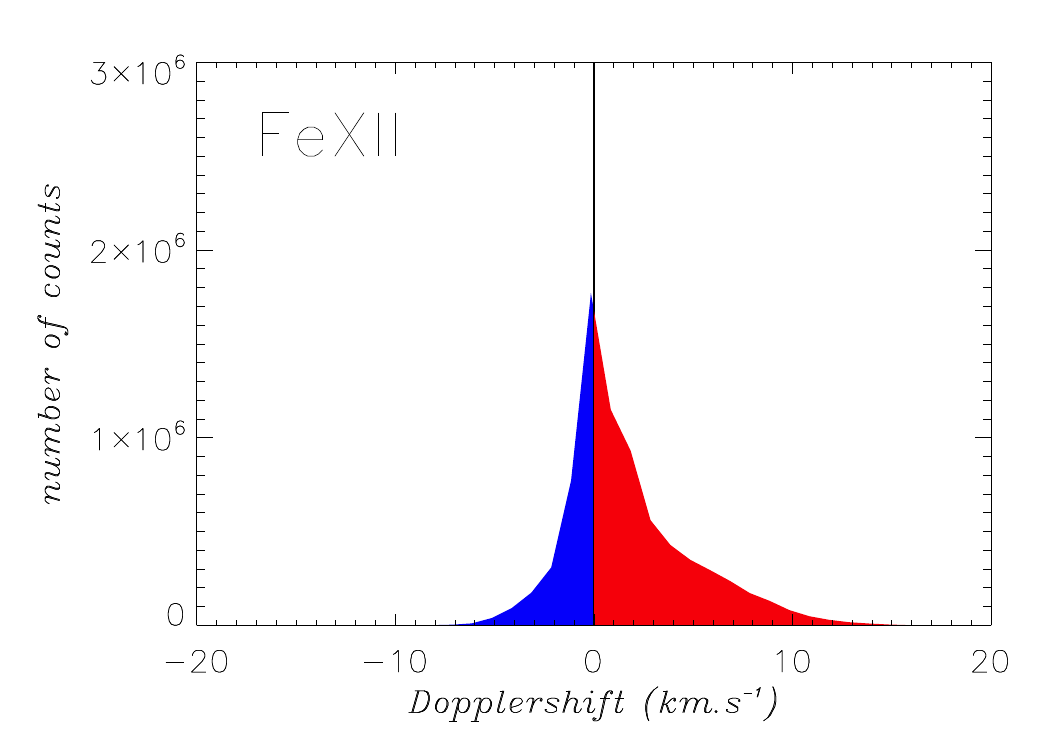}} \\
 
\includegraphics[width=0.18\linewidth, bb= 10 0 560 360]
	{\mydir{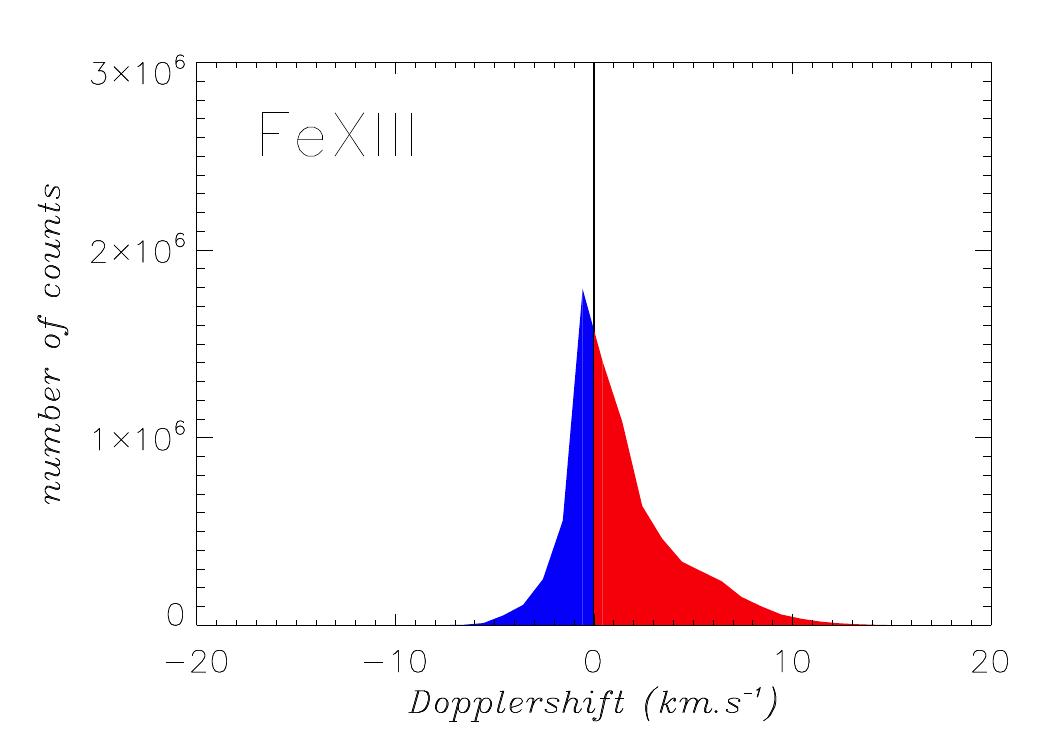}}
&
\includegraphics[width=0.18\linewidth, bb= 10 0 560 360]
	{\mydir{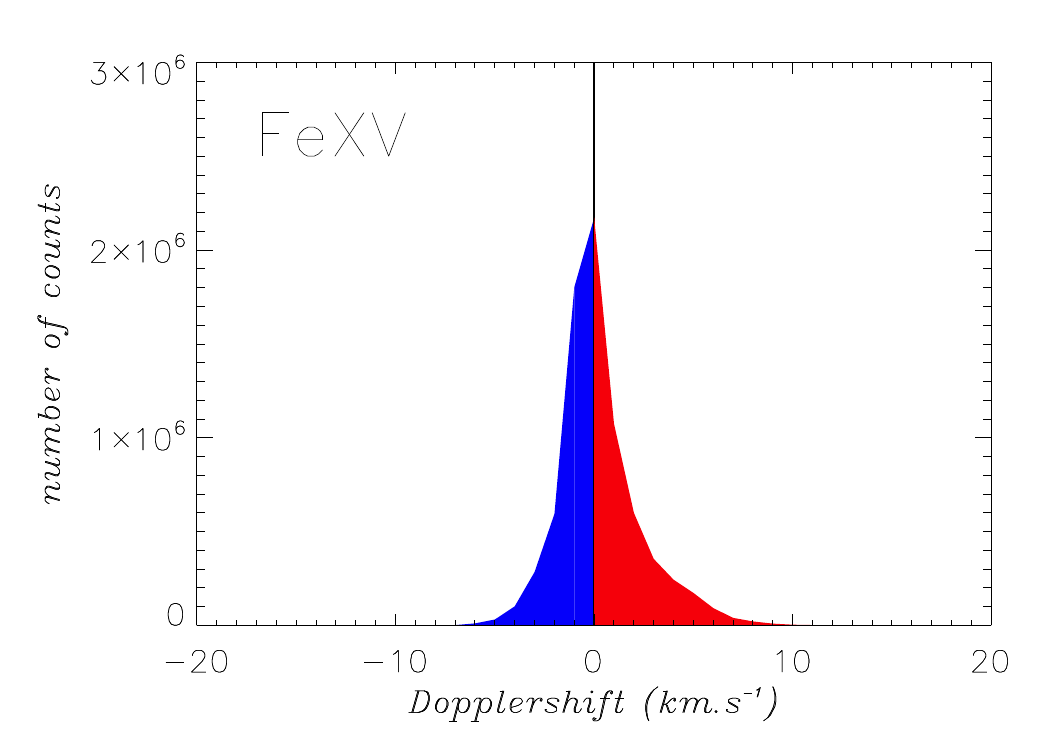}}
&
\includegraphics[width=0.18\linewidth, bb= 10 0 560 360]
	{\mydir{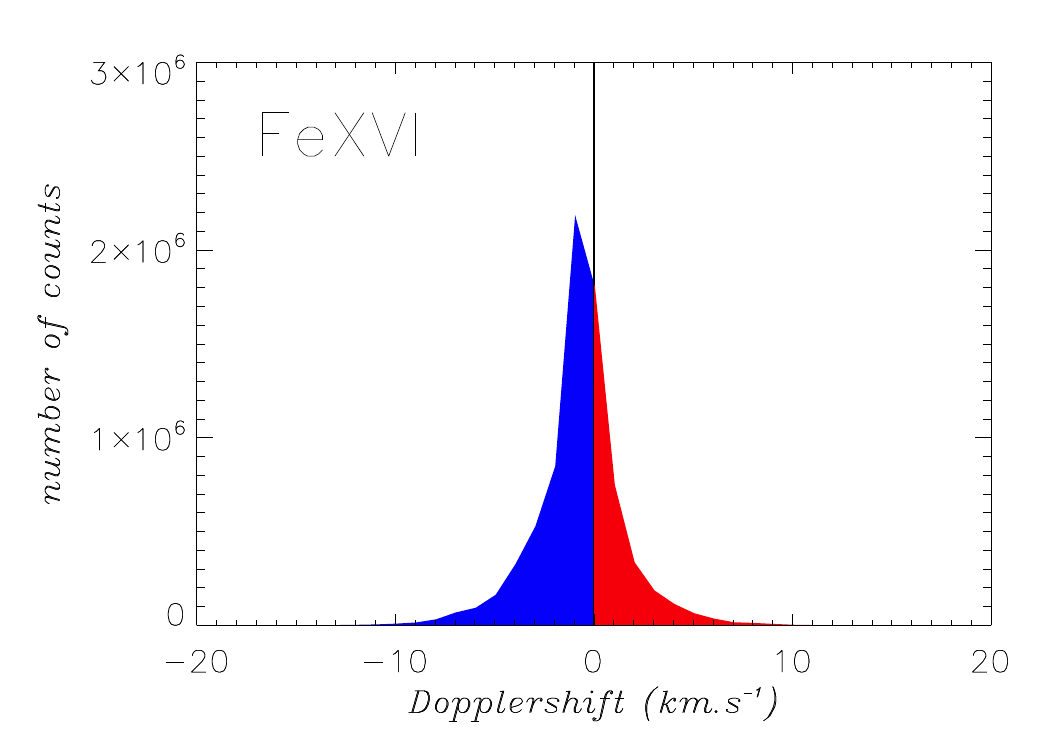}}
&
\includegraphics[width=0.18\linewidth, bb= 10 0 560 360]
	{\mydir{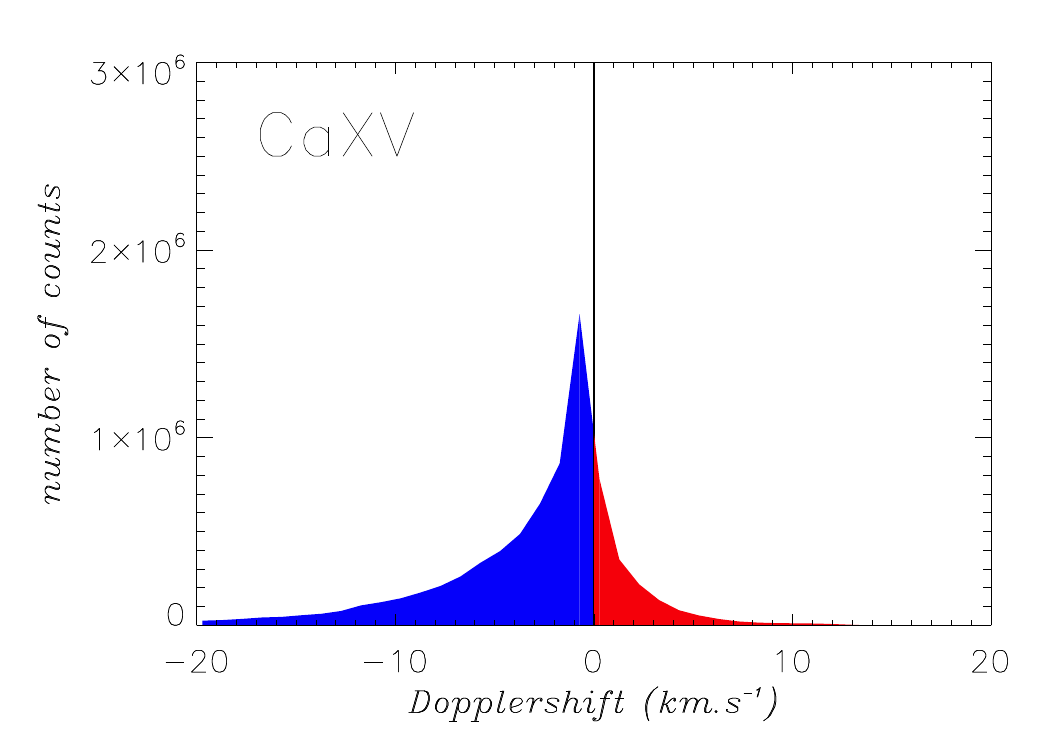}}
&
\includegraphics[width=0.18\linewidth, bb= 10 0 560 360]
	{\mydir{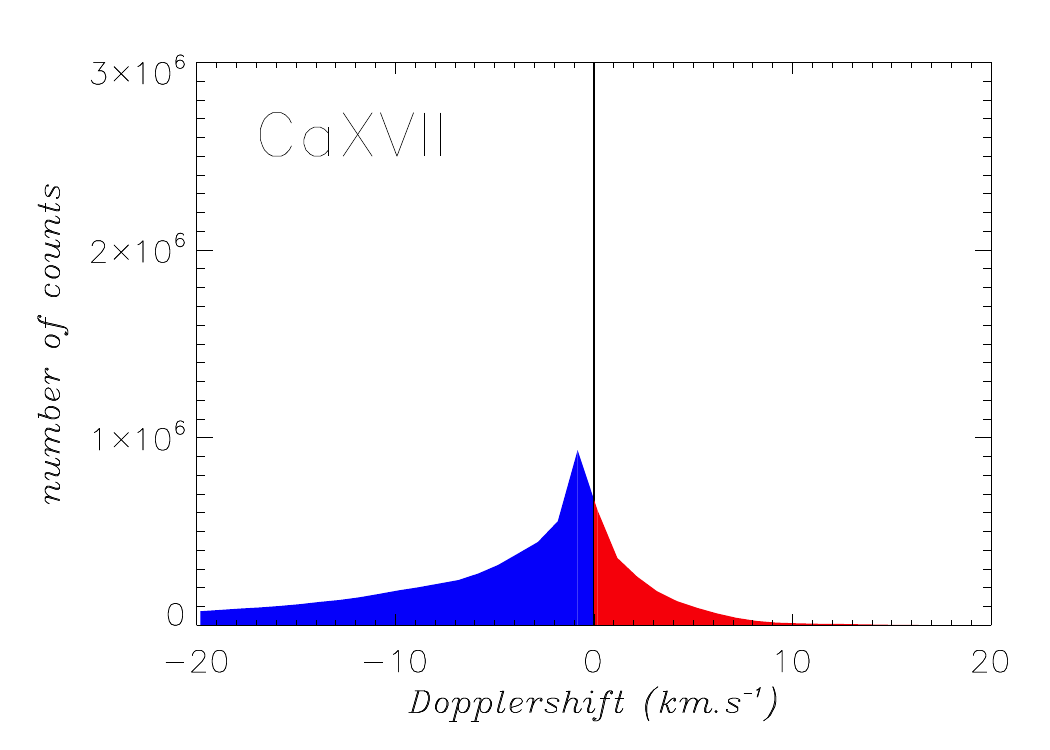}}
\end{tabular}
\caption{Statistical distribution of Dopplershift velocities for {\it Loop {\sc ii}} 
({\em fp}
heating) once the equilibrium is established for all ten wavelengths from O {\sc v} to
Ca {\sc xvii} (see Table~\ref{tab:lines}). The observer direction makes an angle
of 20$^\circ$ with respect to the vertical direction.}
\label{fig:distr_angle20}
\end{figure}

\begin{figure}[!ht]
\centering
\begin{tabular}{ccccc}
\includegraphics[width=0.18\linewidth, bb= 10 0 560 360]
	{\mydir{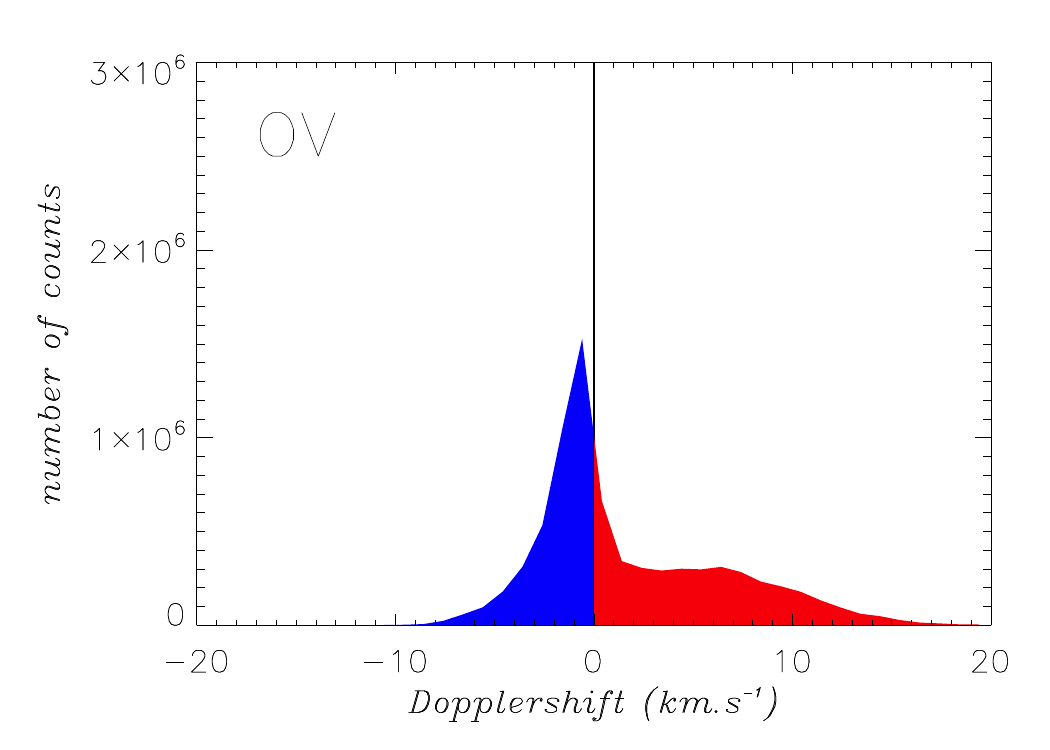}}
&
\includegraphics[width=0.18\linewidth, bb= 10 0 560 360]
	{\mydir{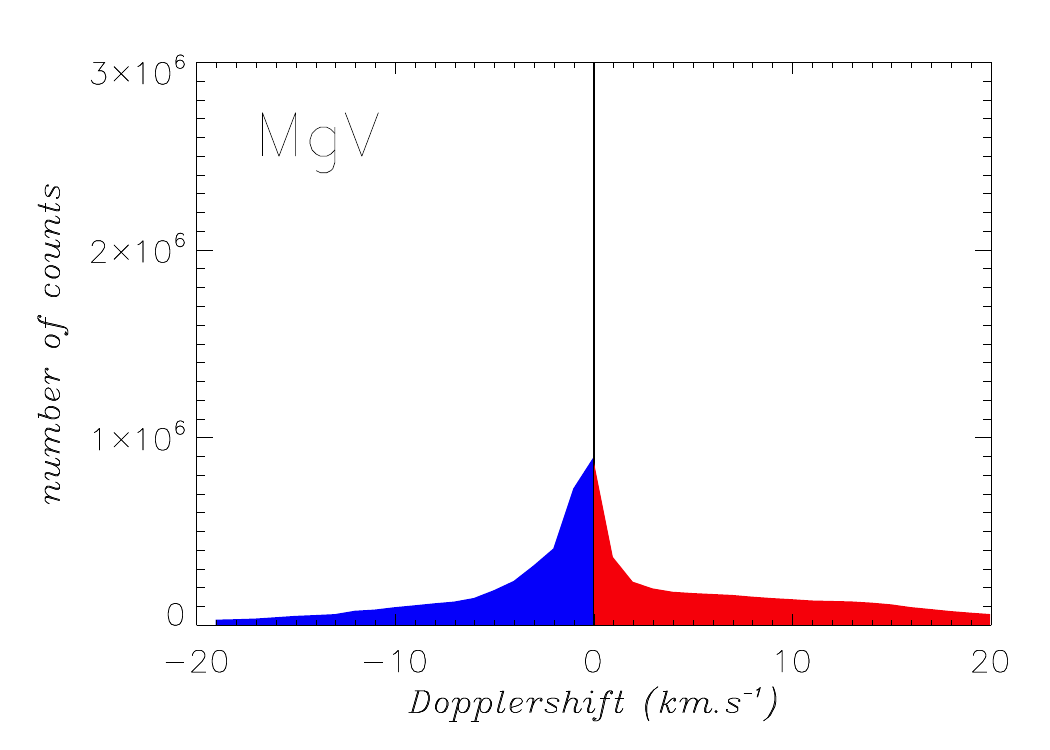}}
&
\includegraphics[width=0.18\linewidth, bb= 10 0 560 360]
	{\mydir{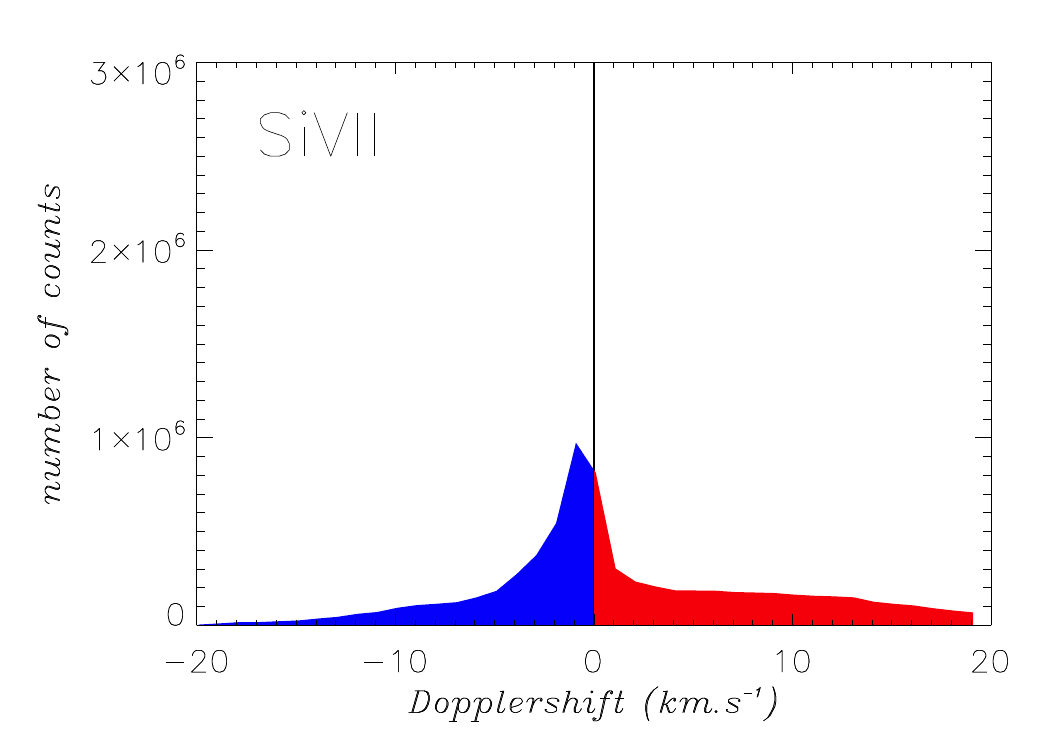}}
&
\includegraphics[width=0.18\linewidth, bb= 10 0 560 360]
	{\mydir{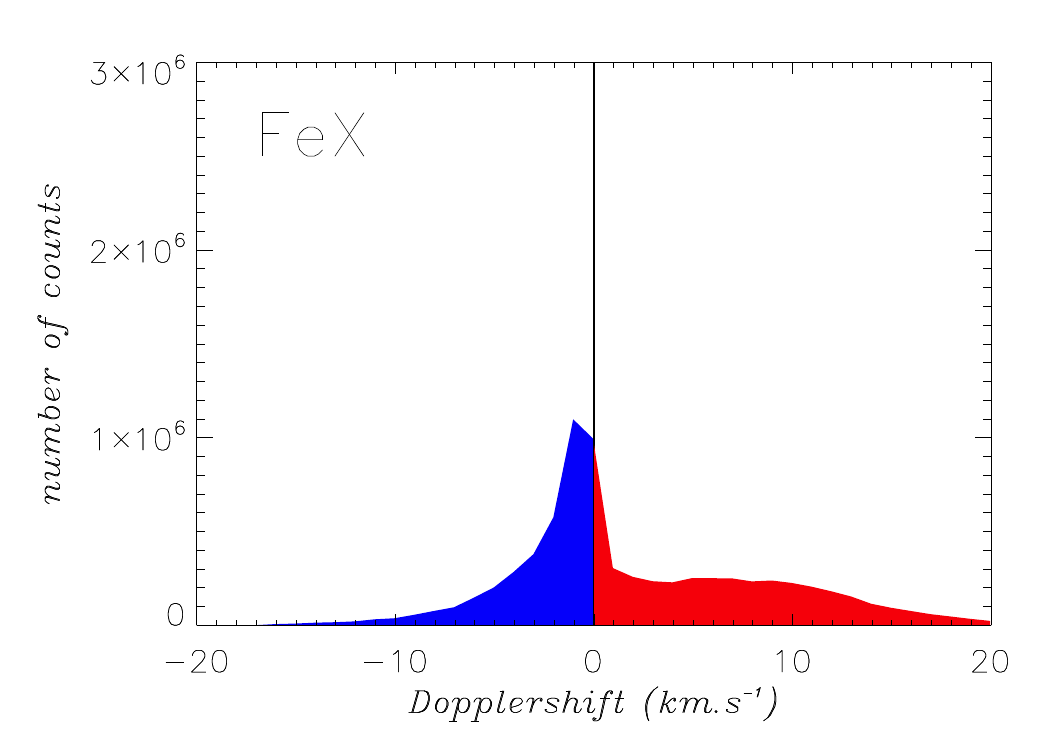}}
&
\includegraphics[width=0.18\linewidth, bb= 10 0 560 360]
	{\mydir{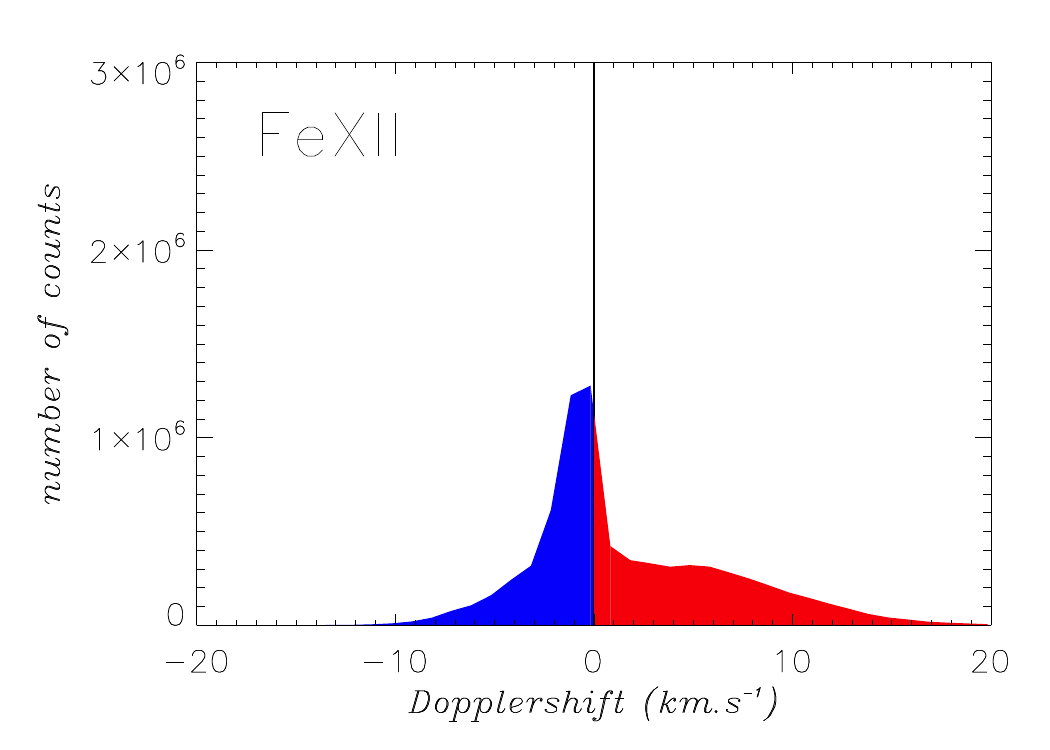}} \\

\includegraphics[width=0.18\linewidth, bb= 10 0 560 360]
	{\mydir{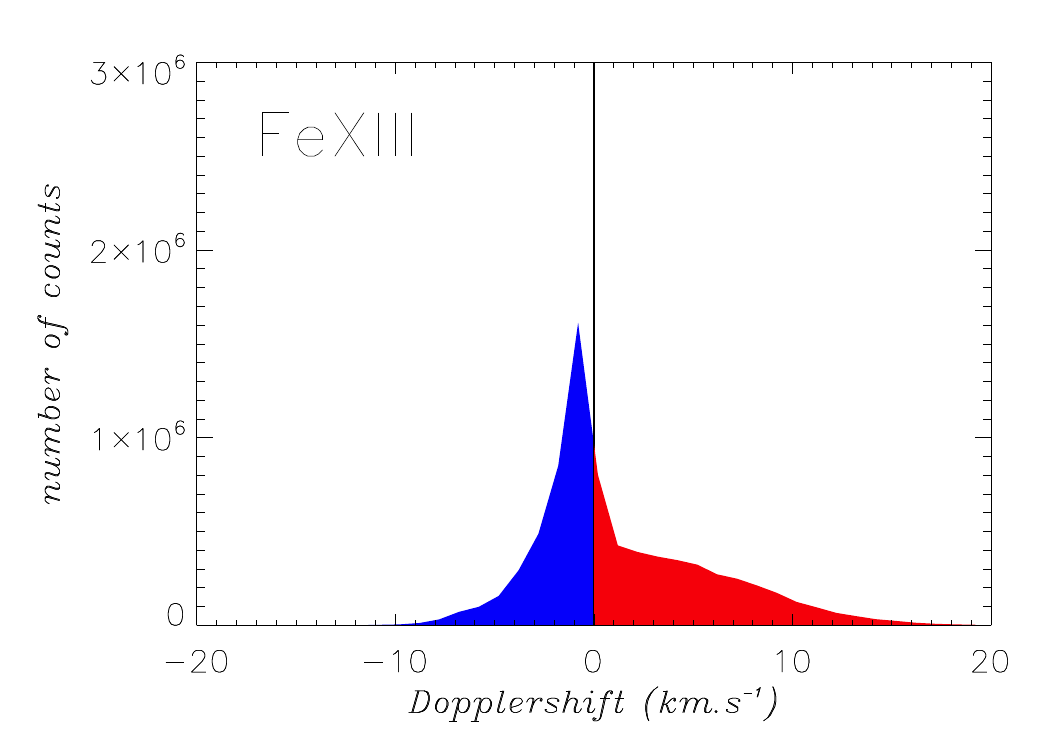}}
&
\includegraphics[width=0.18\linewidth, bb= 10 0 560 360]
	{\mydir{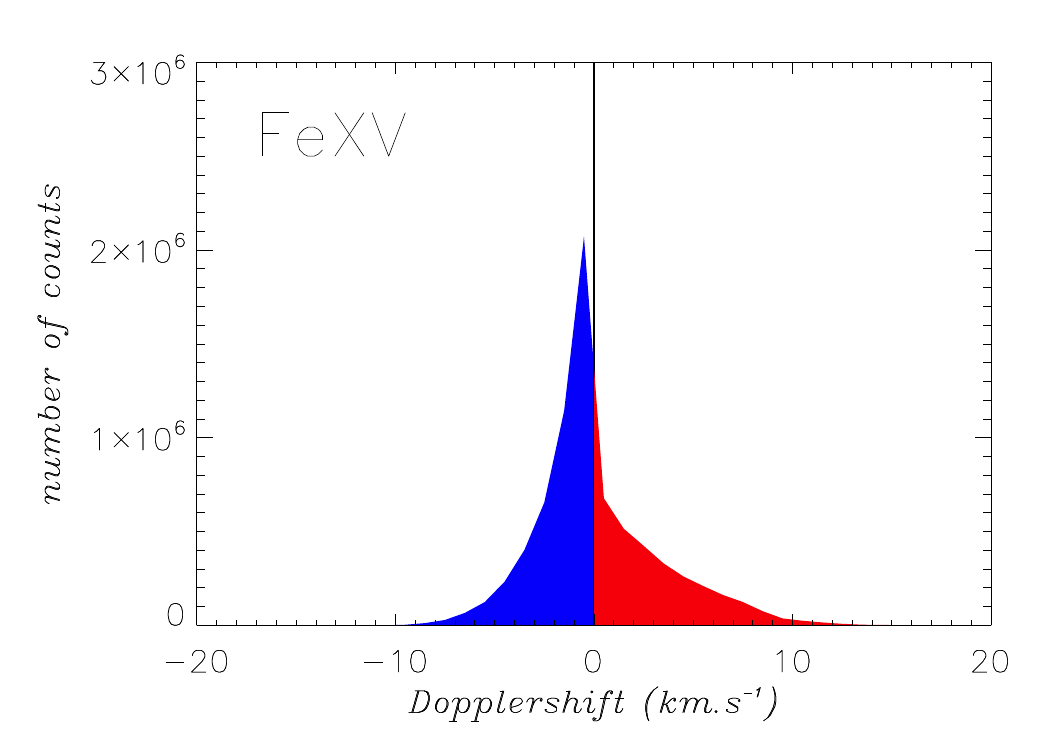}}
&
\includegraphics[width=0.18\linewidth, bb= 10 0 560 360]
	{\mydir{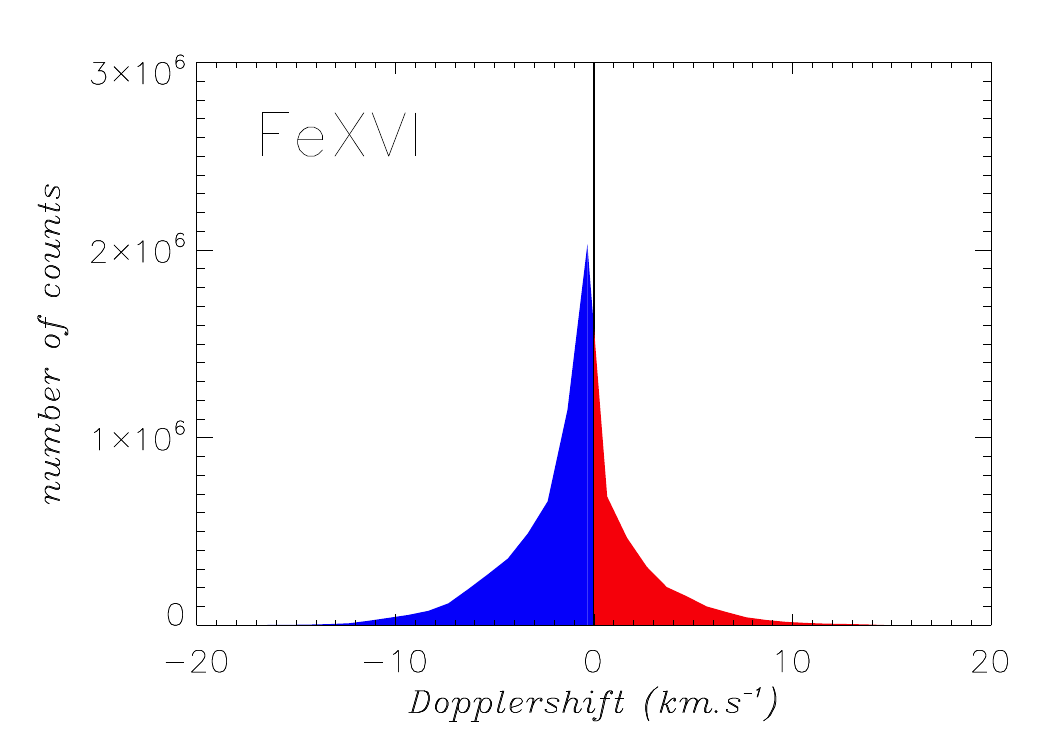}}
&
\includegraphics[width=0.18\linewidth, bb= 10 0 560 360]
	{\mydir{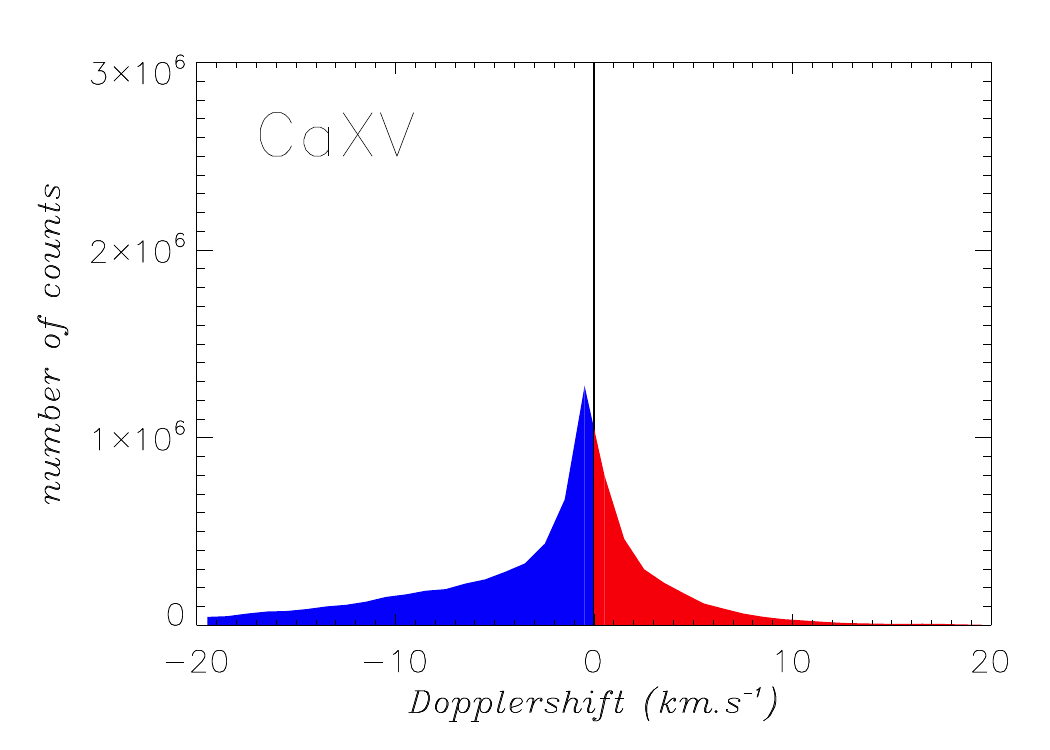}}
&
\includegraphics[width=0.18\linewidth, bb= 10 0 560 360]
	{\mydir{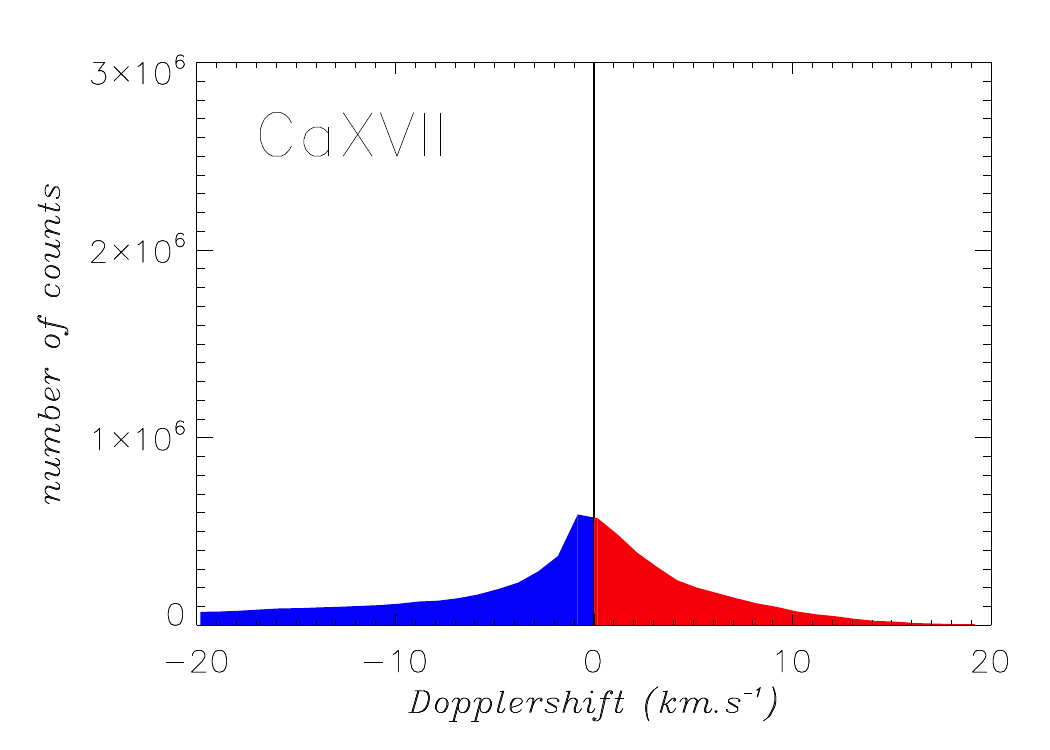}}
\end{tabular}
\caption{Same as Figure~\ref{fig:distr_angle20} for an angle of 45$^\circ$ 
with respect to the vertical direction.}
\label{fig:distr_angle45}
\end{figure}

\begin{figure}[!ht]
\centering
\begin{tabular}{ccccc}
\includegraphics[width=0.18\linewidth, bb= 10 0 560 360]
	{\mydir{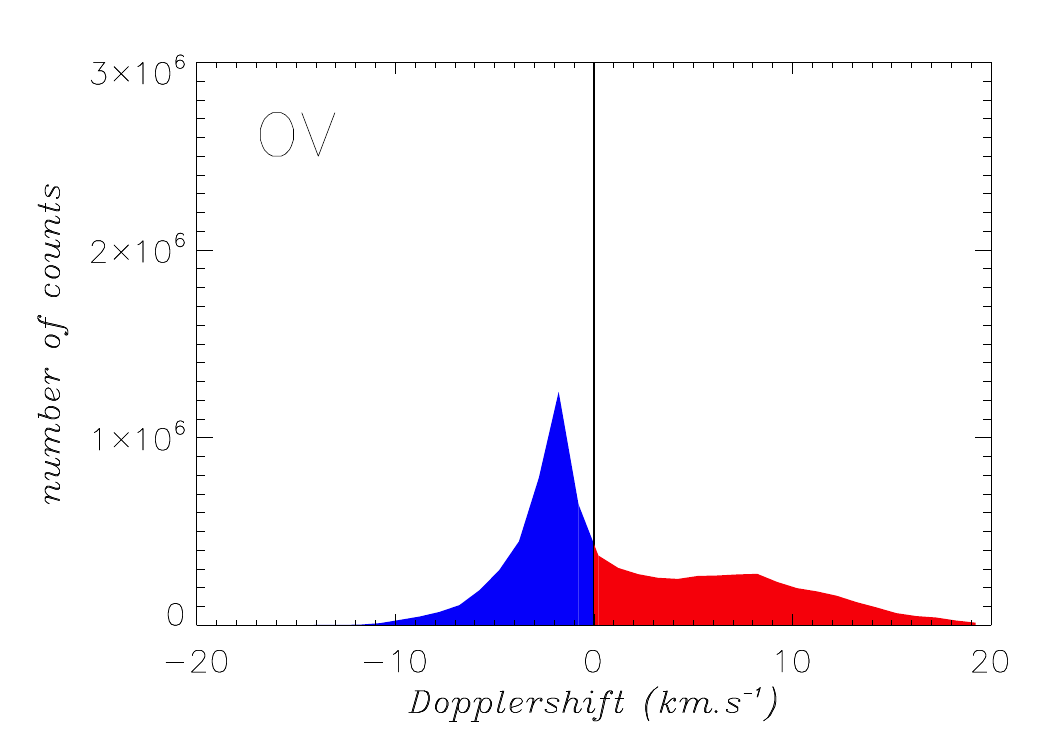}}
&
\includegraphics[width=0.18\linewidth, bb= 10 0 560 360]
	{\mydir{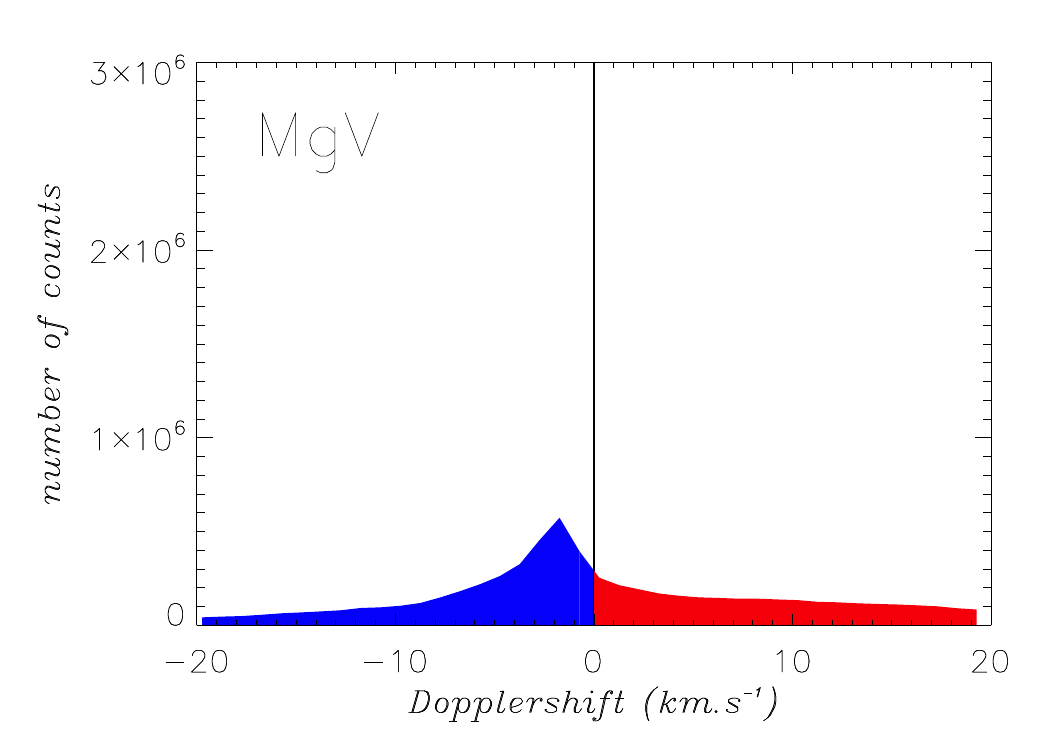}}
&
\includegraphics[width=0.18\linewidth, bb= 10 0 560 360]
	{\mydir{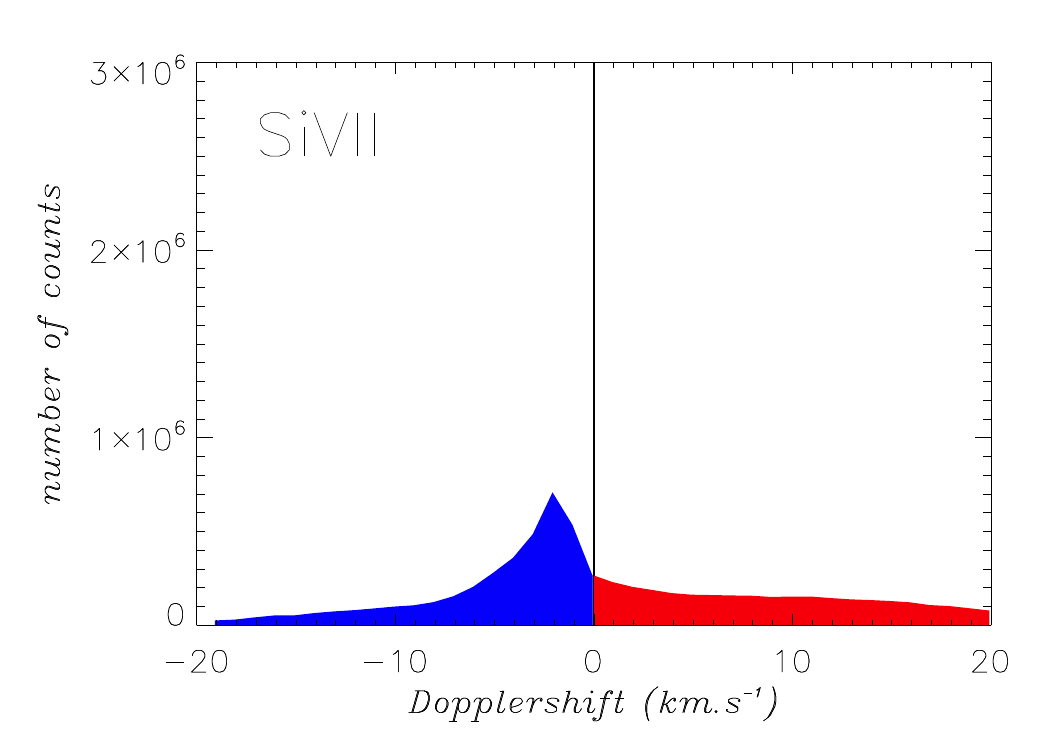}}
&
\includegraphics[width=0.18\linewidth, bb= 10 0 560 360]
	{\mydir{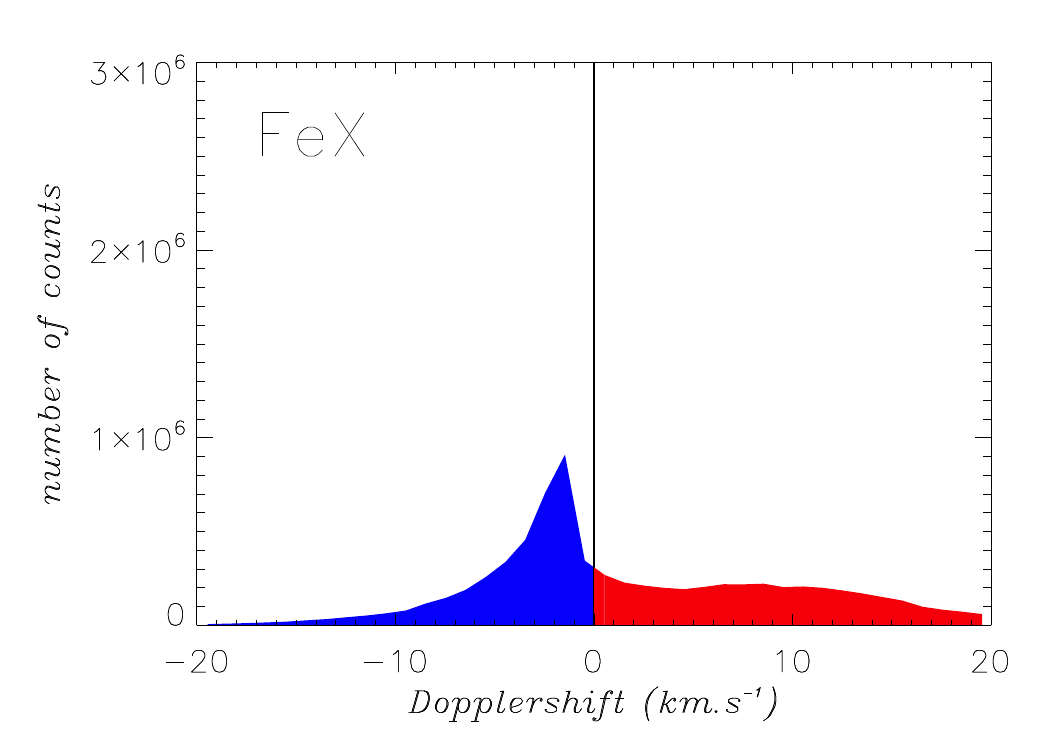}}
&
\includegraphics[width=0.18\linewidth, bb= 10 0 560 360]
	{\mydir{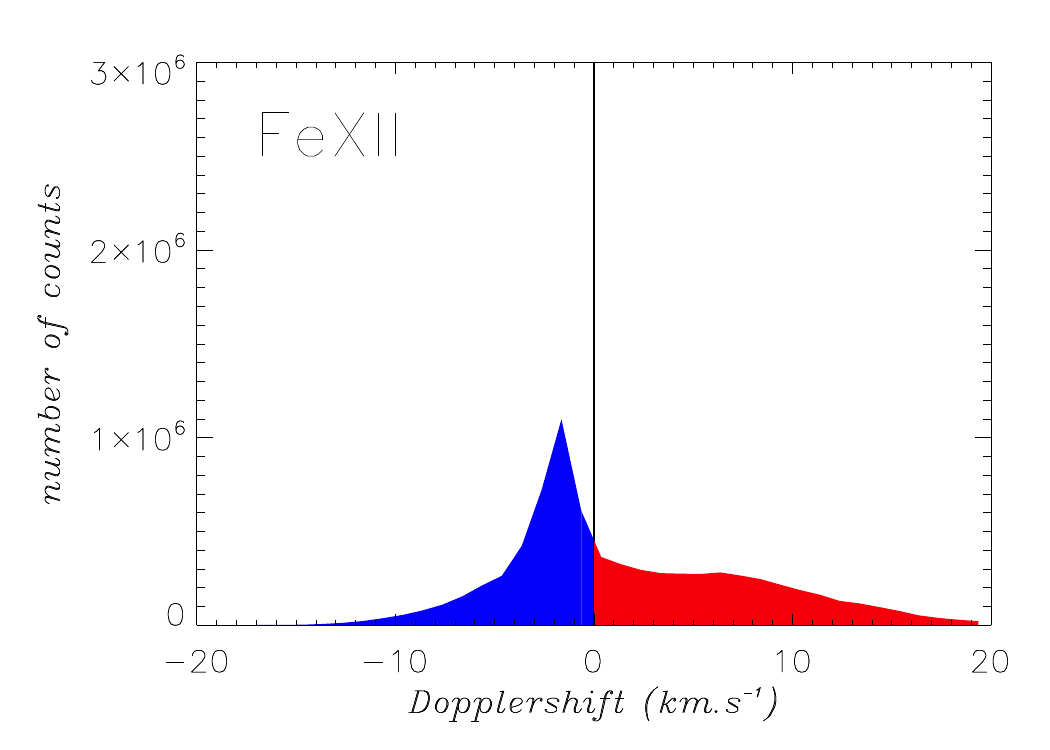}} \\

\includegraphics[width=0.18\linewidth, bb= 10 0 560 360]
	{\mydir{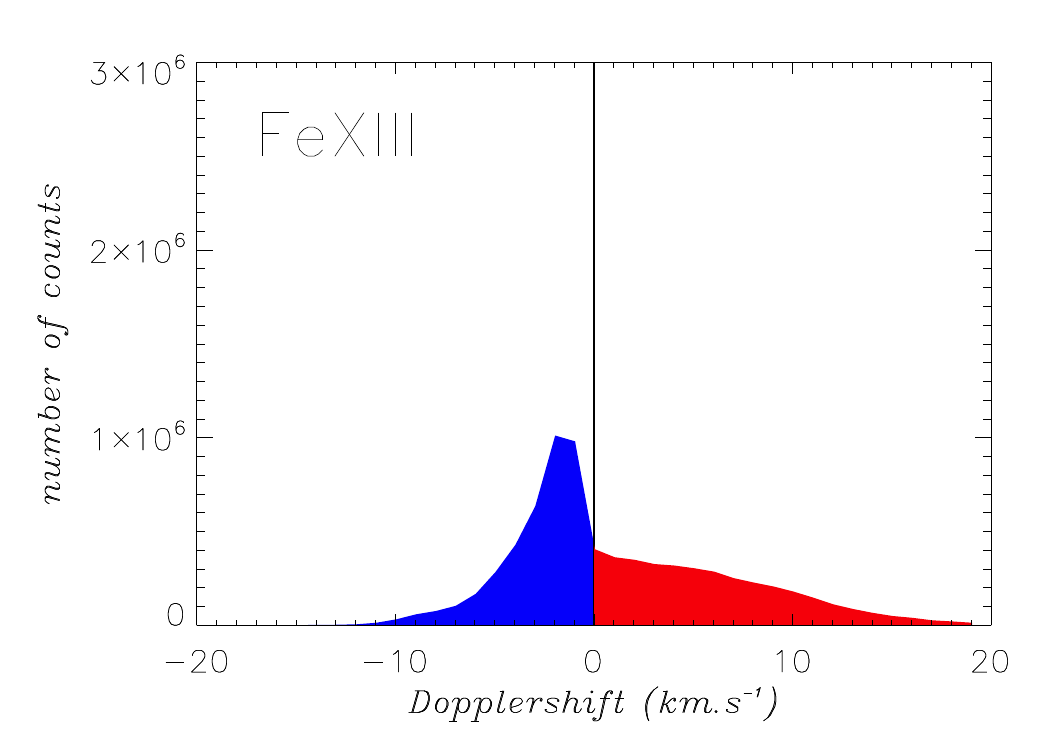}}
&
\includegraphics[width=0.18\linewidth, bb= 10 0 560 360]
	{\mydir{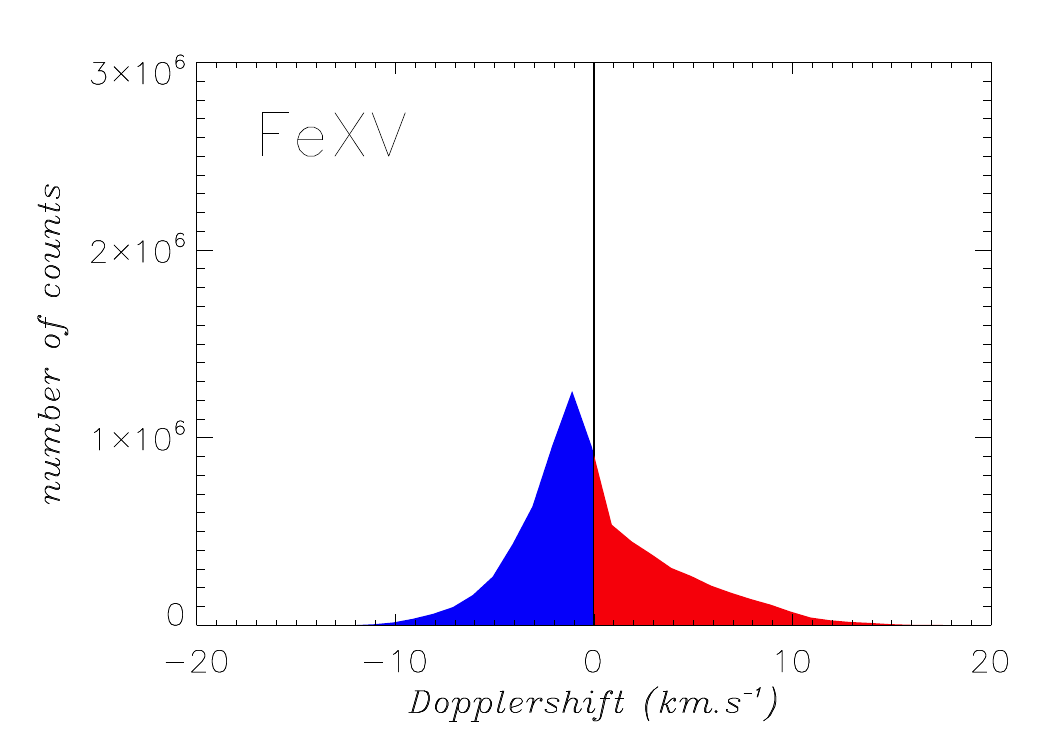}}
&
\includegraphics[width=0.18\linewidth, bb= 10 0 560 360]
	{\mydir{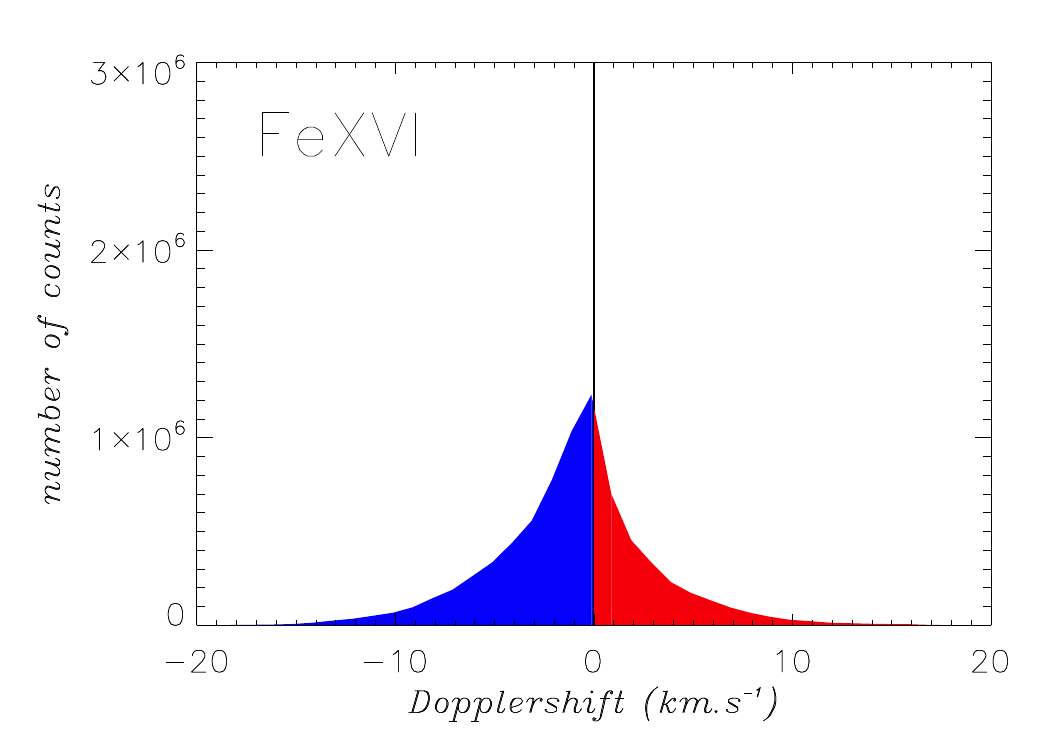}}
&
\includegraphics[width=0.18\linewidth, bb= 10 0 560 360]
	{\mydir{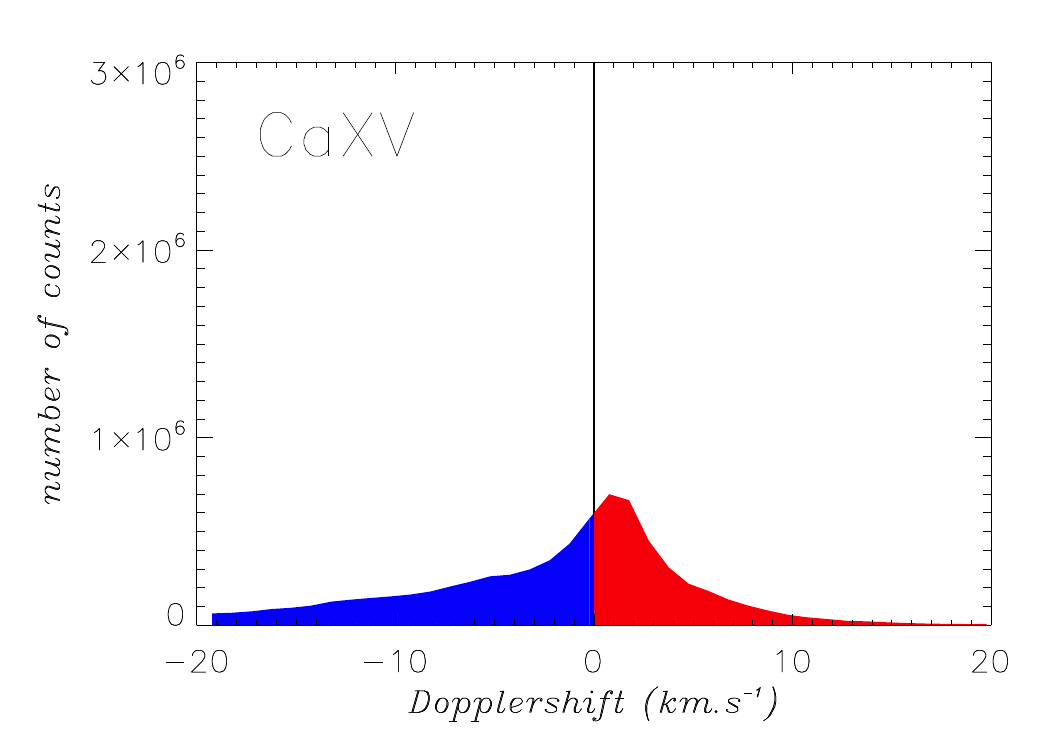}}
&
\includegraphics[width=0.18\linewidth, bb= 10 0 560 360]
	{\mydir{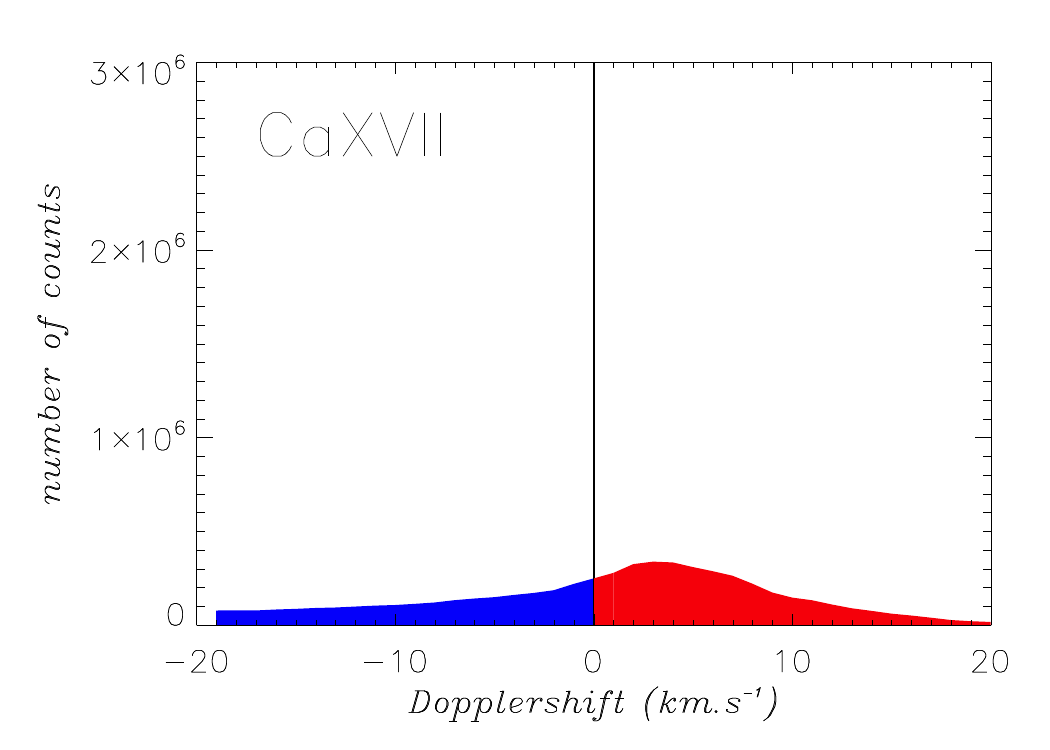}}
\end{tabular}
\caption{Same as Figure~\ref{fig:distr_angle20} for an angle of 70$^\circ$ 
with respect to the vertical direction.}
\label{fig:distr_angle70}
\end{figure}

In Section~\ref{sec:doppler}, we studied the distributions of Dopplershifts when
looking at the loop from the top. We now study the distributions when the
observer's direction is making an angle of 20$^\circ$, 45$^\circ$ and 70$^\circ$
with respect to the vertical direction (see Figure~\ref{fig:setuploop}). In Figures~\ref{fig:distr_angle20},
~\ref{fig:distr_angle45} and~\ref{fig:distr_angle70}, we plot the Dopplershifts statistical distributions for the ten
selected spectral lines (see Table~\ref{tab:lines}). 

%\end{document}

Comparing the distributions of Dopplershifts for the different viewing angles,
we clearly see the increase of blueshifts for spectral lines with increasing
peak temperature for angles between 0$^{\circ}$ and 45$^{\circ}$, while this
behaviour is not obvious for a viewing angle of 70$^{\circ}$. The peak of the
distribution of blueshifts is shifted towards larger values when the angle
increases: for instance, from about 0 to -2 \kms\ for the Fe {\sc xii} line.
Whilst the distribution of redshifts is flattened to reach a more uniform
velocity distribution. 

Refering to the schematic Gaussian distributions of Figure~\ref{fig:vel_distrib}, the 20$^\circ$ case is similar to the 0$^\circ$ case exhibiting a double-peak distribution which is redshifted for cooler lines and blueshifted for hotter lines. As discussed in Section~\ref{sec:doppler}, these double-peak distributions are characteristic of plasma condensation and evaporation. For the 45$^\circ$ and 70$^\circ$ cases, the distributions are often double peaked (see for instance the distribution for Fe {\sc xii} at a 70$^\circ$ viewing angle) with both a blueshifted peak and a redshifted peak. This scenario is different from the one discussed in Section~\ref{sec:doppler} and appears when both plasma condensation and evaporation are at play in the integrated line-of-sight.  

%In Figure~\ref{fig:y} top, we plot the percentage of blueshifts and redshifts
%regardless of their actual values for the three different viewing angles. As for
%the distribution, there is a clear increase of the amount of blueshift with the
%peak temperature of spectral lines for viewing angles between 0$^{\circ}$ and
%45$^{\circ}$, while the amount of Dopplershifts is getting more and more evenly
%distributed with the percentage of blueshift and redshift being around 50\% for
%all the spectral lines.  

\begin{figure*}[!ht]
\centering
\begin{tabular}{ccc}
\multicolumn{1}{l}{(a)} & \multicolumn{1}{l}{(b)} & \multicolumn{1}{l}{(c)} \\
\includegraphics[width=0.32\linewidth]
	{\mydir{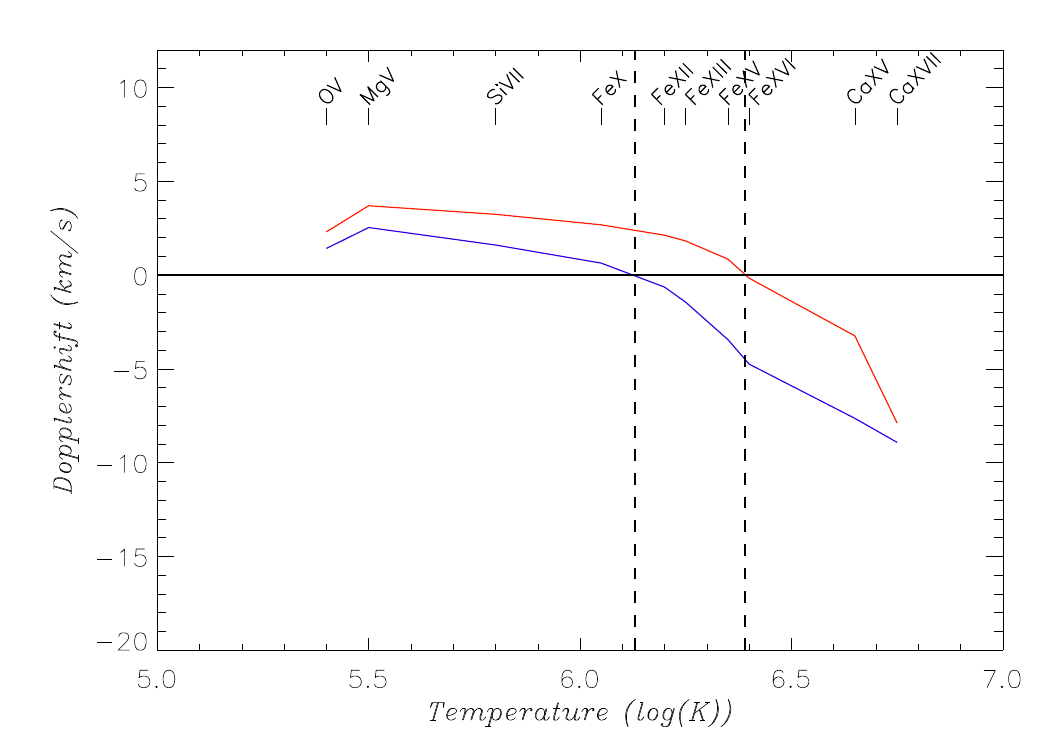}}
& 
\includegraphics[width=0.32\linewidth]
	{\mydir{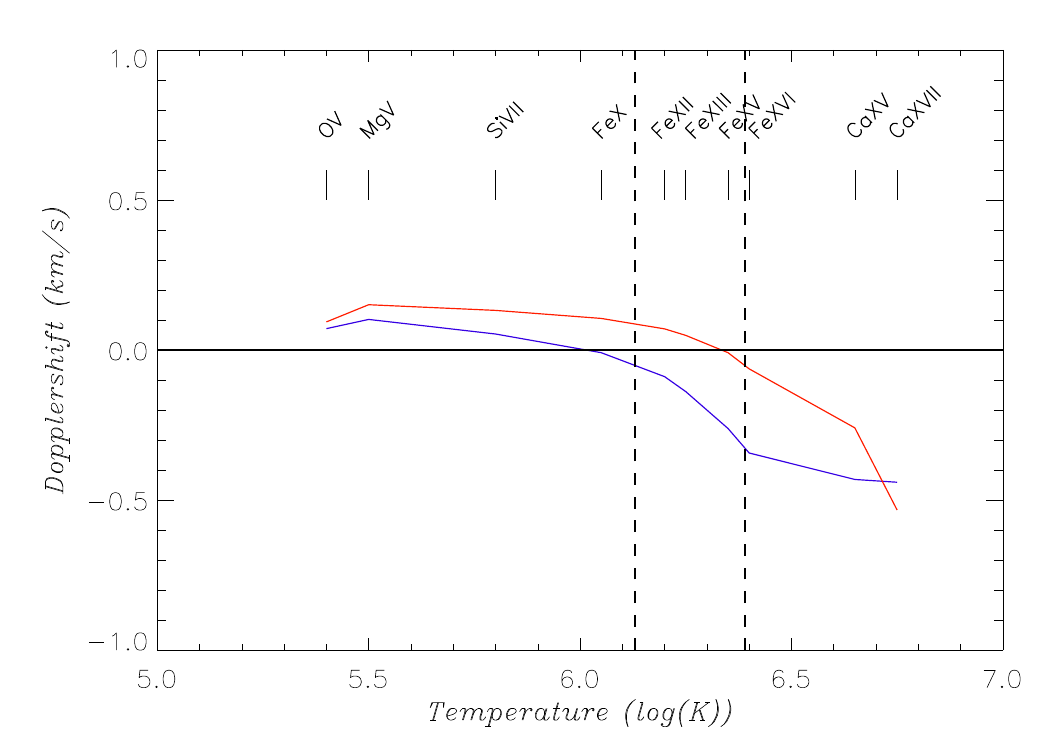}}
& 
\includegraphics[width=0.32\linewidth]
	{\mydir{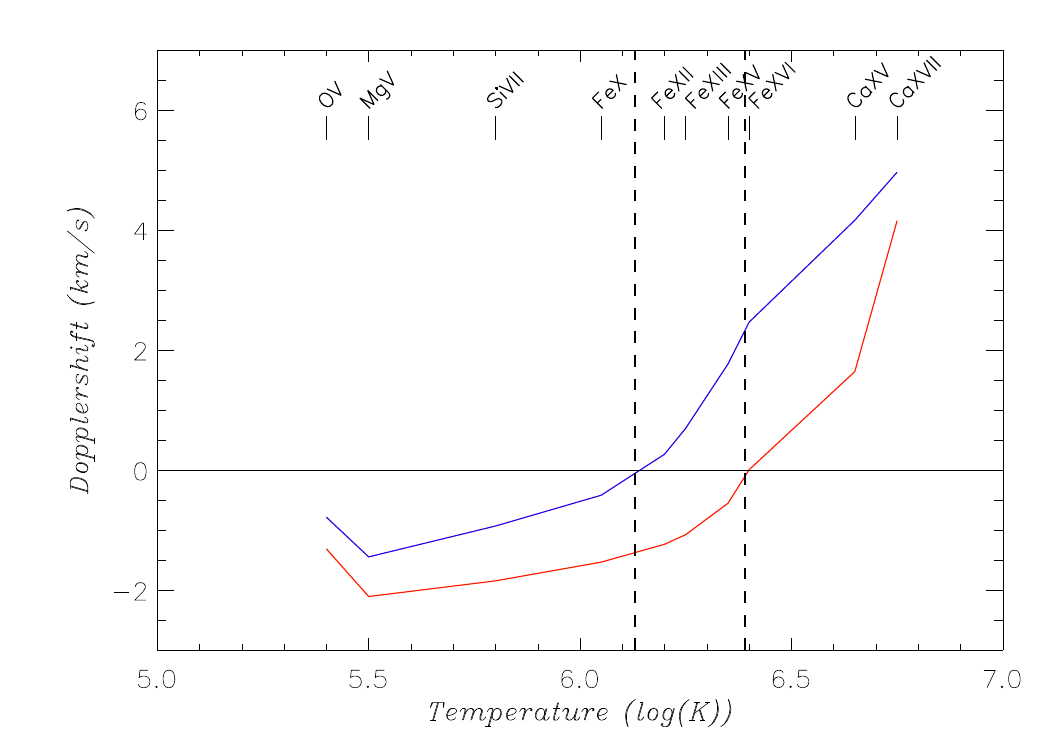}}
\end{tabular}
\caption{Temperature diagnostic from the average Dopplershifts at the footpoints of {\it Loop {\sc i}} (blue) and {\it Loop {\sc ii}} (red) at different viewing angle: (a) 20$^{\circ}$, 
(b) 45$^{\circ}$ and (c) 70$^{\circ}$.}
\label{fig:y}
\end{figure*}

In Figure~\ref{fig:y}(a)-(c), we test our temperature against the change in
viewing angle. We thus plot the average Dopplershift for a given spectral line
or temperature. The temperature diagnostic that we have defined states that  a
vanishing average Dopplershift marks the temperature of the modelled coronal
loop. The average temperature of the modelled loops is indicated by the dash
lines. For a viewing angle of 20$^{\circ}$, the curves of average Dopplershift
are similar to those for a viewing angle of 0$^{\circ}$ (see
Figure~\ref{fig:temp_diag} left). The temperature diagnostic at 20$^{\circ}$
provides the same temperature as the diagnostic at 0$^{\circ}$. For a viewing
angle of 45$^{\circ}$, the average Dopplershift has been significantly reduced
even if the curve has kept the same shape. The average Dopplershift is below 0.5
\kms\ in absolute value. In addition the vanishing average Dopplershift is for a
temperature of 1.25 MK for {\it Loop {\sc i}} and  2.24 MK for {\it Loop {\sc ii}}. This
implies a change of about 8\% in the estimated temperature of the loops. For a
viewing angle of 70$^{\circ}$, the curves of average Dopplershift are reversed
with a dominant blueshift velocity for the low peak temperature spectral lines.
Nevertheless, the temperature diagnostic at 70$^{\circ}$ is similar to the
diagnostic at 0$^{\circ}$. 

From this analysis of the behaviour of Dopplershifts as a function of the
viewing angle, we conclude that our temperature diagnostic is robust for a wide
range of viewing angles; however, a moderate viewing angle (like 45$^{\circ}$)
leads to an increase of the errors on the estimated temperature. On the
observational point-of-view, we note that the average Dopplershift for a viewing
angle of 45$^{\circ}$ is small ($<$ 0.5 \kms), and thus difficult to
observe with the current instrumentation.   

\end{document}